\documentclass[a4paper,11pt]{article}

\usepackage{graphicx}  
\usepackage{dcolumn}   
\usepackage{bm,relsize}        
\usepackage{amssymb, amsmath}
\usepackage{textcomp}
\usepackage{wasysym}
\usepackage{slashed}
\usepackage{caption, subcaption}
\usepackage{multirow}
\usepackage{gensymb}
\usepackage{subcaption}
\usepackage{colortbl}
\usepackage{booktabs}
\usepackage{tabularx}
\definecolor{headergray}{gray}{0.9}
\definecolor{rowgray}{gray}{0.97}

\usepackage[nolist, nohyperlinks]{acronym}

\usepackage{lipsum, color}
\usepackage[dvipsnames,svgnames,table,x11names]{xcolor}
\usepackage{braket}
\usepackage{jheppub} 
\usepackage{booktabs}
\usepackage{pdflscape}
\usepackage[utf8]{inputenc} 

\usepackage{booktabs}
\usepackage{tabularx}

\usepackage{mathrsfs}
\usepackage{bm,amssymb,slashed,graphicx,multirow,soul,mathtools,xspace,array,tikz,amsmath, gensymb}
\usepackage{siunitx} 
\usepackage{float}   
\usepackage{cancel}
\allowdisplaybreaks
\usepackage{ bbold }
\usepackage{caption,subcaption}
\usepackage{hyperref}
\usepackage{colortbl}
\usepackage{tcolorbox}

\usepackage{comment}

\usepackage[capitalise, english]{cleveref}

\definecolor{nicered}{rgb}{0.7,0.1,0.1}
\definecolor{nicegreen}{rgb}{0.1,0.5,0.1}
\definecolor{violet}{rgb}{0.7,0.3,0.3}
\hypersetup{colorlinks,citecolor= nicegreen,linkcolor= nicered}

\newcommand{\lp}{\left(}
\newcommand{\rp}{\right)}

\newcommand{\g}{\gamma}
\newcommand{\be}{\begin{equation}}
\newcommand{\ee}{\end{equation}}

\newcommand{\eventname}{KM3-230213A}
\newcommand{\kmn}{KM3NeT}
\newcommand{\ic}{IceCube}

\newcommand{\TeV}{\si{\tera\electronvolt}}

\newcommand{\PeV}{\si{\peta\electronvolt}}

\newcommand{\dd}[0]{\mathrm{d}}

\newcommand{\beq}{\begin{equation} }
\newcommand{\eeq}{\end{equation}} 
\newcommand{\bi}{\begin{itemize} }
\newcommand{\ei}{\end{itemize} }
\newcommand{\EeV}{\mathrm{EeV}}
\newcommand{\TT}{\mathcal{T}}

\newcommand{\deriv}[2]{\frac{\partial #1}{\partial #2}}

\definecolor{Red}{rgb}{1.,0.,0.}
\definecolor{Grn}{rgb}{0.,0.75,0.}
\definecolor{Blu}{rgb}{0.,0.,1.}
\definecolor{Pink}{rgb}{1,0.08,0.58}

\usepackage[T1]{fontenc} 

\usepackage{amsmath,amssymb,epsfig,color,slashed}
\allowdisplaybreaks  

\newcommand{\gcm}{{\rm g}/{\rm cm}^3} 

\newcommand{\km}{\si{\kilo\metre}}

\newcommand{\py}{\mathcal{P}(y)}

\newcommand{\rev}[1]{#1}

\newcommand{\revnew}[1]{#1}

\begin{document} 


\title{Exploring ultra-high energy neutrino experiments through the lens of the transport equation}

\author[a,b,c]{Stefano Palmisano,}
\author[c]{Diego Redigolo,}
\author[a,c]{Michele Tammaro,}
\author[c]{Andrea Tesi}
\affiliation[a]{Galileo Galilei Institute for Theoretical Physics, Largo Enrico Fermi 2, I-50125 Firenze, Italy}
\affiliation[b]{Università degli Studi La Sapienza, Piazzale Aldo Moro 5, 00185 Roma, Italy}
\affiliation[c]{INFN Sezione di Firenze, Via G. Sansone 1, I-50019 Sesto Fiorentino, Italy}

\emailAdd{stefano.palmisano@uniroma1.it}
\emailAdd{diego.redigolo@fi.infn.it}
\emailAdd{michele.tammaro@fi.infn.it}
\emailAdd{andrea.tesi@fi.infn.it}

\date{\today}

\preprint{}

\abstract{We develop a first-principles formalism, based on the transport equation in the line-of-sight approximation, to link the expected number of muons at neutrino telescopes to the flux of neutrinos at the Earth's surface. We compute the distribution of muons inside Earth, arising from the up-scattering of neutrinos close to the detector, as well as from the decay of taus produced farther away. This framework allows one to account for systematic uncertainties, as well as to clarify the assumptions behind definitions commonly used in the literature, such as the effective area. We apply this formalism to analyze the high-energy muon event recorded by KM3NeT, with a reconstructed energy of $ 120^{+110}_{-60} \, \mathrm{PeV}$ and an elevation angle of $\left(0.54\pm 2.4\right)\degree$, in comparison with the non-observation of similar events by IceCube. We find a $3.1\,\sigma$ tension between the two experiments, assuming a diffuse neutrino source with a power-law energy dependence. Combining both datasets leads to a preference for a very low number of expected events at KM3NeT, in stark contrast to the observed data. The tension increases both in the case of a diffuse source peaking at the KM3NeT energy and of a steady point source, whereas a transient source may reduce the tension down to $1.6\,\sigma$. The formalism allows one to treat potential beyond-the-Standard-Model sources of muons, and we speculate on this possibility to explain the tension.}

\maketitle

\newpage

\section{Introduction}
\label{sec:intro}
Neutrinos provide the cleanest observation channel for extragalactic ultra-high energy (UHE) astrophysical processes, primarily due to their extremely weak interactions with ordinary matter. This property allows neutrinos produced in distant environments to propagate undisturbed on intergalactic scales, whereas other Standard Model (SM) particles, such as high-energy photons or charged fermions, interact with the cosmic background.

UHE neutrino telescopes ($\nu$T) aim to detect UHE neutrinos, of either cosmic or astrophysical origin, by observing the passage of associated charged particles, such as leptons produced from charged current (CC) interactions or hadronic showers produced in neutral current (NC) scatterings. Currently, three $\nu$Ts are operational. IceCube~\cite{IceCube:2016zyt}, located at the South Pole, has been running continuously for nearly 14 years. KM3NeT~\cite{Margiotta:2014gza}, under construction in the Mediterranean Sea, is composed of two modules~\cite{KM3Net:2016zxf}: ARCA, dedicated to UHE $\nu$ observations, and ORCA, optimized for studying neutrinos produced in cosmic ray showers. KM3NeT has already collected almost 3 years of data, albeit with a reduced detector volume. Baikal-GVD~\cite{Baikal-GVD:2023beh}, under construction in Lake Baikal, Russia, shares a similar setup to KM3NeT/ARCA, and has been collecting data since 2018. All these experiments are structured around the same principle: strings of optical modules are immersed in water or ice to detect the Cherenkov light emitted by high-energy charged particles.

IceCube has measured over 160 muon neutrino events with energies above 20 TeV, with the highest energy event recorded at $E_\nu = 4.8$ PeV~\cite{IceCube:2016umi}. These events are well-fitted by a diffuse flux with a single power-law over a broad energy range, from $10\,\TeV$ to a few $\PeV$~\cite{IceCube:2013low, IceCube:2020acn, Abbasi:2021qfz, IceCube:2023sov,Silva:2023wol, IceCube:2024fxo}. Additionally, IceCube has found evidence for point sources (PSs) of neutrinos from a nearby galaxy~\cite{IceCube:2019cia, IceCube:2022der}, and recorded an event of electron neutrino with energy of roughly 6 PeV~\cite{IceCube:2021rpz}. No events consistent with being originated by tau neutrinos have been observed at these energies.

On February 13th of 2023, at 01:16:47 UTC, ARCA recorded the passage of a charged particle, consistent with a muon produced in the surroundings of the detector, with energy $E_\mu = 120_{-60}^{+110}\PeV$, where uncertainties correspond to a $68\%$ confidence interval. This event has been dubbed \eventname{} \cite{KM3NeT:2025npi},
\rev{and is located in the region}
indicated by the red circle in \cref{fig:Sources}. The \kmn{} Collaboration has performed a search for a point-like source in the region, but no catalogued source was found in the 99\% confidence interval region~\cite{KM3NeT:2025npi,KM3NeT:2025ccp,KM3NeT:2025vut,KM3NeT:2025aps,KM3NeT:2025bxl}. The reconstructed neutrino energy is $E_\nu = 220~ \PeV$,\footnote{Note that the reconstructed neutrino energy can depend non-trivially on the assumed prior for the neutrino flux energy spectrum. As a reference, we report here the central value inferred in Ref.~\cite{KM3NeT:2025npi} by assuming a diffuse flux \revnew{with a quadratically decaying spectrum}.} constituting the detection of the most energetic fundamental particle ever observed. This measurement presents a clear challenge to the IceCube results. First, IceCube has been taking data for a significantly longer period, with an exposure approximately $15$ times larger at the time of the event observation. Second, the reconstructed source of this neutrino is located in a region of the sky \rev{from which neutrinos are less attenuated at IceCube than at KM3NeT, as can be seen in \cref{fig:Galactic:transparency}.}

\begin{figure}[t]
    \begin{center}
    \includegraphics[width=\textwidth]{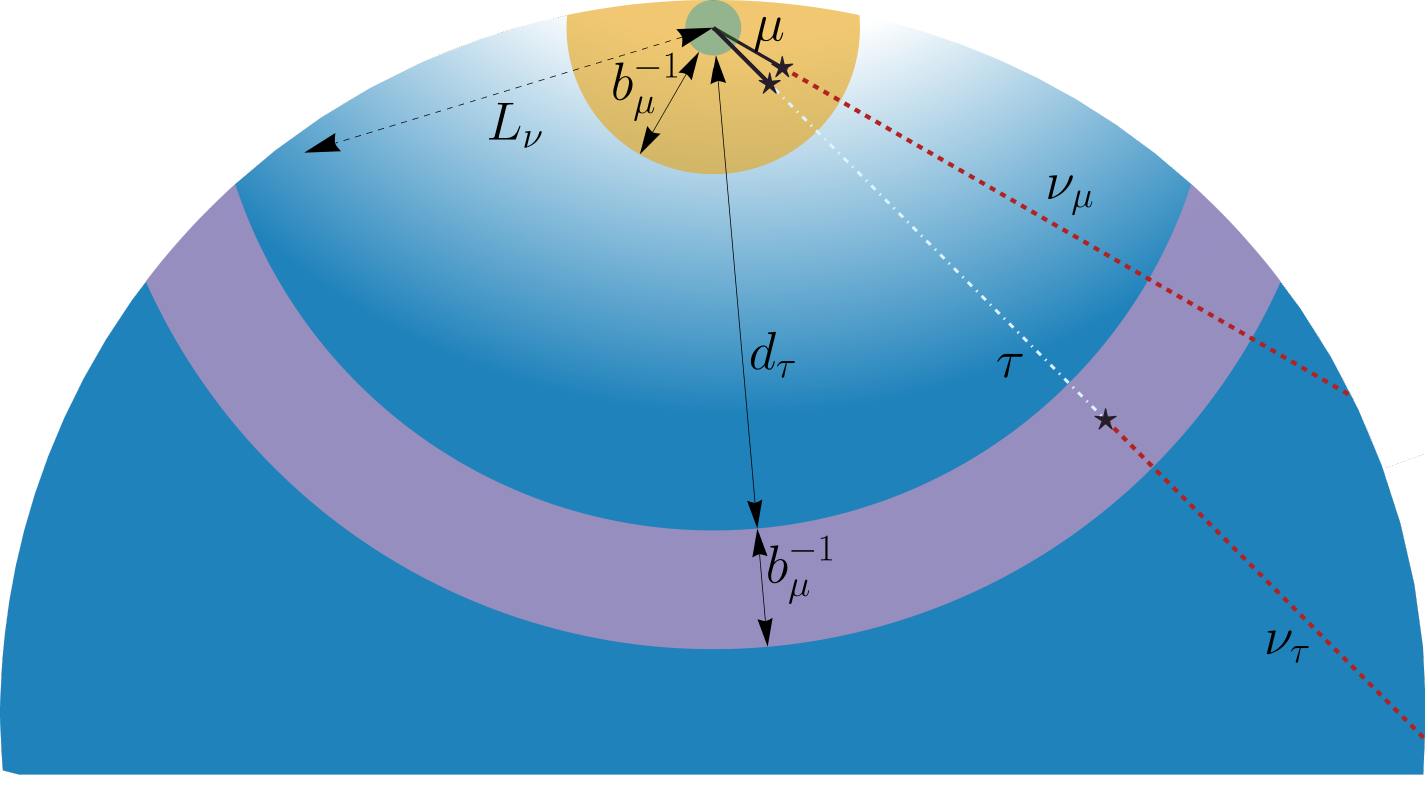}
    \caption{\label{fig:sketch-earth}Muons can be produced either from up-scatterings at distances of order $b_\mu^{-1}$ (brown region) from the detector (in green), or from the decay of particles produced in a shell of radius of order their decay length, whose thickness is again $b_\mu^{-1}$ (purple region). In the SM these two cases are represented by muon neutrinos and tau leptons, as shown in the figure. Note that the various length scales are out of scale. We also show the typical attenuation distance $L_\nu$ of $100\,\PeV$ neutrinos. The shading of the filling gives a hint of the values of $f_\mu$, building up through Earth in competition with the neutrino attenuation, and reaching an approximately constant value in the vicinity of the detector.}
           \end{center}
\end{figure}

In this work, we analyze the status of UHE neutrino measurements in light of the recent KM3NeT result. We take the approach of deriving the expected number of events from first principles, using the powerful tool of Boltzmann (transport) equations for muon production inside the Earth, \rev{which allows for an analytical description of the physics, from the impinging of the neutrinos on the surface to the detection of the muons. As described in the following, we employ several approximations that can, however, be relaxed at the cost of increased complication of the numerical computation, making our treatment ideally only short of a full quantum field theory computation.}


\rev{Similar approaches already exist in the literature, see ~\cite{Beacom_2002,Dutta_2002,Dutta_2005}. In this work we improve on these treatments by considering the set of full Boltzmann equations, namely tracking the distribution of muons (and other particles) in three-dimensional phase-space. A feature of this approach is the ability to fix the overall normalization of the number of events to the integral over the full phase-space of such functions (see \cref{eq:numero}).}

As we discuss in great detail in the rest of the article, the ultimate goal for making predictions for $\nu$T is the computation of the muon distribution at the location of the detector, $f_\mu$, through which a prediction for the expected number of events differential in time, energy, and direction can be made as 
\begin{equation}
\left\langle \frac{\mathrm{d}N}{\mathrm{d}t \mathrm{d}E_\mu \mathrm{d}\Omega_p}\right\rangle \sim \left(\text{area}\right)\times\left(\text{muon energy}\right)^2 \times \left(\text{velocity}\right) \times f_\mu\big|_{\text{detector}} \,. 
\end{equation}
In the SM, muons can be produced either by the up-scattering of muon neutrinos or by the decay of tau leptons, which are in turn produced by up-scatterings of tau neutrinos. They may either be produced inside the detector 
or in its vicinity
. Since muons lose energy as they traverse Earth, they cannot be produced too far away from the detector: the stopping length $b_\mu^{-1}$ determines the effective radius, and thus the effective volume, of the detector. If muons are produced through scattering close to the detector, their production is sensitive to the density of the material surrounding the detector. If they are produced through decays of long-lived particles such as taus, the production of the latter depends on the density of Earth at distances of the order of their boosted decay length from the detector. As a consequence, the thin shell of $\mathcal{O}(b_\mu^{-1})$ in which muons may be produced is transferred at farther distances from the detector. \cref{fig:sketch-earth} summarizes this picture and all the length scales discussed above, as well as the attenuation length of the neutrinos, $L_\nu$ (which is related to their mean free path). 

In the language of transport equations, the distributions are sourced by several other distributions through collision terms. As summarized in \cref{fig:Distributions}, in order to compute the muon distribution, one must supply the distribution of muon neutrinos and of taus inside Earth. The former is directly linked to the astrophysical flux of neutrinos, sourcing neutrinos at the surface of Earth that undergo attenuation at larger depths inside the planet. The latter is instead sourced by the (tau) neutrino distribution itself. The production of muons and taus from neutrinos is controlled by the density of materials as described above, and by the neutrino-nucleon deep inelastic scattering (DIS) cross section, while the production of muons from tau decays is controlled by the boosted decay rate of the tau.

This method offers several advantages: \textit{i)} it allows one to directly compute the observed number of muon events at the detector; \textit{ii)} it allows one to identify the sources of systematic uncertainties, providing a quick way to assess the impact of nuisance parameters; \textit{iii)} with further simplifying assumptions our method allows for an analytical approximation which captures the main physics effects; \textit{iv)} it allows for a direct inclusion of new physics scenarios.

We use such a method to quantify the alleged tension between \ic{} data and the \kmn{} event when compared within a given model for the neutrino flux impinging on the crust of Earth. After all, the only free parameters are the ones controlling the neutrino flux magnitude, directionality and spectral shape. We investigate several models for the neutrino flux, allowing various flavour compositions. Our results are summarized in \cref{tab:anticipation-summary}, where we collect the main outcomes of Bayesian inference on different classes of models, and the respective tension between IceCube and KM3Net.

\begin{table}[!ht]
\begin{center}
\begin{tabular}{l c c }
\toprule
\textbf{Model} &  \textbf{Bayes factor} & $\Delta\sigma$  \\
\midrule
\rowcolor{white}
Diffuse (power-law) & 21 & $3.1$ \\
\midrule

PS & $\approx 35$ &$3.8$  \\

PS (transient) &  $\approx 2.2$ & $1.6$ \\

Diffuse (energy localized) & $\approx 3.4$ & $2.4$ \\
\bottomrule
\end{tabular}
\caption{\label{tab:anticipation-summary} Summary of the results of the statistical analyses of \cref{sec:stat}. We find a tension compatible with that claimed by \cite{titans} for the case of a diffuse power-law flux. The tension increases when one considers PSs, unless transient sources are considered, in which case the tension can be completely reabsorbed down to $1.6\,\sigma$ (see footnote in \cref{sec:stat}).}
\end{center}
\end{table}

The rest of this article is organized as follows.
In \cref{sec:TransportEquation}, we explain our approach in detail, illustrating the interplay between the incoming neutrino flux and the muons produced directly through CC deep inelastic scattering (DIS) or indirectly through tau decays after the taus have been produced through CC DIS. In \cref{sec:TheoryInputs} we go through the physics inputs that enter in the transport equation and their uncertainties. \cref{sec:TheoryOutputs,sec:Results} contain the main results of our work. In \cref{sec:TheoryOutputs} we define observables in terms of expectation values on the muon distribution, and we provide master formulae for predictions at $\nu$T experiments.  In \cref{sec:Results} we consider both the case of a diffuse power-law flux as well as that of a localized PS to assess the incompatibility between the \kmn{} event and \ic{} results. In the latter case, we further discuss the possibility of a flux dominated by tau neutrinos, and one dominated by beyond the Standard Model (BSM) particles.

\section{Transport equation}
\label{sec:TransportEquation}
In this section we aim to compute the rate of muon events in a $\nu$T. We are interested in the differential number of muons in energy and solid angle, recorded by the experiment over its full data-taking window. It has dimensions of
\begin{equation}\label{experimental-rate}
\left\langle    \frac{{\rm d}N_\mu}{{\rm d}t\,{\rm d}E_\mu\,{\rm d}\Omega_p} \right\rangle =\frac{\text{number of muons}}{\text{time}\times \text{energy}\times\text{solid angle}}\,.
\end{equation}
This quantity also captures phenomena that are transient in time, such as astrophysical sources active in a given time interval. As we shall discuss, the muon rate is sourced by an astrophysical neutrino flux $\phi^\oplus_\nu$ at the surface of Earth, with dimensions of
\begin{equation}\label{eq:nuflux}
\phi^\oplus_\nu(E_\nu,\Omega_\nu,t) =\frac{\text{number of neutrinos}}{\text{time}\times\text{area}\times\text{energy}\times\text{solid angle}}\,.
\end{equation}
The neutrino flux depends on energy, direction, and time for transient sources. Unless explicitly stated otherwise, we assume an equal flux of the three neutrino flavors, as expected for distant high-energy neutrino sources~\cite{Athar:2000yw}. Since neutrinos and anti-neutrinos are indistinguishable in UHE experiments, we focus on the neutrino flux, assuming equal populations of their anti-particles.

Our approach is to compute \cref{experimental-rate} from first principles, by writing down the relevant transport equation for muons, which takes as input a flux of neutrinos at the surface of the Earth. With this initial setup, the muons arriving at the detector can be produced in the SM via: 
\begin{enumerate}
    \item \rev{scattering of ${\nu_\mu}$ off nuclei},
    \item \rev{decay ${\tau}$ leptons,  produced in turn from $\nu_\tau$ scattering off nuclei}.
\end{enumerate}
As becomes apparent from the discussion in the next sections, one can account for muons produced by the decay of beyond-the-Standard-Model (BSM) particles produced in neutrino scatterings in a similar fashion.
Additionally, scattering of $\nu_e$ will produce electrons; we assume that these events are completely distinguishable from muon ones, as we also discuss in \cref{sec:EnergyLoss}.

\subsection{The muon distribution function}\label{sec:muonprod}
We compute observables by averaging over the phase space distribution of muons, $f_\mu(t,\vec{x},\vec{p})$, with the appropriate weights. This distribution depends on the time, position and momentum of the muon itself, and encodes statistical information that allows us to calculate expectation values of quantities relevant for the experiments. Throughout this work, $f_\mu$ is normalized so that its integral over momentum $\vec p$ and volume $V$ gives the instantaneous total number of muons
\begin{equation}\label{eq:numero}
\left\langle N_\mu \right\rangle = \int\limits_V  d^3x \int \frac{d^3p}{(2\pi)^3} f_\mu(t, \vec{x}, \vec{p})\,.    
\end{equation}
With this normalization, the distribution $f_\mu$ is determined by solving the corresponding transport equation, which governs the muon distribution in the presence of sources. These sources can be classified into two types: i) production via scattering of particles in the Earth medium on their way to the detector, ii) production via decays of particles. The particles involved in these processes are either flux particles or those produced from scattering or decay of flux particles.

The full transport equation for the muon distribution can be written as
\beq
\begin{aligned}\label{eq:transport}
& \deriv{f_\mu}{t} + \dot{\vec{x}}\cdot\deriv{f_\mu}{\vec{x}} + \dot{\vec{p}}_\mu\cdot\deriv{f_\mu}{\vec{p}_\mu} = K_{\rm{scattering}}^{\nu_\mu}(t, \vec{x},\vec{p}_\mu)+K_{\rm{decay}}^{\tau}(t, \vec{x},\vec{p}_\mu)\,,
\end{aligned}
\eeq
where the dot indicates the time derivative, and the two kernels $K_{\rm{scattering}}^{\nu_\mu}$ and $K_{\rm{decay}}^{\tau}$ are the sources mentioned above.  Knowledge of the source terms allows one to solve this equation.  In writing \cref{eq:transport} we have neglected muon conversion into neutrinos via scattering off nuclei, as well as muon NC interactions, which are subdominant processes, being suppressed by $f_\mu$. This approximation makes \cref{eq:transport} independent of $f_\mu$, which calls for a simple solution to the equation.

Let us now discuss the form of the kernel functions. They can be computed from first principles, as detailed in \cref{app:boltzmann}, knowing the fundamental interactions. In general, their expressions can be put in the following form
\beq \label{eq:kernel-decay}
K_{\rm{scattering}}^a(t, \vec x, \vec p_b) \equiv\int \dd^3 p_a f_a(t, \vec x, \vec p_a) \, n_N(\vec x)  v\,   \frac{\dd\sigma\left(a\,N \to b \, \ldots\right)}{\dd^3p_b}\,,
\eeq
\beq\label{eq:kernel-scattering}
K_{\rm{decay}}^a(t, \vec x, \vec p_b)\equiv\int \dd^3 p_a f_a(t, \vec x, \vec p_a)    \frac{\dd\Gamma \left( a\to b\,\ldots\right)}{\dd^3p_b}\,.
\eeq
Here we remained as generic as possible for the scattering of $a$ off targets into $b$ (or decay of $a $ into $b$). For example, $a$ refers to particles that, like muon neutrinos, produce muons through DIS scattering and $b$ stands for particles, like muons, in the final states. The target $N$ is always constituted by nucleons in this work, and $n_N \sigma v$ and $\Gamma$ are the scattering and boosted decay rates. Note that, with a slight abuse of notation, $K^a_{\text{decay}}(\vec p_a)$ is the total boosted decay rate times the distribution of the decaying particle.
A sum over all particles and processes involved is intended. 
The boundary condition for this equation is that at all times, and for all energies, the muon distribution on the Earth's crust vanishes
\beq\label{eq:boundary-condition}
f_\mu(t,\vec{x}\in \oplus,\vec{p}_\mu) = 0 \quad\text{(at any $t$ and for any }\vec{p}_\mu\text{)}\,.
\eeq

In a pure SM picture\rev{, as per \cref{eq:transport}},
we need to use the muon/tau-neutrino distribution $f_{\nu_{\mu/\tau}}$ and the $\tau$ distribution $f_\tau$ in \cref{eq:kernel-scattering,eq:kernel-decay}, respectively. They satisfy their own transport equations, which are coupled by weak interactions. The system of equations is similar to the one for muons and can be written in general as 
\begin{equation}\label{eq:system-transport-full}
\left\{\begin{aligned} 
\deriv{f_{\nu_\mu}}{t} + \dot{\vec{x}}\cdot\deriv{f_{\nu_{\mu}}}{\vec{x}} + \dot{\vec{p}}_{\nu_\mu}\cdot\deriv{f_{\nu_\mu}}{\vec{p}_{\nu_\mu}}  &= C_{\rm flavor-ind.}(t,\vec x, \vec{p}_{\nu_\mu}) + K_{\rm{decay}}^\tau(t, \vec x, \vec p_{\nu_{\mu}})\,,\\
\deriv{f_{\nu_\tau}}{t} + \dot{\vec{x}}\cdot\deriv{f_{\nu_{\tau}}}{\vec{x}} + \dot{\vec{p}}_{\nu_\tau}\cdot\deriv{f_{\nu_\tau}}{\vec{p}_{\nu_\tau}} &= C_{\rm flavor-ind.}(t,\vec x, \vec{p}_{\nu_\tau})+ K_{\rm{decay}}^\tau(t, \vec x, \vec p_{\nu_{\tau}}) \,,\\
\deriv{f_{\tau}}{t}+\dot{\vec{x}}\cdot\deriv{f_\tau}{\vec{x}} + \dot{\vec{p}}_\tau\cdot\deriv{f_\tau}{\vec{p}_\tau} &= K_{\rm{scattering}}^{\nu_\tau}(t,\vec x, \vec p_\tau) - \Gamma_\tau(\vec p_\tau) f_\tau \,.
 \end{aligned}\right. 
\end{equation} 
The collisional term $C_{\rm flavor-ind.}$ corresponds to the flavor universal contributions coming from CC and NC interactions, which is discussed in greater detail in \cref{eq:neutrino-att}, see for example \cite{Vincent:2017svp,Safa:2019ege}.
Recall that the cross section contributing to the neutrino energy losses receives contributions from both charged and NC while the production of charged leptons 
goes through CC only. It is important to notice that the coupled transport equations for tau leptons and neutrinos can be solved prior to their use in \cref{eq:transport}, so that $f_{\nu_\mu}$ and $f_\tau$ can be taken as inputs for the purpose of computing the muon distribution.  \cref{eq:system-transport-full} describes a coupled and closed system of transport equations, that can be solved with an increasing level of sophistication.

\begin{figure}[t]
    \begin{center}
    \includegraphics[width=0.8\linewidth]{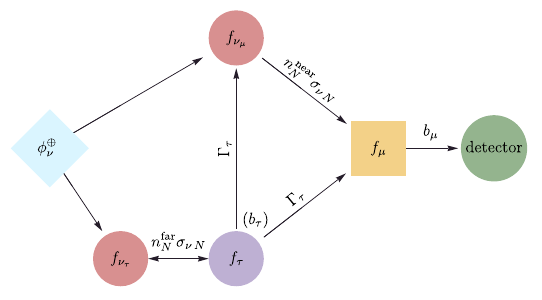}
    \caption{    \label{fig:Distributions}Visual representation of the system of transport equations in \cref{eq:system-transport-full}. The edges are labelled with the inverse time or length scales relevant for the processes linking the various distributions, which are discussed in \cref{sec:TheoryInputs}.
    }
    \end{center}

\end{figure}

The interplay of these distributions and the relevant rates is sketched in \cref{fig:Distributions}. The main ingredients needed for the calculation are discussed in \cref{sec:TheoryInputs} while approximated analytical solutions for the relevant observables are derived in \cref{sec:TheoryOutputs}. Before diving into the specifics, let us discuss the general regime in which we are solving \cref{eq:transport}. Of interest in this study are muons of very high energy, for which a number of simplifications occurs in the Boltzmann system. 

As we discuss in \cref{sec:TheoryInputs} the mean free path associated to electroweak (EW) interactions is always much longer than the rate of energy losses due to the different quantum electrodynamics (QED) processes (see \cref{fig:RelevantScales} for a summary of the relevant scales). As a consequence, in one EW interaction length we can average over many QED interactions and account for their effect on the charged particle trajectory in the medium as a background effect. \revnew{We comment upon this approximation in the next section.} 

Concretely, for muons and taus, the term proportional to $\dot{\vec{p}}$ in \cref{eq:transport} should account for the average energy losses in the medium (and therefore it is not included as a source term in the right-hand side). Since the energy loss is dominated by QED process with low momentum exchange, we can always write $\vec{p}_\ell=E_\ell\hat{n}_\ell$ and assume that the charged lepton direction does not change much with time, $\dot{\hat{n}}_\ell=0$. The energy loss can then be parametrized as
\begin{equation}
-\left\langle\frac{d E_\ell}{dx}\right\rangle= a_\ell(E_\ell)+b_\ell(E_\ell) E_\ell\ ,\label{eq:energylossgeneral}
\end{equation}
where $\ell=\mu,\tau$, and the coefficients $a_\ell(E_\ell)$ and $b_\ell(E_\ell)$ account for energy losses through ionization and radiation, respectively, and can be computed from first principles in the microscopic theory, as described in \cref{sec:EnergyLoss}. At the energies of interest in this study, we can always take $E_\ell \gg b_\ell(E_\ell)/a_\ell(E_\ell)$ and the energy loss through ionization can always be neglected, while the coefficient $b_\ell(E_\ell)$, controlling the energy loss through radiation, becomes only mildly energy-dependent. In this approach, for neutral particles such as  neutrinos the mean energy loss is zero and $\dot{\vec{p}}_{\nu_i}=0$ in \cref{eq:system-transport-full}. \rev{Note that in the above formula we treat the muon energy loss as a continuous effect, thus neglecting its stochastic nature (see \cite{Lipari:1991ut}). This is implicit in the approximation under which we neglect the QED timescales with respect to all other timescales of the problem.

Last but not least, we would like to express the solution $f_\mu$ in terms of the only unknown: the neutrino flux at the Earth's surface \cref{eq:nuflux}. Physically, the muon neutrinos will promptly produce a population of muons that then propagates to the detector, while the $\tau$ neutrinos will start producing $\tau$'s that eventually can decay close to the detector in a secondary population of muons (and muon neutrinos). As such, the boundary condition for $f_{\nu_\mu}$ is to match to \cref{eq:nuflux} on the surface of the Earth, while $f_\tau$ must be zero at the crust, as for muons. The relation between the neutrino distribution and the flux is\footnote{Notice that here the flux is defined according to $f_\nu$. This normalization differs from the one of Ref.~\cite{titans} by a factor of $(2\pi)^3$. We absorb this unphysical normalization into our definition of the flux, or equivalently we posit $d\Omega_p \to \left(2\pi\right)^3\, d\Omega_p$.}
\beq\label{eq:flux}
4\pi \phi_\nu=f_\nu E_\nu^2 v\,.
\eeq

\subsection{Line-of-sight approach to \texorpdfstring{$f_\mu$}{fmu}}
Thanks to the previous discussion, it should be clear that the kernels in \cref{eq:kernel-scattering} and \cref{eq:kernel-decay}, can be simply treated as input functions defining a total source for the transport equation of the muon, which in the SM is $K\equiv K_{\rm{scattering}}^{\nu_\mu}(t, \vec{x},\vec{p}_\mu)+K_{\rm{decay}}^\tau(t, \vec{x},\vec{p}_\mu)$. We then rewrite the muon transport equation as
\beq
\begin{aligned}\label{eq:transport2}
\deriv{f_\mu}{t} + \dot{\vec{x}}\cdot\deriv{f_\mu}{\vec{x}} + \dot{\vec{p}}\cdot\deriv{f_\mu}{\vec{p}} = K(t, \vec{x},\vec{p})\,. 
\end{aligned}
\eeq
It is an inhomogeneous linear transport equation, that may be solved with the method of characteristics, also known as line-of-sight (LOS), provided the boundary condition \cref{eq:boundary-condition}. For ease of notation, we henceforth label the muon momentum with $\vec p$ instead of $\vec p_\mu$.

The method consists in the identification of trajectories $(t(\xi), \vec x(\xi), \vec p(\xi))$  along which the differential equation becomes ordinary. Here $\xi$ is a dummy variable with dimensions of length that tracks  the motion of the neutrino/muon along the LOS, spanning from $\xi = 0$ on the crust, to $\xi = L$ at the depth of the detector, which depends on the angle at which neutrinos impinge on the crust, as shown in \cref{fig:sketch-los}. 

The trajectories are found by solving the differential equation obtained by equating ${\rm d}t/{\rm d}\xi$, ${\rm d}\vec x/{\rm d}\xi$ and ${\rm d}\vec p/{\rm d}\xi$ to the respective coefficients of the differential operator of the left-hand side. In the system under consideration, such trajectories describe the evolution of time, position and momentum along the LOS. 

For concreteness, fixing a convention to be followed from now on, let us consider a muon with momentum $\vec p \equiv p \, \hat n_p$. Assuming that the direction of the momentum $\hat{n}_p$ does not change as the particle travels along $\xi$, the characteristic equations can be written as
\begin{equation}\label{eq:caratteristiche}
     \left\{\begin{array}{l}
      {\rm d}t/{\rm d}\xi = 1/v \\
     {\rm d}\vec{x}/{\rm d}\xi = \hat{n}_p \\
     {\rm d}\vec{p}/{\rm d}\xi = - b_\mu(\xi) \vec{p}
     \end{array}  \right.
     \implies
     \left\{\begin{array}{l}
     t(\xi) = t +  (\xi-L_{x,p})/v \\
      \vec x(\xi) = \vec{x} - (L_{x,p}-\xi)\hat n_p\\
      \vec p (\xi) = p\,\hat{n}_p \,\exp(\int^{L_{x,p}}_\xi {\rm d}\xi' b(\xi') )
     \end{array}  \right.\,,
\end{equation}
where in the last equation we are approximating the energy loss for high-energy muons as discussed below \cref{eq:energylossgeneral}. We defer a complete discussion on the energy loss of charged particles to \cref{sec:EnergyLoss}.

\revnew{The dissipative term $\dot{\vec{p}}_\mu \neq 0$ arises from collisions in the medium in the limit of small energy losses. This has two consequences. First, especially at the high energies in which we are interested, larger losses of energy per collision cannot be ignored \cite{Lipari:1991ut,ParticleDataGroup:2020ssz} (see \cref{sec:EnergyLoss}); second, by using \cref{eq:caratteristiche} in \cref{eq:transport}, the equation does not conserve the total number of muons. Starting from the full Boltzmann equation, one can systematically expand the part of the collisional integral featuring QED interactions with the medium in the limit of small energy losses. At first order in this expansion, that is, neglecting diffusion, one obtains the above dissipative term, as well as another term, which would compensate for the particle number loss. One way to avoid both issues would be to solve the full, integro-differential, Boltzmann equation, which would complicate the treatment. The two effects, however, compete:  underestimating the muon energy loss leads to a larger number of muons above threshold, while the dissipation leads to an underestimation of the total number of muons. We assume that the two effects can be captured by a systematic error on the energy-loss parameter, $b_\mu$, to be treated as a nuisance parameter. We leave a more detailed treatment, both at the level of the equation and at the level of the statistical analysis, to future work, and in this work we employ the approximate Boltzmann equation introduced above.}
\begin{figure}[t]
    \centering
    \includegraphics[width=\textwidth]{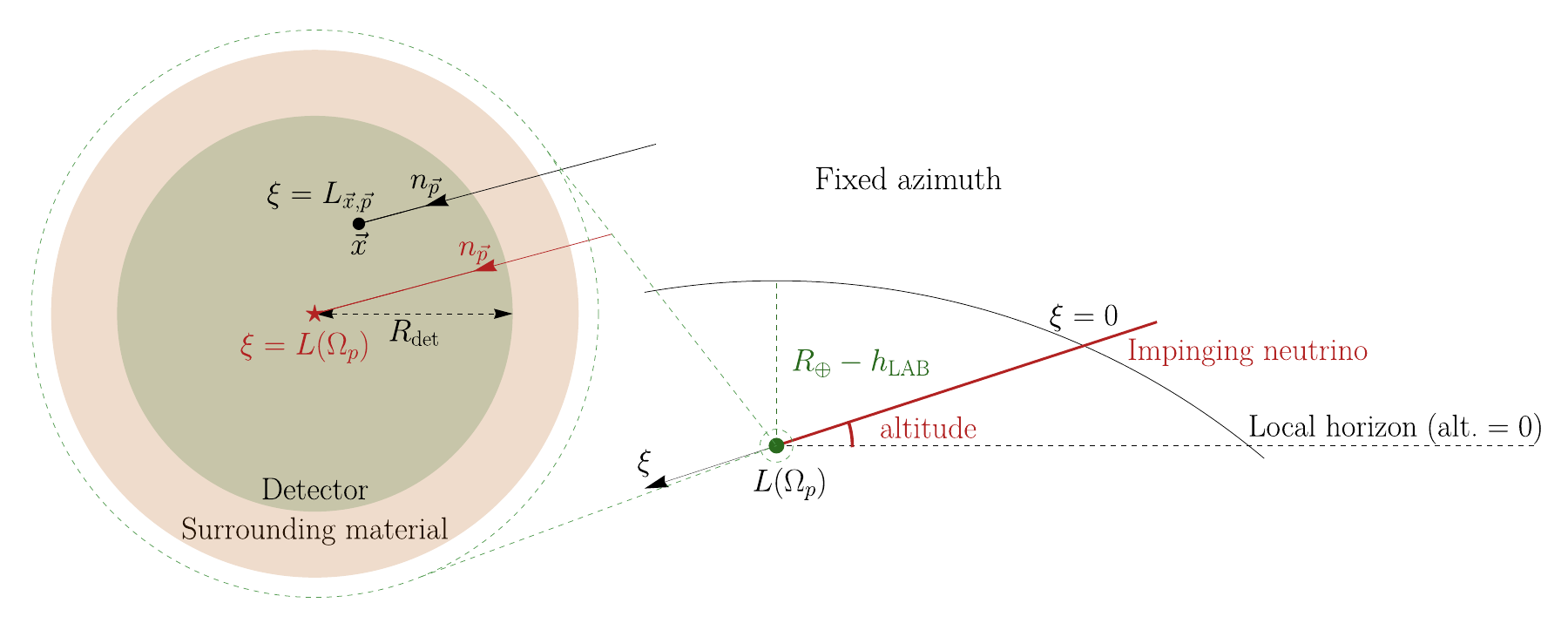}
    \caption{Two-dimensional sketch of a neutrino detector elucidating the LOS approximation and the notation used throughout the article. On the right, the black curve represents the Earth's crust, on which neutrinos impinge from a given direction, identified by the altitude and azimuth. On the left, a zoom on a spherical detector of radius $R_{\text{det}}$. As shown by the black line, one should in principle account for the path of each muon track, assuming that the muon is detected as $\vec{x}$. For simplicity, in practice we assume that all tracks pass through the center of the detector, where the muon is detected, so that the length of the path only depends on $\Omega_p$.}
    \label{fig:sketch-los}
\end{figure}

In \cref{eq:caratteristiche}, $L_{x,p}$ is the distance traveled from the crust to the point $\vec{x}$ in the detector along the direction $\hat{n}_p$ or angle $\Omega_p$, and $p$ is the momentum of the muon as measured at $\vec{x}$. The solution to the transport equation can be written as
\beq\label{eq:full-solution}
f_\mu(t,\vec x, \vec p)=\int\limits_{0}^{L_{x,p}} \frac{{\rm d}\xi}{v} K\lp t + \frac{(\xi-L_{x,p})}{v}, \vec x-(L_{x,p}-\xi)\hat n_p, p\,\hat{n}_p\, \exp\left[ \int^{L_{x,p}}_{\xi} {\rm d}\xi' b(\xi') \right] \rp\,.
\eeq
The muon distribution must vanish outside Earth as well as on the crust, for all times and momenta (\cref{eq:boundary-condition}), as indicated by the lower extreme of integration. On the other hand, $L_{x,p}$ depends on the relative distance between $\vec x$ and the point on the crust selected by a ray originating from $\vec x$ with direction $-\hat n_p$. As shown in \cref{fig:sketch-los}, in practice, we take the approximation that $\vec{x}$ always coincides with the center of the detector, in such a way that $L_{x,p} = L(\Omega_p)$.\footnote{Neglecting the detector size compared to the distance spanned by these lines, an approximate expression for $L_{x,p}$ is
$ L(\Omega_p) = (R_\oplus - h_{\text{LAB}}) \big(\sqrt{\sin^2\delta -(1-R_\oplus^2/(R_\oplus-h_{\rm LAB})^2)} -\sin\delta\big)$, where $\delta\in\left[-\frac{\pi}{2},\frac{\pi}{2}\right]$ is the altitude as in \cref{fig:sketch-los}, while $R_\oplus$ is the radius of Earth and $h_{\text{LAB}}$ is the distance between the crust and the center of the detector.}}  

The expression of $f_\mu$ in \cref{eq:full-solution} is what we need to compute our observable of interest, \cref{experimental-rate}.  However we can further improve on it. By replacing the kernel with the expressions in \cref{eq:kernel-scattering,eq:kernel-decay}, we can factor out the input neutrino flux; furthermore, we use spherical coordinates to write ${\rm d}^3p = E^2{\rm d}E{\rm d}\Omega$. This leads to the following form for $f_\mu$:
\be\label{eq:transport-to-detector}
f_\mu(E_\mu,\Omega_p)|_{\rm detector}= \frac{1}{E_\mu^2}\int\limits_0^{L(\Omega_p)} \frac{{\rm d}\xi}{v}  \int {\rm d}E_\nu {\rm d}\Omega_\nu \TT(E_\mu,\Omega_p| E_\nu, \Omega_\nu;\xi)\, \phi_\nu^\oplus(E_\nu,\Omega_\nu)\,,
\ee
where we defined the transport function $\TT(E_\mu,\Omega_p| E_\nu, \Omega_\nu;\xi)$. It has units of inverse length, energy and solid angle and it takes into account all the transport effects that bring a neutrino with energy $E_\nu$ on the crust to a muon at the detector with energy $E_\mu$. In the following, we show how the kernel in  \cref{eq:full-solution} can be cast in this form in concrete computations.

\subsection{Explicit solution to \texorpdfstring{$f_\mu$}{fmu}}
We are now in a position to find an explicit solution for \( f_\mu \). To proceed, we need approximate forms for \( f_{\nu_{\mu/\tau}} \) and \( f_\tau \), which generally solve the system in \cref{eq:system-transport-full}. For simplicity, we neglect the re-population of neutrinos from \( \tau \) decays (see Ref.~\cite{Safa:2019ege} for details), which conservatively reduces the number of muons arriving at the detector. As a result, the neutrino transparency is only due to neutral and charged EW interactions.
Assuming universality across different flavors, we factor it out and define it as the ratio of the neutrino flux at a given location to the flux on the crust:
\begin{equation}
D(\vec{x}, E_\nu, \Omega_\nu) \equiv \frac{\phi_\nu(\vec{x}, E_\nu, \Omega_\nu)}{\phi_\nu^\oplus(E_\nu, \Omega_\nu)}\, .  \label{eq:transparency}  
\end{equation}
This is computed in \cref{neutrinoxsec}.

\rev{The expression for $f_\tau$ is obtained in a manner analogous to the muon case, \cref{eq:full-solution}, with the main difference that the $\tau$ has a finite lifetime. To simplify the treatment and remove one length scale from the problem, we make the approximation $ 0 \approx b_\tau \ll b_\mu $, which allows us to assume energy conservation for the $\tau$ along the line of sight, i.e., $ \dot{p}_\tau \approx 0 $. At not extremely high energies, at which the $\tau$ travels distances shorter than its stopping range before decaying, this approximation may be justified, but at higher energies one should treat it carefully.}
Accounting for the finite stopping length $b_\tau$ (which is smaller than \( b_\mu \) by less than one order of magnitude, as we will discuss in \cref{sec:EnergyLoss}), will amount to solving a system like the one in \cref{eq:caratteristiche} for the muon. Here we assume $b_\tau \approx 0$ for simplicity, which allows us to write an approximate solution for $f_\tau$ as
\begin{equation}\label{eq:ftau}
f_\tau(t, \vec{x}, \vec{p}_\tau) = e^{-\Gamma_\tau(p_\tau) t} \int\limits_0^{L_{x,p_\tau}} \frac{\mathrm{d}\xi}{v} \, e^{\Gamma_\tau(p_\tau) t(\xi)} K_{\nu_\tau}(t(\xi), \vec{x}(\xi), \vec{p}_\tau)\, ,    
\end{equation}
where \( t(\xi) \) and \( \vec{x}(\xi) \) follow the same form as for muons, given by \cref{eq:caratteristiche}.

Having two expressions for $f_\nu$, \cref{eq:transparency}, and $f_\tau$, \cref{eq:ftau}, we can write the kernel functions explicitly. Therefore, by using the LOS approach, the full solution for the muon phase space distribution $f_\mu$ is given by 
\be\label{eq:fmu-completa}
\begin{split}
f_\mu(t,\vec x, \vec p)&= \int\limits_0^{L_{x,p}} \frac{\mathrm{d}\xi} {v}  \int \dd^3p_{\nu_\mu} D(\vec x(\xi),p_{\nu_\mu}) \frac{4\pi \phi_\nu^\oplus(\vec p_{\nu_\mu})}{E_{\nu_\mu}^2} \frac{\dd\sigma(p(\xi))}{\dd^3p(\xi)} n_N(\vec x(\xi))\\
&+\int\limits_0^{L_{x,p}} \frac{\mathrm{d}\xi} {v} \int \dd^3p_\tau \bigg[e^{-\Gamma_\tau(p_\tau) t(\xi)} \int\limits_0^{L_{x(\xi),p_\tau}} \frac{\dd\xi'}{v}\, e^{\Gamma_\tau(p_\tau) t(\xi')}\\ 
&\times \int \dd^3p_{\nu_\tau} D(\vec x(\xi'),p_{\nu_\tau}) \frac{4\pi \phi_\nu^\oplus(\vec p_{\nu_\tau})}{E_{\nu_\tau}^2} \frac{d\sigma(p_\tau)}{d^3p_\tau} n_N(\vec x(\xi'))\bigg]\frac{\dd\Gamma_{\tau\to \mu}(p(\xi))}{\dd^3p(\xi)}\,.
\end{split}
\ee
We emphasize that this is all we need to compute the number of expected events in the detector. The same expression can be cast in the form of \cref{eq:transport-to-detector} upon defining the two transport functions for the case of $\nu_\mu$ and $\nu_\tau$ respectively,
\begin{equation}\label{eq:transport-function-mu}
 \TT|_{\mu\leftarrow\nu_\mu} = 4\pi  E_\mu^2 D(\xi,E_{\nu_\mu}) \frac{d\sigma(p(\xi))}{d^3p(\xi)} n_N(\xi)\,,
\end{equation}
\begin{equation}\label{eq:transport-function-tau}
\TT|_{\mu\leftarrow\nu_\tau} = 4\pi  E_\mu^2 \int \dd^3p_\tau \bigg[\int\limits_0^{L(\xi)} \frac{d\xi'}{v}\, e^{\frac{\xi'-\xi}{d_\tau}}  D(\xi',E_{\nu_\tau}) \frac{d\sigma(p_\tau)}{d^3p_\tau} n_N(\xi')\bigg]\frac{d\Gamma(p(\xi))}{d^3p(\xi)}\,.
\end{equation}
Here we have introduced $d_\tau=\beta \gamma c \tau_\tau$ for the boosted decay length of tau leptons. Notice also that the transport function from $\nu_\tau$ is the convolution of the transport from $\nu_\tau$ to $\tau$ up to location $\xi$ (square bracket) and then transported via decay to the detector. 

As it stands, it is difficult to make any simple use of \cref{eq:fmu-completa}, although it contains all the ingredients for a numerical computation; we will perform this in \cref{sec:TheoryOutputs}. First, we describe all the theory inputs that go into the above integrals in the next section.

\section{Theory inputs}
\label{sec:TheoryInputs}
In this section, we go through all the theoretical inputs needed for the calculation of the event rate. In \cref{sec:EarthDensity}, we specify the modelling of the nucleon number density $n_N(\vec{x})$, which controls the charged lepton production rates, as well as their energy losses and the neutrino transparency. In \cref{neutrinoxsec} we define the different quantities controlling the neutrino physics: i) our parametrization of the neutrino flux at the Earth's surface, ii) the cross section of interactions of neutrinos with nucleons, iii) the neutrino transparency. In \cref{sec:EnergyLoss} we will elaborate on \cref{eq:energylossgeneral} giving further details on the energy loss of UHE charged particles.  

\subsection{Earth density}
\label{sec:EarthDensity}
\begin{figure}[t]
    \centering
    \includegraphics[width=.85\linewidth]{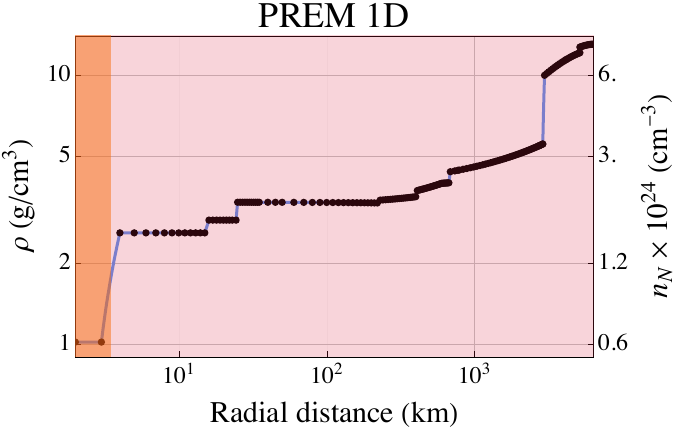}
    \caption{One-dimensional PREM density profile of the Earth. The {black points} are taken from Ref.~\cite{10.1785/0220120032}, and the {blue curve} shows the polynomial interpolation used in this work. The left vertical axis shows the mass density in $\mathrm{g/cm^3}$, while the right vertical axis indicates the corresponding nucleon number density in $\mathrm{cm^{-3}}$, computed as $n_N(r) = \rho(r)/m_N$. The {orange shaded} area on the left indicates the radial distances spanned by \ic{} for a PS at the location of the \eventname{} event, while the {red shaded} region is the respective range for \kmn{} (see \cref{sec:PointSource} for more details).}
    \label{fig:PREM}
\end{figure}

\cref{eq:transport-function-mu,eq:transport-function-tau} both depend on the number density $n_N$ of nucleons as a function of position inside the Earth, which can be written as the mass density divided by the nucleon mass
\begin{equation}
    n_N(\vec{x}) \equiv \frac{\rho(\vec{x})}{m_N}.
\end{equation}
To model the variation of the nucleon number density from the surface of Earth to its core, we follow the preliminary reference Earth model (PREM)~\cite{Dziewonski:1981xy}. In particular, we adopt the one-dimensional polynomial fit provided by IRIS-DMC~\cite{10.1785/0220120032}. The resulting mass density profile is shown in \cref{fig:PREM} as a function of the radial distance from crust, with the original tabulated data points in black and our interpolation in blue. In this one-dimensional model, the Earth is treated as a set of concentric spherical shells, and the density is assumed to depend only on the radial distance from the center.

In practice, due to energy loss, only muons produced near the detector yield a significant contribution (see below). Near IceCube, this region consists almost entirely of ice, with mass density \(\rho_{\rm ice} = 0.917\,\mathrm{g/cm^3}\), yielding a local nucleon number density of
\begin{equation}
    n_N^{\rm near}\big|_{\rm IC} = \frac{\rho_{\rm ice}}{m_N} \approx 5.48 \times 10^{23}\,\mathrm{cm^{-3}}, \label{eq:nearIC}
\end{equation}
where we take the average nucleon mass to be \(m_N = 1.674 \times 10^{-24}\,\mathrm{g}\). In contrast, KM3NeT is surrounded by water\footnote{Note that this is only true in the PREM, and the treatment can be refined by considering that from certain directions there are other materials in the proximity of KM3NeT (see \cite{KM3NeT:2025npi}).} with \(\rho_{\rm water} = 1.00\,\mathrm{g/cm^3}\), corresponding to
\begin{equation}
    n_N^{\rm near}\big|_{\rm KM3NeT} \approx 5.98 \times 10^{23}\,\mathrm{cm^{-3}}. \label{eq:nearKM3NET}
\end{equation}

Including muons from the decay of heavier particles (e.g., taus or BSM states) introduces a dependence on materials located farther from the detector, at distances 
given by the decay length of the mother particle, which may have a very different composition than the material surrounding the detector. We mention here an important geometric effect:
considering a point source compatible with the \eventname{} observation, and noting that the IceCube observatory is located at the South Pole, the radial distances relevant for neutrino propagation at IceCube are always within a few kilometers from the Earth's crust (see \cref{sec:PointSource} for more details). This effectively fixes the nucleon number density for IceCube to the value of \cref{eq:nearIC}. Conversely, KM3NeT spans lengths up to about the Earth diameter, $2R_{\oplus}\approx10^4$ km, probing regions with significantly larger densities than \cref{eq:nearKM3NET}. This effect may result in $n^{\rm{far}}_N\vert_{\rm{KM3}}\gg n^{\rm{far}}_N\vert_{\rm{IC}}$ and can, in principle, enhance the effective detector volume of KM3NeT relative to IceCube as we will discuss in \cref{sec:Results}.

\subsection{Neutrinos}\label{neutrinoxsec}
Neutrinos constitute the particles that $\nu$T are ultimately interested in detecting, thus it is important to link theoretical predictions to the free parameters that describe the neutrino flux at the crust, $\phi_\nu^\oplus$. In the following we introduce our conventions for the flux, and we emphasize the importance of knowing the high-energy behavior of neutrino interactions with matter as well as their propagation inside Earth at such energies.

\subsubsection*{Neutrino flux}
The unknown neutrino flux at the surface of Earth is denoted by $\phi_{\nu}^\oplus$, and it depends in general on both the neutrino energy \( E_\nu \) and impinging direction \( \Omega_\nu \). The flux may also depend on time, if the neutrino source has a transient behavior, a possibility we briefly address in \cref{sec:PointSource}, but ignore for the time being. We assume that these dependencies are factorized 
\be
\phi_\nu^\oplus(E_\nu, \Omega_\nu) = \phi_\nu^\oplus(E_\nu)\times \text{(directionality)}(\Omega_\nu)\,.
\ee
We bear two types of scenarios in mind: $i)$ a single power-law spectrum; $ii)$ an energy-localized flux. The former can be parametrized as follows
\beq\label{eq:NeutrinoFlux}
\phi_\nu^\oplus(E_\nu) = \phi_0 \lp 10^{18}\,\si{\giga\electronvolt\centi\metre\squared\second\steradian}\rp^{-1} \lp \frac{E_\nu}{E_\ast} \rp^{-\g}\,,
\eeq
with $\gamma > 0$. The latter case may be of several types -- monochromatic, Gaussian, power-law with either positive or negative spectral index (in which case a high-energy cutoff must be introduced).
The normalization in brackets is the order of magnitude expected for a diffuse power-law flux consistent with observations at \ic{} \cite{IceCube:2013low, IceCube:2020acn, Abbasi:2021qfz, Silva:2023wol, IceCube:2024fxo}.

For the directionality of the flux we consider two cases: $i)$ a diffuse source; $ii)$ a PS at an angle $\Omega_{\rm PS}(t)$. In formulae,
\be
\text{(directionality)}(\Omega_\nu) \propto \left\{\begin{array}{cc}
     1 & \text{diffuse,} \\
     \delta^{(2)}(\Omega_\nu-\Omega_{\rm PS}(t))& \text{PS.} 
\end{array} \right.
\ee
Due to the rotation of the Earth, the PS exhibits a time dependence with a daily period, on top of possible intrinsic time scales of the astrophysical/cosmological source.

\subsubsection*{Neutrino cross section}
\begin{figure}[t]
   \begin{center}
    \includegraphics[width=.7\textwidth]{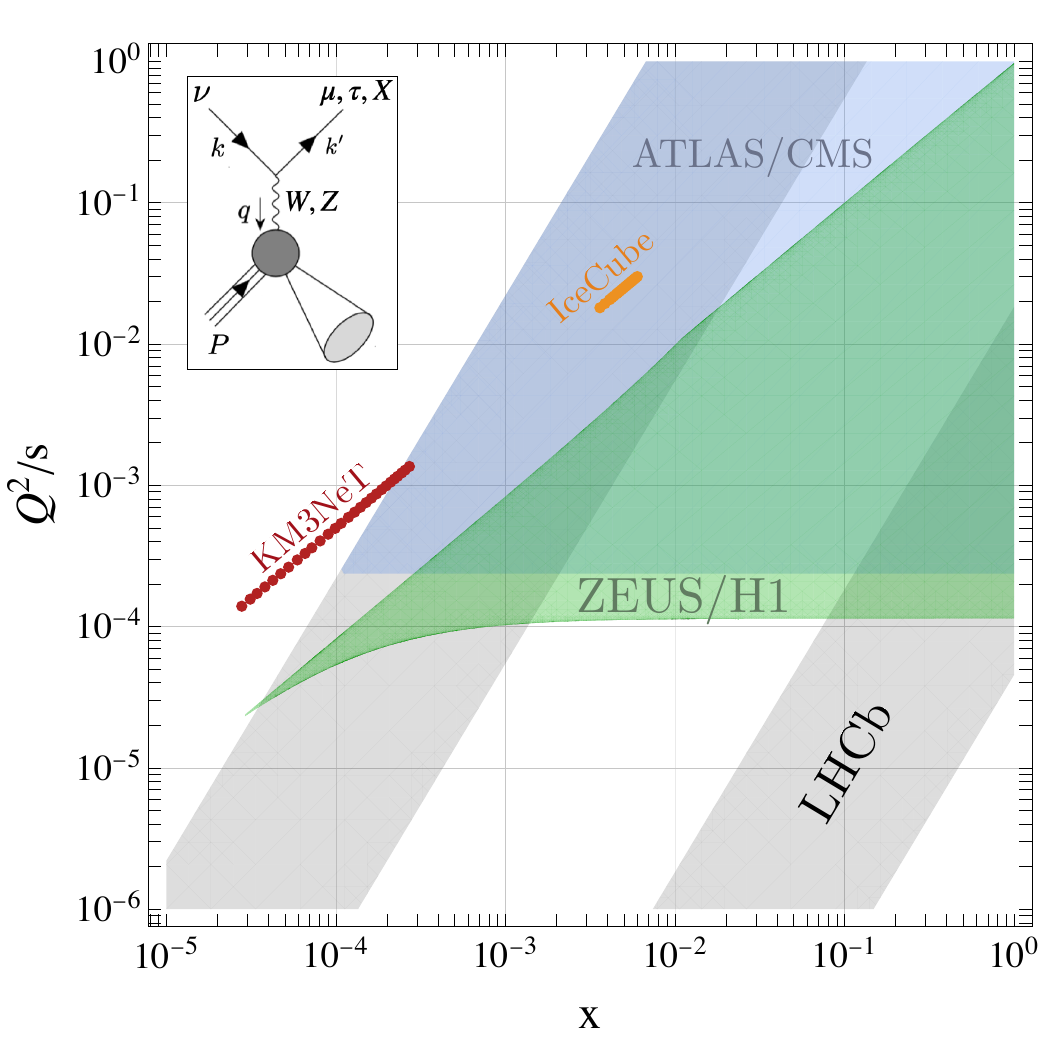}
    \caption{Kinematical accessible region at different experiments: ZEUS and H1 at HERA are shown in green, while ATLAS and CMS are in blue; the two bands accessible by LHCb are indicated in gray. The burgundy (orange) line is obtained from the \kmn{} (\ic{}) fit, where the spread in $Q^2$ corresponds to the different possible values of the inferred neutrino energy, assuming a diffuse spectrum (see \cref{sec:Diffuse} for more details). The kinematics of the different experiments is detailed for completeness in \cref{app:kin}.}
    \label{fig:DIS}
    \end{center}
\end{figure}

We now discuss the cross section of the neutrino-nucleon scattering. Remarkably, the center-of-mass energy of the neutrino–nucleon collision for neutrino energies $E_\nu\gtrsim 100\,\rm{PeV}$ exceeds that of the Large Hadron Collider (LHC). However, as we discuss below, the energy available for the fundamental neutrino–quark ($\nu q$) subprocess is significantly smaller, as illustrated in \cref{fig:DIS}. The key kinematic variables governing neutrino deep inelastic scattering (DIS) off nuclei are the center-of-mass energy $s = (k + P)^2$, the squared momentum transfer $Q^2 = -q^2 = (k - k')^2$, the Bjorken scaling variable $x = \frac{Q^2}{2 P \cdot q}$, and the inelasticity $y = \frac{P \cdot q}{P \cdot k}$, where the momenta are defined in \cref{fig:DIS}. They are related by $Q^2 = x y s$. For ultra-high-energy neutrinos with $E_\nu \gtrsim 100~\mathrm{PeV}$ the center-of-mass energy reaches $\sqrt{s} \approx \sqrt{2 m_N E_\nu} \gtrsim 14~\mathrm{TeV}$. At these high energies, the Fermi theory stops being accurate and the presence of the W-boson propagator limits $Q^2$ from above $Q^2\lesssim M_W^2$, favoring $Q^2$ at the edge of this upper limit. Since the cross section is dominated by events with moderate inelasticity, $\left\langle y\right\rangle \sim \mathcal{O}(0.1)$, this implies that the cross section is peaked at very small Bjorken-$x$, where the upper bound scales as
\beq
x \lesssim\frac{M_W^2}{y s}=3\cdot 10^{-4}\left(\frac{100\,\rm{PeV}}{E_\nu}\right)\left(\frac{0.1}{\left\langle y\right\rangle}\right)\ . 
\eeq
Consequently, the energy carried by the struck parton is bound from above by
$E_{\rm{parton}} \lesssim x \sqrt{s} \approx 7\,\rm{GeV}\left(E_\nu/100\,\rm{PeV}\right)$, which dangerously approaches scales where the quark parton model is expected to break down.  

Concretely, the doubly differential cross section of neutrino CC scattering off of an isoscalar nucleon $N=(n+p)/2$ can be written as 
\be
\frac{d\sigma_{\nu N}}{dx dy}=\frac{G_F^2 s}{\pi [Q^2+M_W^2]^2/M_W^4}\, \, \left[x q(x,Q^2)+(1-y)^2 x \bar{q}(x,Q^2)\right]\,,\label{eq:diff1}
\ee
where $q$ and $\bar{q}$ are the quark and anti-quark PDFs. The NC cross section can be obtained by replacing $M_W\to M_Z$ and replacing the PDFs above with the ones associated to the NCs. 

\begin{figure}[t]
	\centering
 \includegraphics[width=.85\linewidth]{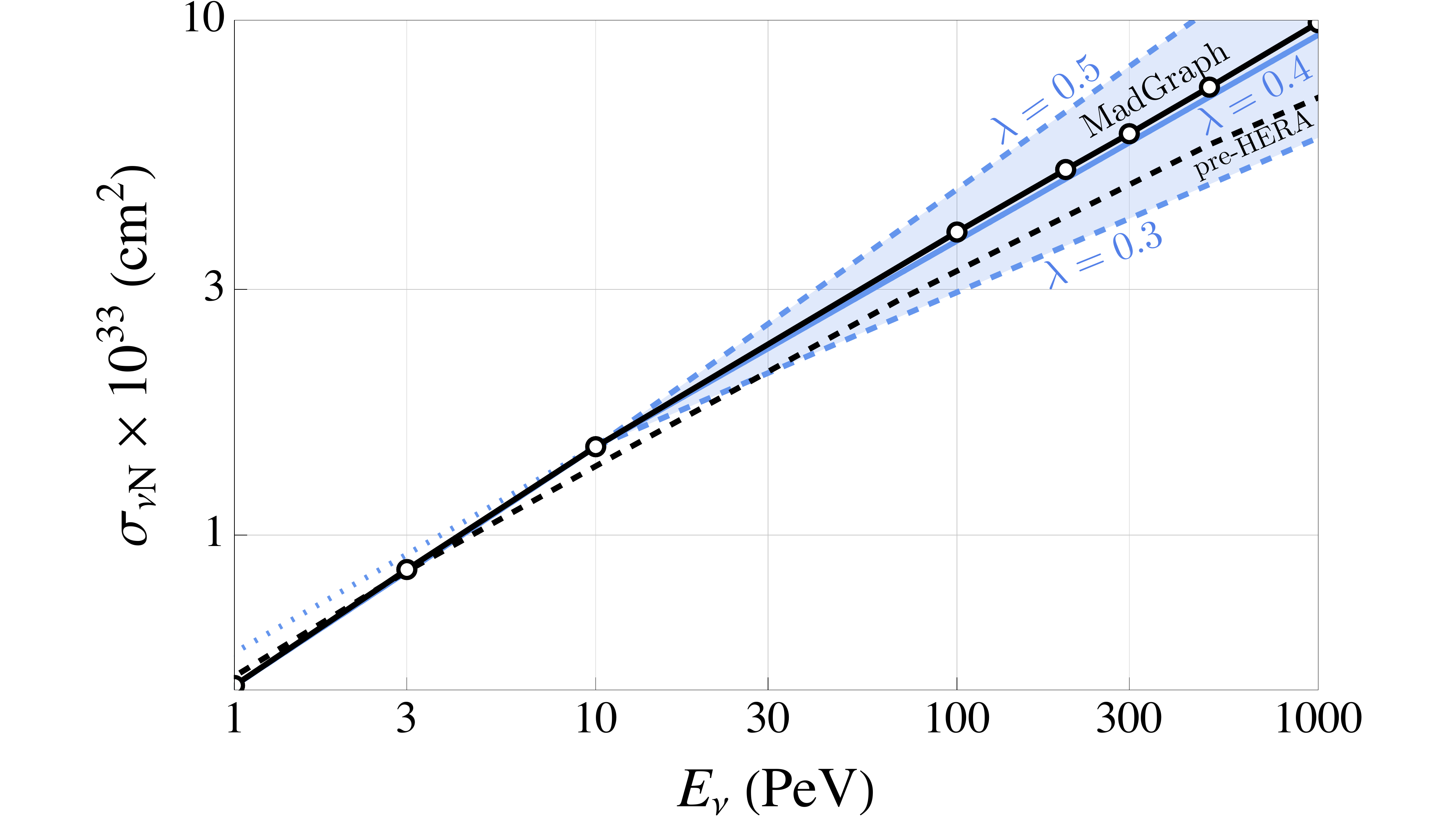}
    \caption{\label{fig:xsectot}The neutrino CC cross section used in this work with the uncertainty band in {light blue}  estimated by varying $\lambda$ for energies $E_\nu\gtrsim 10\,\rm{PeV}$ (see text for details). Our envelope roughly agrees with  the one derived in Ref.~\cite{Gandhi:1995tf}, while Ref.~\cite{Cooper-Sarkar:2011jtt} gives $\lambda\in(0.3-0.4)$. The {solid black} line shows the {\tt MadGraph} prediction obtained with standard LHA-PDF~\cite{Buckley:2014ana}, the {dashed black} corresponds to the old pre-HERA result of Ref.~\cite{Quigg:1986mb}.}
\end{figure}

In Ref.~\cite{Gandhi:1998ri}, HERA data is used to fit the PDFs down to small $x$, updating the previous results from Ref.~\cite{Quigg:1986mb}, and more recent updates in Refs.~\cite{Cooper-Sarkar:2011jtt,Bertone:2018dse,Valera:2022ylt} further refine these results. As shown in \cref{fig:DIS}, the neutrino energy corresponding to the \eventname{} event is slightly beyond the HERA region, leading to some uncertainty due to the need for extrapolation at small $x$ and small $Q^2/s$. The inclusion of LHCb data in PDF sets could reduce this uncertainty and allow direct measurements in part of the kinematic region relevant for UHE neutrinos.

In order to estimate the uncertainty at $E_\nu>10\,\rm{PeV}$ we implement a procedure which allows us to match the different  uncertainty bands reported in the literature. Given the cross section above we can write the low-$x$ behavior of the quark (and anti-quark) PDF as
\begin{equation}\label{eq:Low-x-Limit}
\lim_{x\to0}x q(x)\simeq \lim_{x\to0}x \bar{q}(x)= x^{-\lambda}\, ,  
\end{equation}
where we assume $\lambda$ to be independent on $Q^2$. 
Letting $\tilde{s}=s/M_W^2$, we can rewrite the differential cross section in \cref{eq:diff1} as
\begin{equation}
\frac{{\rm d}^2\sigma_{\nu N}}{{\rm d}x {\rm d}y}\approx\frac{G_F^2M_W^2}{\pi x^\lambda}\left[\frac{(1+(1-y)^2)\tilde{s}}{(1+x y \tilde{s})^2}\right]\,.    
\end{equation}
We are interested in understanding the analytical scaling of the cross section in the limit $\tilde{s} x\ll1$, as well as that of the mean Bjorken variable $x$ and the mean inelasticity $y$,
\begin{equation}
\left\langle x\right\rangle=\frac{1}{\sigma_{\nu N}}\int\limits_0^1 dx  x \frac{d\sigma_{\nu N}}{dx}\quad ,\qquad \left\langle y\right\rangle=\frac{1}{\sigma_{\nu N}}\int\limits_0^1 dy  y \frac{d\sigma_{\nu N}}{dy}\ . 
\end{equation}
In this limit analytical expressions can be easily obtained. For example
\beq
\begin{aligned}
&\frac{{\rm d}\sigma_{\nu N}}{{\rm d}y} \approx \frac{G_F^2M_W^2}{\pi} \tilde{s}^\lambda \lp  y^{\lambda-1} \frac{2 + y (y-2)}{1-\lambda} \rp\Gamma(2-\lambda)\Gamma(1+\lambda)\,,\\
&\frac{{\rm d}\sigma_{\nu N}}{{\rm d}x} \approx \frac{G_F^2M_W^2}{x^{\lambda+1}\pi} \left[ \frac{\zeta (2 + \zeta(3+2\zeta)) - 2 (1+\zeta)^2\log(1+\zeta)}{\zeta^2 (1+\zeta)} \right] \,,
\end{aligned}
\eeq
where we defined $\zeta = x 2m_N E_\nu/M_W^2$.

\begin{figure}[t]
	\centering
    \includegraphics[width=.49\linewidth]{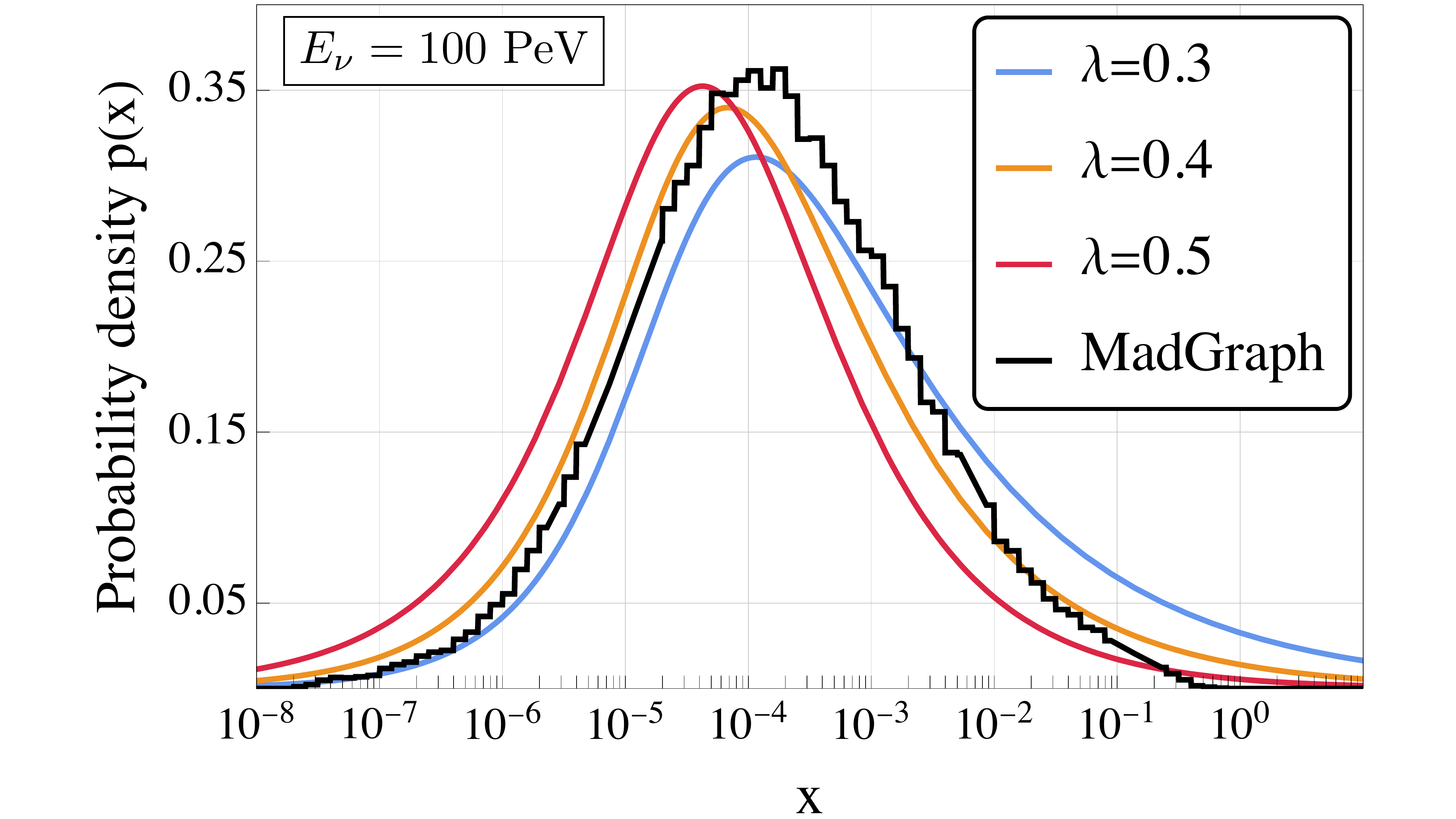}\hfill
    \includegraphics[width=.49\linewidth]{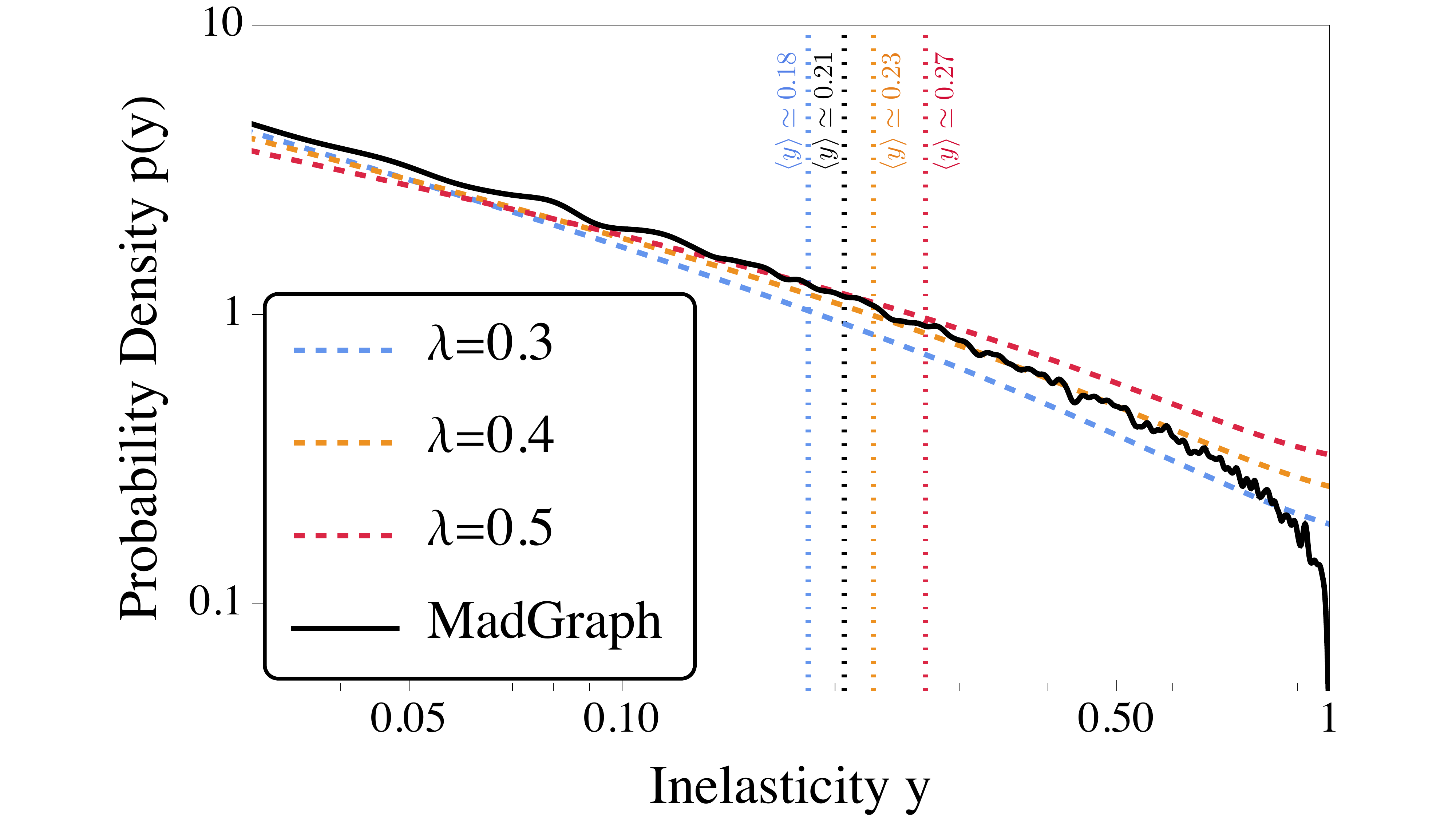} 
    \caption{{\bf Left:} Probability density function ${\cal P}(x)$ of the Bjorken-$x$ in DIS of neutrinos off of protons at rest. The black line has been obtained by simulating the processes with {\tt MadGraph}, with initial energy of the neutrino fixed to $E_\nu = \SI{100}{PeV}$. The blue, orange and red lines show our result computed with the values $\lambda = 0.3, 0.4, 0.5$. {\bf Right:} Probability density function $\py$. The color code is the same as the left plot. The vertical lines show the expected values $\left\langle y \right\rangle$, as obtained from our analytical estimate.}
    \label{fig:p(y)}
\end{figure}

The total cross section for neutrino CC DIS at $E_\nu>10\,\rm{PeV}$ can be written as\footnote{The total cross section for neutrino NC DIS can be obtained from the CC one given that $\sigma^{\rm NC}/\sigma^{\rm CC}=g_L^2+r g_R^2\approx 0.32$, where \(g_L^2 = \tfrac{1}{2} - \sin^2\theta_W + \tfrac{5}{9}\sin^4\theta_W\) and \(g_R^2 = \tfrac{5}{9}\sin^4\theta_W\), with $\sin\theta_{\rm w}^2\simeq0.22$ and \(r = \sigma(\bar\nu\,N \to \mu^+ X)/\sigma(\nu\,N \to \mu^-\,X) \approx 0.5\). }
\begin{equation}\label{eq:crossection-Above10PeV}
\sigma_{\nu N}^{\rm CC}(E_\nu)= \sigma_0 \lp \frac{E_\nu}{E_0}\rp^\lambda = 1.48\times10^{-33}~{\rm cm}^2 \times \lp \frac{E_\nu}{10~{\rm PeV}} \rp^\lambda\,.
\end{equation}
The normalization $\sigma_0$ is fixed by matching our cross section to the result obtained with {\tt MadGraph v3.5.1}~\cite{Alwall:2014hca} at $E_\nu=E_0=10\,\rm{PeV}$, where the default LHA-PDF~\cite{Buckley:2014ana} are used. \cref{fig:xsectot} shows the behavior of \cref{eq:crossection-Above10PeV} for three different choices of $\lambda = 0.3,~0.4,~0.5$. Notice that the {\tt MadGraph} result closely follows the $\lambda=0.4$ line for $E_\nu>10\,\PeV$, while deviating at smaller energies, as the approximation in \cref{eq:Low-x-Limit} fails. Thus at energies $E_\nu<10\,\PeV$ we use the {\tt MadGraph} cross section results, and keep the simple analytical expression in \cref{eq:crossection-Above10PeV} for larger energies.

One might worry that the power-law growth of the total cross section will eventually violate the Froissart bound~\cite{Froissart:1961ux}, which, assuming only analyticity and unitarity requires the structure functions, proportional to the $x q$ and $x\bar{q}$ in the perturbative quark model, to not grow more rapidly than $\log^2(1/x)$ at small $x$, independently on the value of $Q^2$. Ref.~\cite{Block:2006dz,Berger:2007vf,Block:2010ud} performed a fit requiring the low-$x$ behavior to match this theoretical expectation. From the result of Ref.~\cite{Block:2010ud}, one can check that the total neutrino cross section obtained with this procedure agrees well with the polynomial fit for $E_\nu\lesssim 1\,\rm{EeV}$, which are the energies of interest for this paper. 

We are now ready to discuss the PDF scaling at small $x$ that have been extracted from data. For instance, Ref.~\cite{Gandhi:1998ri} compiled a range of small-$x$ power-law behaviors from various PDF sets, finding that $\lambda$ can range between $\lambda \in [0.2, 0.5]$. More recent studies such as Ref.~\cite{Cooper-Sarkar:2011jtt} have narrowed down this uncertainty to $\lambda \in [0.3, 0.4]$. In the same small-$x$ approximation we can derive analytical expression for the mean of $x$ and $y$. We show the behavior of these analytical expression compared to the {\tt Madgraph} result in \cref{fig:p(y)}, using again as reference values $\lambda = 0.3,~0.4,~0.5$.

The transport equations of \cref{sec:TransportEquation} feature the differential cross section in the muon momentum ${\rm d}\sigma_{\nu N}/{\rm d}^3p_\mu$ in the laboratory frame, which we parameterize as 
\begin{equation}\label{eq:pdf_of_p}
 \frac{1}{\sigma_{\nu N}} \frac{{\rm d}\sigma_{\nu N}}{{\rm d}^3p_\mu} = \delta^{(2)}(\Omega_\nu - \Omega_p) \frac{(1 - y)}{E_\mu^3}  \py\,.
\end{equation}
Here we assume that, at such high energies, the direction of the muon momentum in the laboratory frame is identical to that of the incoming neutrino. The muon energy depends on the inelasticity distribution $\py \equiv (\dd\sigma/\dd y)/\sigma$, which is only mildly sensitive to the incoming neutrino energy and can be obtained via Monte Carlo simulation. In particular, we extract it by simulating neutrino-proton scattering using \texttt{MadGraph}, as shown in the right panel of \cref{fig:p(y)}.  
Recall that in the rest frame of the nucleon, the inelasticity reduces to $y = 1 - \frac{E_\mu}{E_\nu}$, thus \cref{eq:pdf_of_p} is normalized to $1$. In the elastic limit (i.e. $\py = \delta(y)$), it is proportional to $\delta(E_\nu - E_\mu)$.

Finally, having fixed the neutrino cross section and the nucleon number density we can define a scale playing an important role, namely the neutrino mean free path 
\begin{equation}
\lambda_\nu(E_\nu)=\frac{1}{n_N\sigma_{\rm tot}(E_\nu)}\approx 1.2\times 10^{4}\, \rm{km} \left(\frac{10\,\rm{PeV}}{E_\nu}\right)^{\lambda} \,,\label{eq:EWmeanfreepath}  
\end{equation}
which is of the same order of the Earth radius $R_\oplus=\SI{6.38e4}{\kilo\metre}$. This implies that for neutrinos energies $E_\nu>10\,\PeV$ attenuation cannot be ignored, and indeed plays an important role in the physics of the \eventname{} event.

\subsubsection*{Neutrino propagation} 
In our approach all the information about the neutrino flux at any position inside the Earth, for both tau and muon neutrinos, is encoded in the distribution $f_\nu(\vec x, \vec p_\nu)$, described by the full system of equations in \cref{eq:system-transport-full}. We take the limit in which we neglect the effect of tau decay in repopulating neutrinos of lower energy, whose treatment can be found in Ref.~\cite{Safa:2019ege}. This way the system \cref{eq:system-transport-full} decouples, and only depends on the kernel $C_{\text{flavor-ind.}}$, controlled by CC and NC interactions. The total cross section (CC plus NC  interactions) contributes to deplete the neutrino flux due to the $\nu N \to \mathrm{(anything)}$ scatterings, while NC also repopulates them at lower energies due to the inelasticity of the $\nu N\to \nu X$ scatterings.
\cref{eq:system-transport-full} for the neutrino flux now becomes
\be\label{eq:neutrino-att}
\frac{\partial\phi_\nu}{\partial \xi}= -\frac{1}{\lambda_\nu(E_\nu)}\left[\phi_\nu(\xi,E_\nu) - \int\limits_0^1 \frac{\dd y}{1-y} \py \frac{\sigma_{\rm NC}}{\sigma_{\rm tot}}\phi_\nu(\xi, \frac{E_\nu}{1-y})\right]\,,
\ee
The second term is the re-population effect from NC elastic scattering, and $\py$ the inelasticity distribution, see for example \cite{Vincent:2017svp,Safa:2019ege}.

\begin{figure}[t]
	\centering
    \includegraphics[width=.85\linewidth]{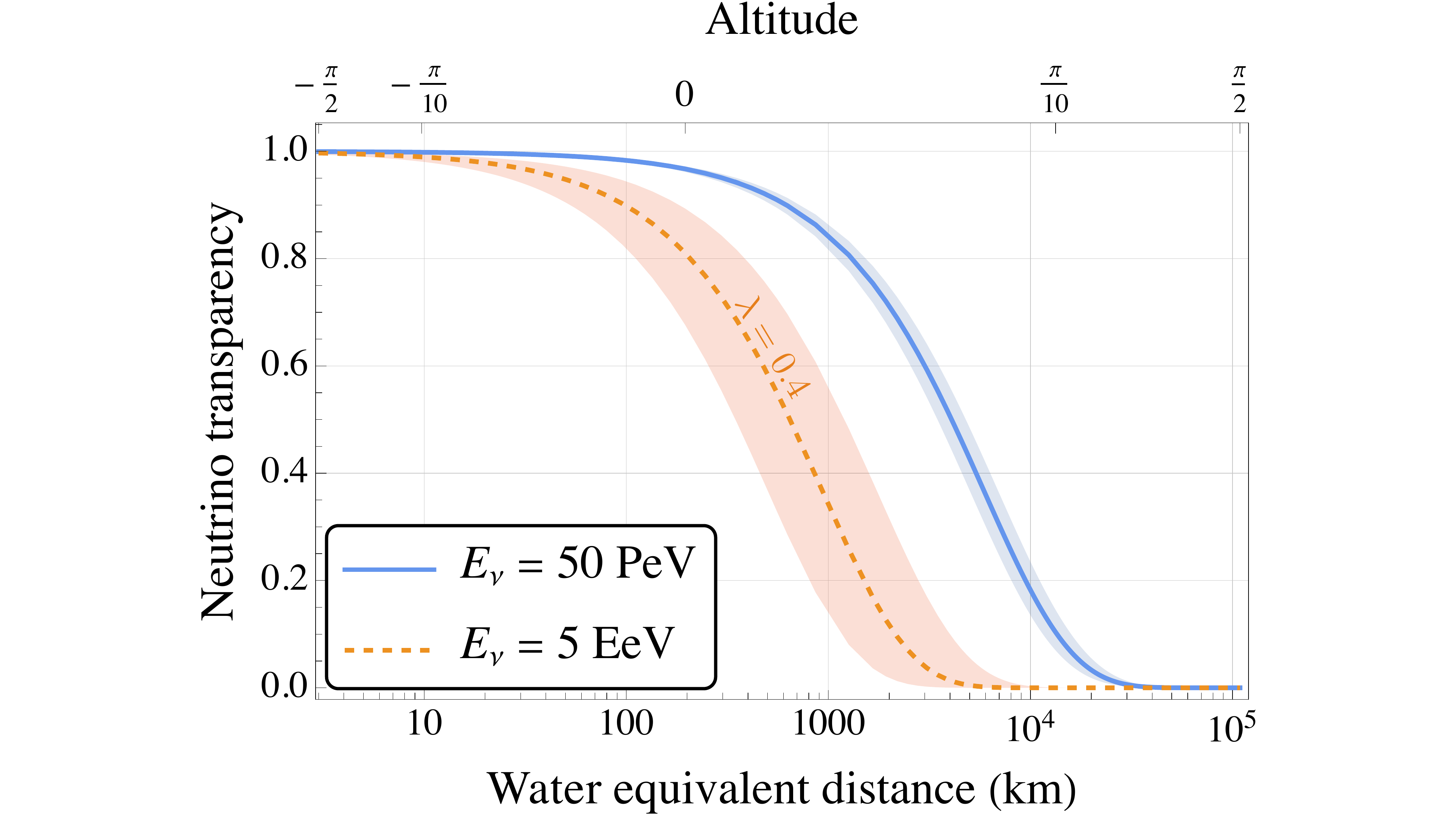}
    \caption{Neutrino transparency as a function of the water equivalent distance from Earth's surface, for two sample energies, $E_\nu = \SI{50}{PeV}$ (blue) and $E_\nu = \SI{5}{EeV}$ (dashed orange). The upper axis shows the correspondent altitude angle with respect to the center point of the \kmn{} detector, $\delta=0$ corresponding to the horizon. The lines indicate the results with $\lambda = 0.4$, while the bands show the variation obtained by scanning in the interval $\lambda \in [0.3,0.5]$.}
    \label{fig:Dfactor}
\end{figure}

In our case, we are only interested in the transparency factor, $D(\xi,E_\nu)$, as defined in \cref{eq:transparency}.  Having computed $\py$, one can solve numerically the differential equation. It is worth noticing that working in the elastic limit (i.e. $\py=\delta(y)$) for the differential NC cross section, the transparency becomes only sensitive to the CC cross section. In this approximation, we get
\begin{equation}
D(\xi,E_\nu)= \exp{\bigg(-\int\limits_0^\xi \frac{d\zeta}{L_{\nu}(\zeta,E_\nu)}\bigg)}=\exp{\lp- \frac{\xi_{\rm we}}{L_{\nu}^{\rm water}(E_\nu)}\rp}\,,
\end{equation}
where $L_\nu$ is the mean free path of CC interactions, and in the second equality we express distances in equivalent kilometers of water: for a given line of sight one  reweighs each Earth segment of density $\rho$ by $\rho/\rho_{\rm water}$,
so that the amount of matter traversed along the line is the same as if the Earth was only constituted of water.
In \cref{fig:Dfactor}, we show the behavior of $D$ for two sample neutrino energies, $E_\nu = 50\, \PeV$ and $E_\nu = 5\, \EeV$, using the water equivalent distance computed in the PREM model.

\subsection{Charged leptons}
\label{sec:EnergyLoss}
Neutrino scatterings via CC results in the production of charged leptons, $\ell= e,\mu,\tau$, which in turn undergo both decay into lighter particles (when they exist), and energy loss in the propagation medium.

\subsubsection*{Decays} 
The lifetimes of the muon and tau leptons are:
\begin{equation}
\tau_\mu = 2.2 \times 10^{-6}~\text{s}, \quad \tau_\tau = 2.9 \times 10^{-13}~\text{s},
\end{equation}
while the electron is stable. In the UHE regime, charged leptons produced via neutrino DIS are highly boosted, leading to large decay lengths:
\begin{align}
d_\mu(E_\mu) &= \frac{E_\mu}{m_\mu} c \tau_\mu = 6.22 \times 10^8~\text{km} \left(\frac{E_\mu}{100~\text{PeV}}\right)\,, \\
d_\tau(E_\tau) &= \frac{E_\tau}{m_\tau} c \tau_\tau = 4.9~\text{km} \left(\frac{E_\tau}{100~\text{PeV}}\right)\,.
\end{align}
At energies relevant for our work, the muon is thus effectively stable on Earth scales, while the tau has a shorter, yet significant, decay length. Taus decay into hadrons most of the time, but can also decay into muons and neutrinos, with a branching ratio of \cite{ParticleDataGroup:2020ssz}
\beq\label{eq:BranchingRatio:tau}
{\cal B}_{\tau\to\mu}\equiv{\cal B}\lp  \tau \to \mu \nu \nu \rp = 17.4\%\,.
\eeq
Due to the small electron mass, the \( \tau \to e \nu \nu \) decay is highly suppressed, making the tau's contribution mainly relevant for the muon yield in neutrino detectors.

\begin{figure}[t]
	\centering
    \includegraphics[width=.85\linewidth]{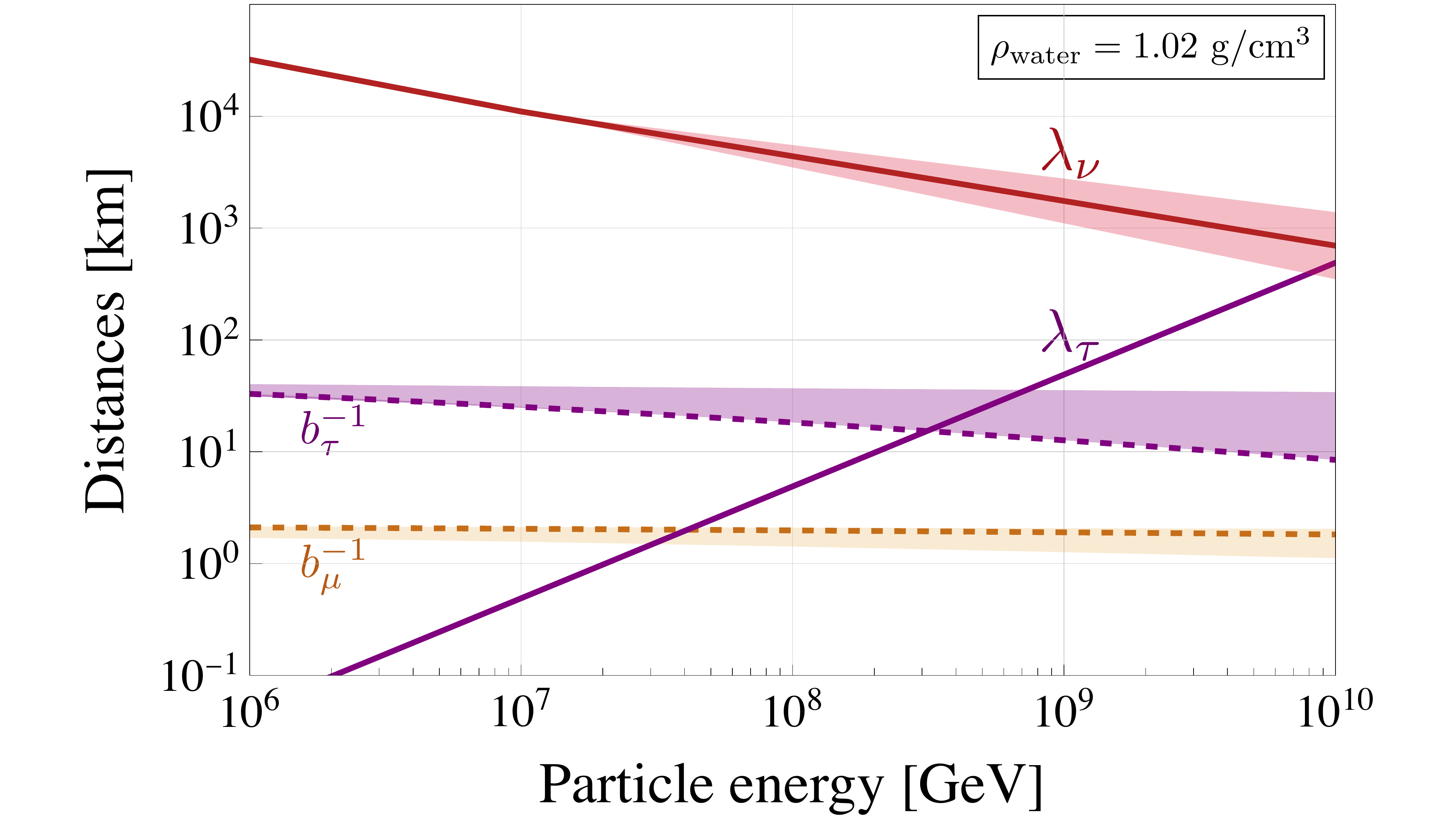}
    \caption{Comparison of the relevant length scales discussed in \cref{sec:TheoryInputs}, as a function of the particle energy. The solid purple and red lines indicate the $\tau$ decay length and the neutrino mean free path, respectively. The purple and brown dashed lines show instead the stopping length, $b^{-1}$, for the tau and muon, respectively. All values are computed assuming the density of water, $\rho_{\rm water} = 1.02\,\gcm$, and the shaded bands indicate the respective uncertainties, see text for details.}
    \label{fig:RelevantScales}
\end{figure}

\subsubsection*{Energy losses} 
Charged leptons lose energy as they travel through the Earth. In the transport equations in \cref{sec:muonprod}, this energy loss is captured in the charged particle geodesic by averaging over many soft QED processes within a single EW mean free path. This approach is valid as long as the ranges of energy loss for charged particles are much shorter than the neutrino mean free path. This is indeed the case for the energy range considered in this study, as shown in \cref{fig:RelevantScales}. 

We repeat the standard parametrization for the charged lepton mean energy loss given by \cref{eq:energylossgeneral}:
\begin{equation}\label{eq:energyloss}
-\left\langle \frac{\dd E_\ell}{\dd x} \right\rangle = a_\ell(E_\ell) + b_\ell(E_\ell) E_\ell\,.
\end{equation}
At UHE energies relevant to this study, the electronic energy loss function \( a_\ell(E_\ell) \) approaches a constant value \( a_\ell \), depending on the electron density of the medium and is independent of the mass of the charged particle, as dictated by the Bethe-Bloch formula~\cite{Groom:2001kq}. In liquid water, for UHE particles, $ a_\ell = 0.3 \, \rm{TeV}/\rm{km} $ for $ \ell = \mu, \tau $. At UHE, the radiative energy loss function also becomes approximately constant, $b_\ell(E_\ell) \approx b_\ell $, but it depends non-trivially on the mass of the charged particle. 

In the UHE limit we can easily integrate \cref{eq:energylossgeneral} and define the average distance traveled by a lepton losing energy from $E_\mu^{\rm{in}}$ to $E_\mu^{\rm{out}}$ 
\begin{equation}
\Delta L(E_\ell^{\rm{in}},E_\ell^{\rm{out}})=\int\limits_{E_\ell^{\rm{out}}}^{E_\ell^{\rm{in}}} \frac{d E}{\left\langle\frac{d E_\ell}{dx}\right\rangle}\approx\frac{1}{b_\ell}\log\left(\frac{E_\ell^{\rm{in}}}{E_\ell^{\rm{out}}}\right)\ ,
\end{equation}
where we expanded for $E_\ell\gg a_\ell/b_\ell$, which is highly justified in the range of energies of interest here.\footnote{$E_{\ell}^{\rm crit}\equiv a_\ell/b_\ell$ is often referred in the literature as ``critical energy'' and corresponds to the energy where ionization energy losses (controlled by $a_\ell$) are equal to the radiative ones (controlled by $b_\ell$). A typical value for water is $E_\ell^{\rm crit.}\approx 0.3\,\mathrm{TeV}$.} 

In general, the inverse range of a charged lepton is the sum of contributions from three distinct processes: \beq
b_\ell = b_\ell^{\rm brehm} + b_\ell^{\rm pair} + b_\ell^{\rm DIS}\,,\label{eq:btot}
\eeq where $b_\ell^{\rm brehm}$ encodes the contribution of on-shell photon Brehmsstrahlung,  $b_\ell^{\rm pair}$ the electron-positron pair production, both through off-shell radiation and through photon fusion, and $b_\ell^{\rm DIS}$ the contribution of DIS, dominated by virtual photon exchange. The cross section for Bremsstrahlung and electron-positron pair production are well-established (see for example Ref.~\cite{KOEHNE20132070} for a summary). The DIS contribution, \( b_\ell^{\rm{DIS}} \), suffers from significant theoretical uncertainties due to the photon-nucleon cross section at low \( Q^2 \) and \( x \), which dominates the energy loss. We parametrize these uncertainties as summarized in Ref.~\cite{KOEHNE20132070}, with the envelope shown in \cref{fig:RelevantScales}. For further details on the theory uncertainties, see Refs.~\cite{Armesto:2007tg}. 

For every interaction, we are interested in the energy $ \Delta E_\ell $ transferred from the charged particle to the photon, compared to the initial energy $ E_\ell $. For this reason, we define $\nu = \Delta E / E$ and write $ b$ in terms of the microscopic cross section as in Sec.~\ref{sec:TheoryInputs} 
\begin{equation}
b^I_\ell = n_N \int\limits_{\nu_{\rm{min}}^I}^{\nu_{\rm{max}}^I} \rm{d}\nu\, \nu \,\frac{\rm{d} \sigma^I}{\rm{d} \nu}  \,,
\end{equation}
where the index $I$ labels the different processes in \cref{eq:btot}, and both the cross section and the minimal and maximal transferred energy depend on the process under consideration. 

\revnew{For muons, the energy loss is dominated by \( b^{\rm{pair}}_\mu + b^{\rm{brehm}}_\mu \), with \( b_\mu^{\rm DIS} \) contributing just at 10\% level.
For example, at UHE in liquid water, the average distance travelled by a muon continuously losing energy as in Eq.~\eqref{eq:energyloss} is  $b^{\rm{pair}}_\mu \approx 1.4 \, b^{\rm{brehm}}_\mu = 0.23 \, \rm{km}^{-1}$~\cite{Groom:2001kq,Lipari:1991ut,ParticleDataGroup:2020ssz}, corresponding to $b_\mu^{-1}\approx2.4\, \rm{km}$ which is the one reported in Fig.~\ref{fig:RelevantScales} and used throughout this paper. 

The crucial assumption behind Eq.~\eqref{eq:energyloss} is that soft processes with small energy exchange dominate the energy loss, so that one can safely assume a deterministic relation between the mean energy and the spatial displacement. One might wonder to what extent this approximation is justified at the energies of interest and what is the size of the expected corrections from stochastic diffusion due to hard scatterings. The latter is controlled by the diffusion coefficient defined as
\begin{equation}
D^I_\ell=n_N\int_{\nu^I_{\rm{min}}}^{\nu^I_{\rm{max}}} d\nu \nu^2 \frac{d\sigma^I}{d\nu}    
\end{equation}
As a simple case study we consider muon Bremsstrahlung, whose differential cross section for high energy muons can be approximated as
\begin{equation}
\frac{\dd\sigma^{\rm{brehm}}}{\dd\nu}=\frac{16Z^2\alpha^3}{3m_\mu^2\nu}\log\left(\frac{184}{Z^{1/3}}\right)\left(1-\nu+3\nu^2\right)\ ,
\end{equation}
The total number of scatterings for a muon of high energy $E_\mu$ after distance $L$ in water is 
\begin{equation}
N(L)\approx \left(25+2\log\frac{E_\mu}{100\,\rm{PeV}}\right) \left(\frac{n}{6\times10^{23}\,\rm{cm}^{-3}}\right)\left(\frac{L}{1\,\rm{km}}\right)\, ,
\end{equation}
where as an infrared cutoff of the log-diverging integral we used $E_{\mu}^{\rm{crit.}}\approx 0.3 \,\rm{ TeV}$.

We can then estimate how many of these scatterings are soft and how much these contribute to the total energy loss. We define a cut in energy loss, $\nu_{\rm{cut}}$, in order to separate the contribution of the "soft" collisions, with $\nu<\nu_{\rm{cut}}$, and of the "hard" collisions, with $\nu>\nu_{\rm{cut}}$, for all the quantities above. 
In particular, we would like to find $\nu_{\rm{cut}}$ such that $D_{\rm{soft}}\ll b_{\rm{soft}}$ to guarantee that the process can be treated classically, and $N_{\rm{soft}}(b_{\rm{soft}}^{-1})\gtrsim N_{\rm{hard}}(b_{\rm{soft}}^{-1})$ to make the number of hard scattering along the classical mean free path of length $b_{\rm{soft}}^{-1}$ small enough. 

What is often done in the literature (see for example Ref.~\cite{Lipari:1991ut}) is to proceed by trial and error to find $\nu_{\rm{cut}}\approx 10^{-2}$, such that $D_{\rm{soft}}/b_{\rm{soft}}\approx 4\times 10^{-3}$ and $N_{\rm{hard}}(b_{\rm{soft}}^{-1})/N_{\rm{soft}}(b_{\rm{soft}}^{-1})<15\%$. The hard scatterings will contribute diffusively to the muon energy loss, as $D_{\rm{hard}}\gtrsim b_{\rm{hard}}$, and are not taken into account by Eq.~\eqref{eq:energyloss}. Given the number of hard scattering per classical mean free path, neglecting their effect is consistent with a theory uncertainty on $b_\mu$ of about $15\%$. 

As a final remark note that the case of Bremsstrahlung is the one where the mean energy loss approximation of Eq.~\eqref{eq:energyloss} is less accurate because of its weak preference for soft photon emission. For taus, $b^{\rm{brehm}}_\tau \approx \left(\frac{m_\mu}{m_\tau}\right)^2 b_\mu $ is suppressed and $b^{\rm{pair}}_\tau \approx 20 \, b^{\rm{brehm}}_\tau = 0.014 \, \rm{km}^{-1}$~\cite{Bigas:2008ff, KOEHNE20132070} so that the energy loss is primarily due to DIS interactions. These are more peaked at low momentum exchange than Bremsstrahlung, making Eq.~\eqref{eq:energyloss} more reliable. The same would be true for electron-positron pair production from high energy muons, which would slightly reduce the theory uncertainty estimated for bremsstrahlung.

The theoretical predictions of the DIS cross section itself is highly uncertain. This uncertainty is reported in Fig.~\ref{fig:RelevantScales} and even further complicates the predictions of the contribution of tau decays to the final muon yield at the detector.
}

\section{Theory outputs}
\label{sec:TheoryOutputs}

In this section, we use the formalism and equations from \cref{sec:TransportEquation}, endowed with the theory inputs described in \cref{sec:TheoryInputs}, to obtain master formulae for predictions of muon events at $\nu$T. We distinguish events originating from the up-scattering of muon neutrinos and those originating from the decay of taus.

\subsection{Matching to observables}
All the observables can be directly matched to expectation values computed with the phase space distribution, as we have already done in \cref{eq:numero} for the total number of expected events, by identifying proper weights for each observable to be sampled with $f_\mu$. We compute the experimental rate of \cref{experimental-rate} as follows
\be\label{eq:anyDetector}
\left\langle \frac{\dd N_\mu}{\dd t\, \dd E_\mu \dd\Omega_p}\right\rangle = \int\limits_V  \dd^3x \, f_\mu(\vec x, E_\mu, \Omega_p)\times E_\mu^2\,\frac{v}{\ell(\vec x, \vec p)} \,,
\ee
where $\ell(\vec x, \vec p)$ is the distance traveled by a muon along a given direction inside the detector. This expression for the expected rate requires the knowledge of the full detector geometry, and it must in principle be endowed with a factor describing the detector efficiency, which depends in general on both the muon energy and the direction of flight, $\Omega_{\vec p}$. If one assumes that the efficiency is maximal when the surface is aligned with the muon direction, and zero otherwise, the rate only depends on the transverse area $A_T$ seen by the muons, and on the depth of the detector in the flight direction $\vec p$. For this reason, the volume element is better expressed as $\dd^3x = \dd\ell_{\vec{p}}\,\dd^2\Sigma_{\vec{p}}$, where $\ell_{\vec{p}}$ ranges from zero to $\ell(\vec x, \vec p)$. This measure may have a complicated expression for an arbitrary detector geometry, but it is well defined and computable. We see explicitly that by integrating along the depth of the detector as seen from direction $\vec n_p$ an effective transverse area $A_T(\vec n_p)$ arises, sampled by the value of $f_\mu$ inside the detector. By properly taking into account this effect, we could in principle distinguish between muons produced outside or inside the detector.\footnote{Let us consider the detector as a sphere of radius $R_{\rm det}$, and let us assume that $f_\mu$ does not depend on the coordinates orthogonal to $\ell_{\vec{p}}$, which is always our case. We have $\ell_{\vec{p}}=2\sqrt{R_{\rm det}^2-x^2-y^2}$ and $\ell_{\vec{p}}\in [-R_{\rm det},R_{\rm det}]$ as well as $d\Sigma=dxdy$ and $A_T(\ell_{\vec{p}})=\pi (R_{\rm det}^2-\ell_{\vec{p}}^2)$. The rate becomes (suppressing the subscript $\vec{p}$)
\begin{equation*}
\left\langle \frac{\rm{d} N_\mu}{\rm{d}t\,\rm{d}E_\mu}\right\rangle =\pi r^2 \int \frac{\rm{d}\Omega_{\vec p}}{(2\pi)^3}\,\int\limits_{-R_{\rm det}}^{R_{\rm det}} \frac{\dd\ell}{R_{\rm det}}\, \left(1-\frac{|\ell|}{R}\right) E_\mu^2 v f_\mu(\ell) \approx \pi R_{\rm det}^2\, \int \frac{d\Omega_{\vec p}}{(2\pi)^3}\,E_\mu^2 v f_\mu\,,
\end{equation*}
where in the last step we assumed $f_\mu$ constant inside the detector.
We see that it is a very good approximation to include the transverse area for all the case where $f_\mu$ does not change much inside the detector, which makes sense for detectors of moderate dimensions. Otherwise, we need to integrate $f_\mu$ with a weight $(1-|\ell_{\vec{p}}|/R_{\rm det})/R_{\rm det}$. We believe that large detectors -- such as the upgrade of \ic{} -- will be able to resolve the radial dependence of the muon distribution function and this effect will become more and more important.
%
%
}
For an ideal spherical detector of section $A_{\rm disk}$, we have
\beq\label{eq:dN:SphericalDetector}
\begin{split}
\left\langle \frac{dN}{\dd t \dd E_\mu \dd\Omega_p}\right\rangle &= A_{\rm disk}\, E_\mu^2 v\, f_\mu\big|_{\text{detector}}\\&=A_{\rm disk}\, \int\limits_0^{L(\Omega_p)} \dd\xi \int \dd E_\nu \dd\Omega_\nu \TT(E_\mu,\Omega_p| E_\nu, \Omega_\nu;\xi)\, \phi_\nu^\oplus(E_\nu,\Omega_\nu)\,.
\end{split}
\eeq
By integrating this on the time of exposure $T$, in the corresponding energy bin, and on the angular distribution of the muons, we arrive at the expected number of events in a given bin, 
which may be compared with experimental measurements. We emphasize that the only free parameters are the ones of the neutrino flux at the crust of Earth, $\phi_\nu^\oplus$. The transport function instead depends on parameters of the SM as well as on environmental quantities.

\rev{Note that combining \cref{eq:transport-to-detector,eq:dN:SphericalDetector} we can define the effective area as commonly used in the literature, in a fashion similar to \cite{Gonzalez_Garcia_2009}. 
The total rate of events in a muon energy and solid-angle bin $\Delta_{E_\mu} \otimes \Delta_{\Omega_{\vec p}}$ can be found from \cref{eq:dN:SphericalDetector} to be\footnote{Note that, relaxing the simplifying assumption that the detector is spherical, $A_{\rm disk}$ should be replaced with an angle-dependent transverse area accounting for the shape of the detector.}
\beq
\left\langle\frac{\dd N}{\dd t }\right\rangle = \int_{\Delta_{E_\mu}} \int_{\Delta_{\Omega_{\vec p}} } A_{\rm disk} \int \dd \xi \int \dd E_\nu \dd\Omega_\nu \TT(E_\mu, \Omega_{\vec p}|E_\nu, \Omega_\nu;\xi) \phi_\nu^\oplus(E_\nu, \Omega_\nu)\,.
\eeq
Comparing to the formula commonly employed in the literature
\beq
\left\langle\frac{\dd N}{\dd t }\right\rangle = \int \dd E_\nu \dd\Omega_\nu A_{\rm eff}^\nu(E_\nu, \Omega_\nu) \phi_\nu^\oplus(E_\nu, \Omega_\nu)\,,
\eeq
we read off the effective area as
\begin{equation}\label{eq:area-full}
    A_{\rm eff}(E_\nu, \Omega_\nu)= A_{\rm disk} \, \int\limits_0^{L(\Omega_{\vec p})} {\rm d}\xi  \int\limits_{\Delta_{E_\mu} \otimes \Delta_{\Omega_{\vec p}}} \dd E_\mu \dd\Omega_p \, \epsilon\left(E_\mu, \Omega_{\vec p}\right) \, \TT(E_\mu,\Omega_{\vec p}| E_\nu, \Omega_\nu;\xi)\,,
\end{equation}
where we introduced the efficiency $\epsilon\left(E_\mu, \Omega_{\vec p}\right)$, encoding the characteristics of the detector as well as of a specific analysis, which may be attained once and for all through dedicated detector simulations.} 
Typically the effective areas quoted in the literature refer to the whole spectrum of observable muons for a given angular bin, i.e. $\Delta_{E_\mu} = \left[E_\mu^{\rm th} , \infty\right]$, where $E_\mu^{\rm th}$ is an energy threshold that depends on the characteristics of the detector \cite{Gaisser:1990vg}. 

\subsection{Muon energy and detector efficiency}\label{sec:muon-master}
As shown in \cref{fig:RelevantScales}, the muon is the ideal lepton to be detected by the $\nu$T, because it loses energy at a rate that allows resolvable tracks in the detector, even if produced a few kilometers away. On the other side, electrons quickly lose all their energy, thus can be detected only if produced inside the detector; conversely, taus can travel long distances and can be detected only if they decay back to muons. 
\rev{As such, throughout this work we only consider events that are reconstructed as muon tracks in the detector, leaving other event topologies to a future study.}

We start then by writing a master formula to compute the expected number of events from a fixed angle $\Omega_p$ associated to the muon direction in the sky, in an energy bin $[E_\mu^{\rm min}, E_\mu^{\rm max}]$. It can be computed as in \cref{eq:dN:SphericalDetector} under the assumption of a detector with spherical symmetry. As can be seen from \cref{eq:anyDetector}, aside from detector geometry and energy binning, all the effects, including detector resolution and efficiencies, are encoded in $f_\mu$. Hereafter we assume maximal efficiency of the detector, and the only effect that we include in the computation of $f_\mu$, is enforcing that the energy of particles produced inside the detector volume is reconstructed without degradation from energy losses. In other words, if the muon is produced inside the detector volume, it is observed with the same energy with which it was produced, while if it is produced outside, the energy measured will be lower as it exponentially loses energy in the surrounding mediums. In formulae,
\be
\frac{\dd E_\mu}{\dd\xi}=\left\{
\begin{aligned}
 0\,\quad \quad &  \text{inside the detector}\\
 - b_\mu E_\mu \quad &  \text{outside the detector} \\
\end{aligned}\right .
\,.
\ee
where $\xi$ spans the LOS of length $L$ as in the previous section. This modifies the LOS of the muon momentum of \cref{eq:caratteristiche}, to yield
\be
g(\xi)\equiv \frac{E_\mu(\xi)}{E_\mu}=\left\{\begin{array}{ll} 1 \,&\text{inside the detector} \\
\exp\left[(b_\mu \left(L-R_{\text{det}} -\xi\right)\right]\, & \text{outside the detector} \end{array}\right.\,,
\ee
where $E_\mu$ is the measured muon energy and $R_{\text{det}}$ is the radius of the detector.

Now all is needed is to combine \cref{eq:fmu-completa} and \cref{eq:dN:SphericalDetector}. As emphasized, experimentally one is sensitive to muon tracks irrespectively of their origin, whether by neutrino up-scattering or tau decays. Nevertheless, we compute the two contributions separately.

\subsection{Events from \texorpdfstring{$\mu$}{mu}-neutrinos}
The calculation of the differential muon rate at the detector reduces to computing the integral of the transport function. For \( \mu \)-neutrinos, the transport function simplifies using the expression for the differential cross section, which can be integrated over \( \xi \), \( E_\nu \), and \( \Omega_\nu \) to obtain the expected number of events in the laboratory. To take the inelasticity into account, we make a change of variable when plugging \cref{eq:transport-function-mu} into \cref{eq:dN:SphericalDetector}, to get
\beq\label{eq:dN:MasterFormula:Nmu}
\begin{aligned}
\left\langle \frac{\dd N}{\dd t \dd E_\mu \dd\Omega_p}\right\rangle_{\nu_\mu} &= 4 \pi A_{\rm disk}\, \int\limits_0^{L(\Omega_p)} \frac{\dd\xi\, n_N(\xi)}{g(\xi)^{2}}\\
&\times\int \frac{\dd y\py}{(1-y)} 
D\left(\xi, \frac{E_\mu(\xi)}{1-y}\right)
\sigma\lp\frac{E_\mu(\xi)}{1-y}\rp \,\phi_\nu^\oplus\lp\frac{E_\mu(\xi)}{1-y}\rp\,, 
\end{aligned}
\eeq
without any approximation. This is the formula we employ in our numerical calculations in the next section.

Let us notice an important interplay of scales that allows us to make some approximations. Since $1/g(\xi)^2$ is exponentially small away from the detector, the first term only has support on a region around the experiment, say of size $R_{\rm det}+1/b_\mu$. As a consequence, the relevant density is that of the material surrounding the detector, $n_N^{\rm near}$. On those scales the neutrino transparency is roughly a function of only $L/L_\nu(E)$ at the corresponding energy. A further simplification consists in taking the elastic limit $\py=\delta(y)$. By plugging the expression for the power-law flux \cref{eq:NeutrinoFlux}, the differential rate reads:
\be\label{eq:muon-approx}
\left\langle \frac{\dd N}{\dd t \dd E_\mu \dd\Omega_p}\right\rangle_{\nu_\mu} \approx 4 \pi A_{\rm disk}\, n_N^{\rm near}\sigma(E_\mu) \phi_\nu^\oplus(E_\mu) \exp{\lp-\frac{L(\Omega_p)}{L_\nu(E_\mu)}\rp}\int \frac{\dd\xi}{g(\xi)^{2+\gamma-\lambda}}\,,
\ee
where last term yields an effective length
\be\label{eq:effectiveradius}
\int \frac{d\xi}{g(\xi)^{2+\gamma-\lambda}} \approx R_{\rm det} +\frac{1}{b_\mu(2+\gamma-\lambda)}\,.
\ee
This formula has two implications. First, only muons produced at distances of about $1/b_\mu$ or shorter contribute to the rate, and second, that the effective volume is larger than simply $A_{\rm disk}R_{\rm det}$, since it receives a correction that depends on $b_\mu$, a parameter of muon physics. Different particles will have different geometrical effects depending on their energy loss properties and their lifetime. Note that \cref{eq:effectiveradius} assumes a diffuse source with a power-law flux of positive spectral index $\phi_\nu \sim E_\nu^{-\gamma}$, as in \cref{eq:NeutrinoFlux}. In general one may obtain different $\mathcal{O}(1)$ coefficients in front of $1/b_\mu$ depending on the impinging flux.

From \cref{eq:muon-approx}, we see that the full angular dependence enters in the neutrino transparency factor $D\approx e^{-L/L_\nu}$. Depending on the directionality of the neutrino flux, the integral over $\Omega_p$ yields different results.  For diffuse sources, averaging over the full sky, we get
\be
\langle e^{-L/L_\nu}\rangle_{\rm sky} =\int\frac{\dd\Omega}{4\pi}e^{-L/L_\nu} = \frac{1}{2}\lp 1 + \frac{L_\nu}{4R_\oplus}(1 - e^{-2R_\oplus/L_\nu})\rp\,.
\ee
The average transparency is the same for different experiments located approximately at the same depth, and it is at worst $\approx 1/2$.

For PSs, instead, we need to average over a day, and now $\Omega_p=\Omega_{\rm PS}(t)$, obtaining
\be
\langle e^{-L/L_\nu}\rangle_{\rm PS, day}=\frac{1}{24\,\mathrm{h}}\int \dd t e^{-L(\Omega_{\rm PS}(t))/L_\nu}\sim \frac{\text{hours w/o attenuation}}{24\,\mathrm{h}}\,.
\ee
This factor is instead strongly sensitive to the location on the sky as seen by the detector, and does not cancel out when taking ratio of expected number of events at two experiments.

\subsection{Events from \texorpdfstring{$\tau$}{tau}-neutrinos}\label{sec:tau-master}
In this case we need to use the $\tau$ transport function \cref{eq:transport-function-tau} in \cref{eq:dN:SphericalDetector}. To simplify the expression we assume that the differential decay rate of $\tau$s into muons is proportional to $\delta(E_\tau-q E_\mu)$, where at high energies $q=3$ to a very good approximation,
\be
\frac{d\Gamma}{d^3p} =\Gamma(E_\tau) \frac{\delta(E_\tau-q E_\mu)\delta^{(2)}(\Omega_\tau-\Omega_p)}{E_\mu^2}\,.
\ee
\rev{Moreover, recall that we are working under the approximation of negligible tau energy loss, viz. $b_\tau \approx 0$. Later on, we shall discuss some of the subtleties of this assumption.}
The expected number of events used in our numerical computation is then
\be\label{eq:dN:MasterFormula:Ntau}
\begin{aligned}
\left\langle \frac{dN_\mu}{dt dE_\mu d\Omega_p} \right\rangle_\tau = 4\pi & A_{\rm disk} n_N^{\rm far}  \mathcal{B}_{\tau\to\mu}\int\limits_0^{L(\Omega_p)} \frac{d\xi}{g(\xi)^2}\ \int \dd y \frac{\py}{(1-y)} \\
& e^{\left(-\xi n_N^{\rm far} \sigma \lp \frac{q E_\mu(\xi)}{1-y}\rp\right)}\left(1-e^{-\frac{\xi}{d_\tau(\xi)}} \right)\sigma\left(\frac{E_\tau}{1-y}\right) \phi_\nu^\oplus\left(\frac{q E_\mu(\xi)}{1-y}\right)\,,
\end{aligned}
\ee
where $\mathcal{B}_{\tau\to\mu}\approx 17.4\%$ (\cref{eq:BranchingRatio:tau}). Here we made the approximation of neglecting the neutrino attenuation rate with respect to the tau decay rate in the first exponential, which introduces an error of at most $10\%$ at the energies of our interest. This way we have a formula that we can immediately compare with \cref{eq:dN:MasterFormula:Nmu}. We notice that on top of neutrino transparency we have a suppression from the finite lifespan $d_\tau$. The above formula is obtained from \cref{eq:transport-function-tau} by performing the integral in $\xi'$ with the approximation of constant $n_N(\xi') = n_N^{\rm far}$. Indeed, just like muon neutrinos must convert in a thin shell near the detector, taus must decay in a shell of similar width, at a distance of about $d_\tau$ from the detector (see \cref{fig:sketch-earth}). The width of both these shells are controlled by $1/b_\mu \sim \text{few }\si{\kilo\metre}$, over which the composition of Earth does not vary significantly.

As for the muon case, the integral in $\xi$ selects a region around the detector, so that both neutrino transparency and tau lifetime are in practice evaluated at $\xi \approx L$. With the same approximations that lead to \cref{eq:muon-approx}, once again for a diffuse power-law flux \cref{eq:NeutrinoFlux}, we can write in this case
\be\label{eq:tau-approx}
\begin{split}
\left\langle \frac{dN}{dt dE_\mu d\Omega_p}\right\rangle_{\tau} &\approx 4\pi A_{\rm disk}\,  n_N^{\rm far}\, \mathcal{B}_{\tau\to\mu}\, \sigma(q E_\mu)\, \phi_\nu^\oplus(q E_\mu) \\
&\times \exp{\lp-\frac{L(\Omega_p)}{L_\nu(q E_\mu)}\rp}\, \left[1-\exp{\lp-\frac{L(\Omega_p)}{d_\tau(q E_\mu)}\rp}\right]\int \frac{d\xi}{g(\xi)^{2+\gamma-\lambda}}\,.
\end{split}
\ee
We see that the angular dependence is very important for the contribution from $\tau$ decays, arising mainly from directions where $L\gg d_\tau$, the others being linearly suppressed by $\approx L/d_\tau$. While this averages out for a diffuse neutrino sources, it can be important for PSs.

\rev{Having neglected the $\tau$ energy loss has removed one length scale from the problem. However, one needs to treat with care all cases where the other scales, $d_\tau$ and $L$, are comparable with $1/b_\tau$.} We comment upon this in the next section, when comparing this effect with data. Let us \rev{insist here}, for the sake of argument, \rev{on taking} $b_{\tau}=0$ (see \cref{sec:BSM}, where we expand on this idea in the context of physics BSM). We see that the integral over the angular distribution yields
\be
\langle e^{-L/L_\nu}\rangle (1-\langle e^{-L/d_\tau}\rangle).
\ee
The neutrino transparency is further suppressed by the tau lifetime, thus we expect the contribution from tau leptons to strongly depend on the geometry; in particular, the ratio of the expected number of tau-induced events at \ic{} and \kmn{} may differ from that of expected number of muon-neutrino-induced events, especially for PSs. However, one expects a smaller total number of the former type of events as compared to the latter, which are not suppressed by any factor of $L/d_\tau$. \rev{Numerically, we find the $\tau$-initiated events \revnew{-- that is, muon tracks -- }to constitute a percent-level fraction of the total number of events \cite{Abbasi:2021qfz,Stettner:811376}.}

\section{Results and discussions}
\label{sec:Results}
We discuss here the main results of this work. We start by presenting the experimental setups of \ic{} and \kmn{} and the respective datasets in \cref{sec:ExpSetup}. We explain our statistical treatment in \cref{sec:stat}. We then discuss the two cases of interest: a diffuse source of neutrinos with a power-law spectrum in \cref{sec:Diffuse}, and a PS with an energy localized spectrum in \cref{sec:PointSource}.

\subsection{Data and experiments}
\label{sec:ExpSetup}
%

%
\ic{} and \kmn{} have a very similar setup. Both are composed of long arrays of photo-multiplier tubes (PMTs), submerged in ice and sea water, respectively. The environment acts both as a target for the neutrino CC scatterings and as a medium for the Cherenkov detector. It is then essential to have large volumes and sufficiently dense material in the surroundings. Here we briefly describe the two experiments and report the relevant values for their features in \cref{tab:ExpSetup}.

\begin{table}[bh]
    \centering
    \begin{tabular}{c c c}
    \toprule
        & \ic{} & \kmn{} \\ 
        \midrule
        Depth (km) & 1.45 - 2.45 & 3.5 \\ 
        Location (lat,lon) & ($89^\circ 59' \rm S$, $63^\circ27' \rm W$) & ($36^\circ 16' \rm N$, $16^\circ06' \rm E$)\\
        Density ($\gcm$) & 0.918 & 1.02 \\
        \midrule
        Volume (km$^3$) & 1.0 & 0.15 \\
        Time of exposure (years) & 12 & 0.79 \\
        \bottomrule
    \end{tabular}
    \caption{Properties of the two experiments considered in this work. Note that the exposure time refers to that at the time of the \eventname{} observation, namely up to February of 2023.}\label{tab:ExpSetup}
\end{table}

\begin{figure}[t]
    \centering
    \includegraphics[width=.85\textwidth]{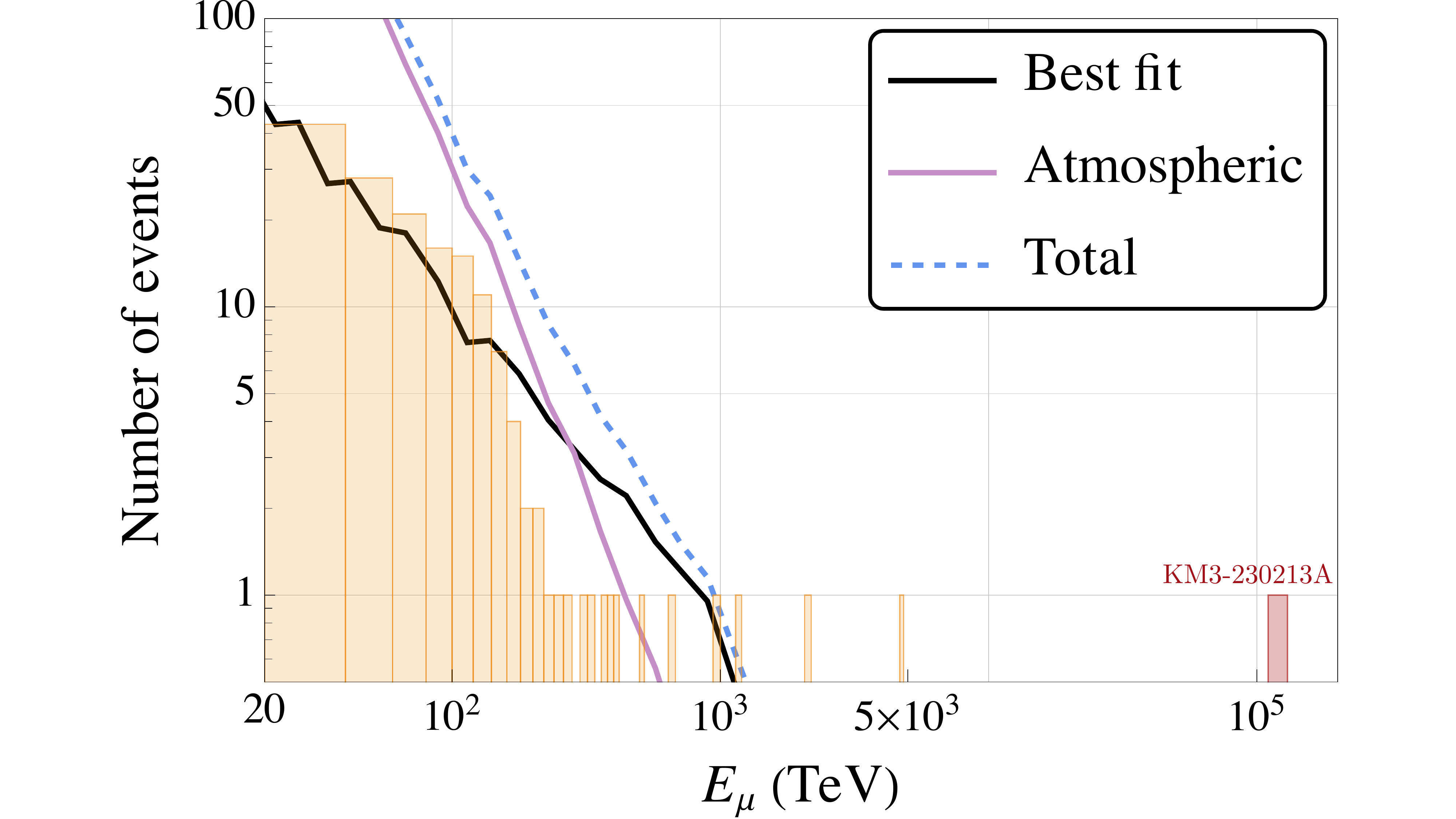}
\caption{\label{fig:IC-HESE}Energy distribution of muon events observed at IceCube (orange), for events with $E_\mu > 20$ TeV. The burgundy bar indicates the \eventname{} energy. The black line shows the best fit distribution for a diffuse power law source (see \cref{sec:Diffuse}), while the purple and dashed blue lines show the atmospheric background prediction and the total event distributions, respectively. }
\end{figure}

\begin{figure}[t!]
    \centering
    \includegraphics[width=1\textwidth]{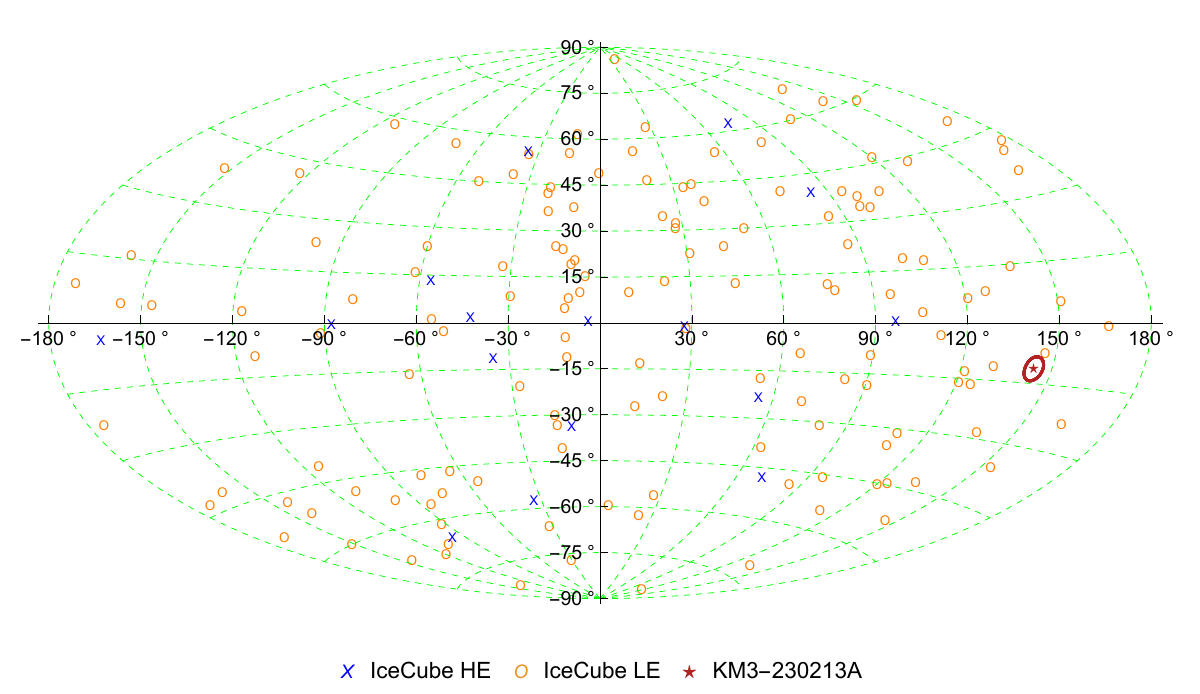}
\caption{High energy neutrino sources inferred by \ic{}: the blue crosses (orange circles) indicate events with $E_\nu > 200$ TeV ($E_\nu < 200$ TeV). The red star represents the \eventname{} reconstructed source, with the circle showing the 90\% confidence level region. Data taken from Ref.~\cite{IceCube:2023sov}.}
 \label{fig:Sources}
\end{figure}

\ic{} is located at the South Pole ($89^\circ 59' \rm S$, $63^\circ27'\rm W$) and frozen in optically clear ice, with an average density of $\rho_{\rm ice} = 0.918 \, \gcm$. The arrays stretch at depths between 1.45 km and 2.45 km, for a total volume of 1 km$^3$, corresponding to $R_{\rm IC} = \SI{0.62}{\kilo\metre}$ under the assumption of a spherical detector. \ic{} has observed about thirty high-energy neutrino events~\cite{IceCube:2023sov} between $\SI{200}{\tera\electronvolt}$ and $\SI{4}{\peta\electronvolt}$, whose energy distribution, shown in \cref{fig:IC-HESE}, is consistent with a power-law (see \cref{sec:Diffuse} for more details). The inferred source locations of these events are shown in \cref{fig:Sources}, where we indicate with blue crosses (orange circles) the events with energy $E_\mu > 200$ TeV ($E_\mu < 200$ TeV), as at this energy the background from muon produced by atmospheric neutrinos becomes subdominant. At the time of \eventname{}{}  detection, \ic{} had a total exposure time of 12 years. 

\kmn{}, specifically the ARCA module~\cite{KM3Net:2016zxf}, is under construction at around 100 km off the coasts of Sicily, in the Mediterranean Sea ($36^\circ 16' \rm N$, $16^\circ06' \rm E$); the average density of sea water is $\rho_{\text{water}} = 1.02 \gcm$. At the time of the \eventname{} observation, the deployed module consisted of 21 strings of PMTs, at an average depth of 3.5 km, for a total volume of 0.15 km$^3$, corresponding to a sphere of radius $R_{\rm KM3} = 0.33\km$. With this setup, ARCA has collected 288 days of data, that is, $0.79$ years.

\subsection{Statistical framework}\label{sec:stat}
We have two datasets, one from \ic{} and one from \kmn{}, and we would like to understand whether they are in tension and see whether they can be combined. The number of observed events at \ic{} is taken from the 12-year data release~\cite{IceCube:2023sov}, as already discussed in the previous section. For \kmn{}, as no dataset is available yet, we assume that the distribution of events observed at $E_\mu\lesssim4$ PeV is consistent with the distribution inferred by \ic{}, that is, a diffuse flux with a power-law energy distribution. 

Since we deal with a very small number of observed events, we model the likelihood in each bin as a Poisson distribution. The individual likelihoods for the two datasets are
\begin{eqnarray}
\label{eq:L-IC}
\mathcal{L}_{\rm IC}&=& \prod_{i=1}^{n_b} {\cal P}\lp N_{i, \rm obs}^{\rm IC} | N_{i, \rm exp}^{\rm IC}(\hat\theta) \rp \,,\\
\label{eq:L-KM}
\mathcal{L}_{\rm KM}&=& {\cal P}\lp N_{\rm obs}^{\rm KM} | N_{\rm exp}^{\rm KM}(\hat\theta) \rp\,,
\end{eqnarray}
where ${\cal P}(k|n) = \frac{n^k e^{-n}}{k!}$. Here $N_{i,\rm obs}$ and $N_{i,\rm exp}(\hat\theta)$ represent the number of observed and expected events in the $i$-th bin in each experiment (for \kmn{} we only consider one ultra-high-energy bin), for a certain set of parameters $\hat\theta$, which are defined by the specific model considered. The number of energy bins is indicated by $n_b$; we consider the energy bins shown in \cref{fig:IC-HESE}, plus one additional bin covering the region $E_\mu > 10$ PeV, where zero events were observed by \ic{}, and one by \kmn{}. 

We employ Bayes factors as a means to compare different hypotheses, taking into account prior knowledge of models and parameters. For each hypothesis (i.e., model) $M_i$, described by the set of parameters $\hat\theta_i$, we infer the posterior distribution $P(\hat\theta_i|D)$ and we compute the ratio of evidences
\be
B_{1\to 2}(D)\equiv \frac{\int P(\hat{\theta}_1|D)\dd\hat{\theta}_1}{\int P(\hat{\theta}_2|D)\dd\hat\theta_1}=\frac{\int \mathcal{L}_D(\hat\theta_1)\pi_1(\hat\theta_1)\dd\hat{\theta}_1}{\int \mathcal{L}_D(\hat{\theta}_2)\pi_2(\hat{\theta}_2)\dd\hat{\theta}_2}\,,
\ee
where in the second step we have introduced the prior probabilities $\pi_i$, in view of Bayes theorem. Following the usual criterion for Bayes factors, $B\gg 1$ ($B\ll1$) is an informed suggestion that the model at the numerator (denominator) is favored as an explanation of the observed data $D$. 

Note that the Bayes factor can be also used to quantify the possible tension of two datasets, $D_1$ and $D_2$, tested against the same model. This can be done when there is prior knowledge that a model is a good fit to $D_1$ for a given (normalized) prior distribution $\pi_{\rm ref_{D1}}(\hat\theta)$ of its parameters (usually called the reference or standard model). A tension between $D_1$ and $D_2$ -- with respect to the reference model -- arises if there is a choice of priors $\pi(\hat\theta)$ that makes the Bayes factor (with respect to the same dataset $D_2$) large:
\be\label{eq:bayes-D1D2}
B_{D_1/D_2}=  \frac{\int \mathcal{L}_{D_2}(\hat\theta|\text{M})\pi(\theta)\dd\hat\theta }{\int\mathcal{L}_{D_2}(\hat\theta|\text{M})\pi_{\rm ref_{D1}}(\hat\theta)\dd\hat\theta}\gg1\,
\ee
In this case, the reference model must be extended to make sense of \( D_1 \) and \( D_2 \). This indicates a tension between the two datasets using a given model—one with free parameters and one prioritized based on a fit to \( D_1 \). In principle, one could take a fit to \( D_2 \) as the reference and compute the analogous quantity \( B_{ D_2/D_1} \). If the datasets are compatible, these two quantities should be similar; however, in the presence of tension, they need not be.%
\footnote{\rev{Note that here we make no mention of the tension in terms of "number of $\sigma$"; this because the likelihood of either dataset cannot be well approximated by a Gaussian distribution. Nevertheless, this approach can give an intuitive picture of how \emph{far apart} the two datasets are. Thus, in \cref{tab:anticipation-summary} and \cref{tab:summary}, we also quote results for the tension in terms of $\sigma$. In order to do so, we assume that the Bayes factor is distributed as a $\chi^2$ with two degrees of freedom. In this approximation, the significance ($\Delta \sigma$) is computed as 
\beq\label{eq:DeltaSigma}
\Delta\sigma = \sqrt{2}\,{\rm erfc}^{-1} \left[ Q\lp\frac{d}{2};2\ln B \rp \right]\,, 
\eeq
where $d$ is the number of degrees of freedom, $Q$ is the regularized $\Gamma$-function, $Q(a,z)\equiv \Gamma(a,z)/\Gamma(a,0)$, with representation $\Gamma(a,z)=\int\limits_z^\infty dt\, t^{a-1} e^{-t}$.}
}

Concretely, in the next sections we use \cref{eq:bayes-D1D2} to quantify the tension between \ic{} and \kmn{}. In the case of a diffuse power-law flux, discussed in \cref{sec:Diffuse}, it is only natural to take the \ic{} fit as a reference model, and use the \kmn{} likelihood under the sign of integral. In the case of a generic source whose flux is localized in energy to the UHE bin, both the fit to \ic{} null result or the \kmn{} measured event can be used as reference models, in principle.

\subsection{Power-law diffuse neutrino spectrum}\label{sec:Diffuse}
In this section we analyze the results of \ic{} and \kmn{} under the assumption of a diffuse neutrino source with a single power-law energy spectrum. By diffuse source, we mean an isotropically distributed source of neutrinos, whose flux is independent on the direction in the sky. In reality, we expect that 
this source is a collection of many different, unresolved PSs, whose distribution can be approximated as continuous in the sky. In this model there are only two free parameters, namely the normalization of the flux, $\phi_0$, and its spectral index, $\gamma$, of \cref{eq:NeutrinoFlux}. \rev{In what follows, we only consider the muon component of the neutrino flux, deferring a proper treatment of the tau-neutrinos and tau decays into muons for later.} \ic{} \cite{IceCube:2013low, IceCube:2020acn, Abbasi:2021qfz, IceCube:2023sov,Silva:2023wol, IceCube:2024fxo} interprets its high energy tail of reconstructed neutrino events with a single power-law diffuse background of the form \cref{eq:NeutrinoFlux}. We proceed to assess whether this model can also fit the \kmn{} event.

Note that, for this case, there is little to no difference between the experimental setups of \kmn{} and \ic{}, meaning that in an exposure time interval $\Delta T$, if the experiments had the same volume, we would expect a similar number of events from a diffuse source to be observed in both detectors. An intuitive understanding of the previous statement can be drawn by inspection of \cref{eq:dN:MasterFormula:Nmu}, under the approximation that the neutrinos are unattenuated above the horizon, and entirely attenuated below it; at energies $\lesssim1$ EeV, this behavior can be roughly seen in \cref{fig:Dfactor}, as above the horizon the neutrino will only incur in the water shell of the PREM, while for larger inclinations the trajectory intersects layers of higher density. Under this approximation the two experiments really only differ for their exposures and surrounding materials, whose densities are however comparable, being water and ice.

\begin{figure}[t]

    \includegraphics[width=.475\textwidth, height=.475\textwidth]{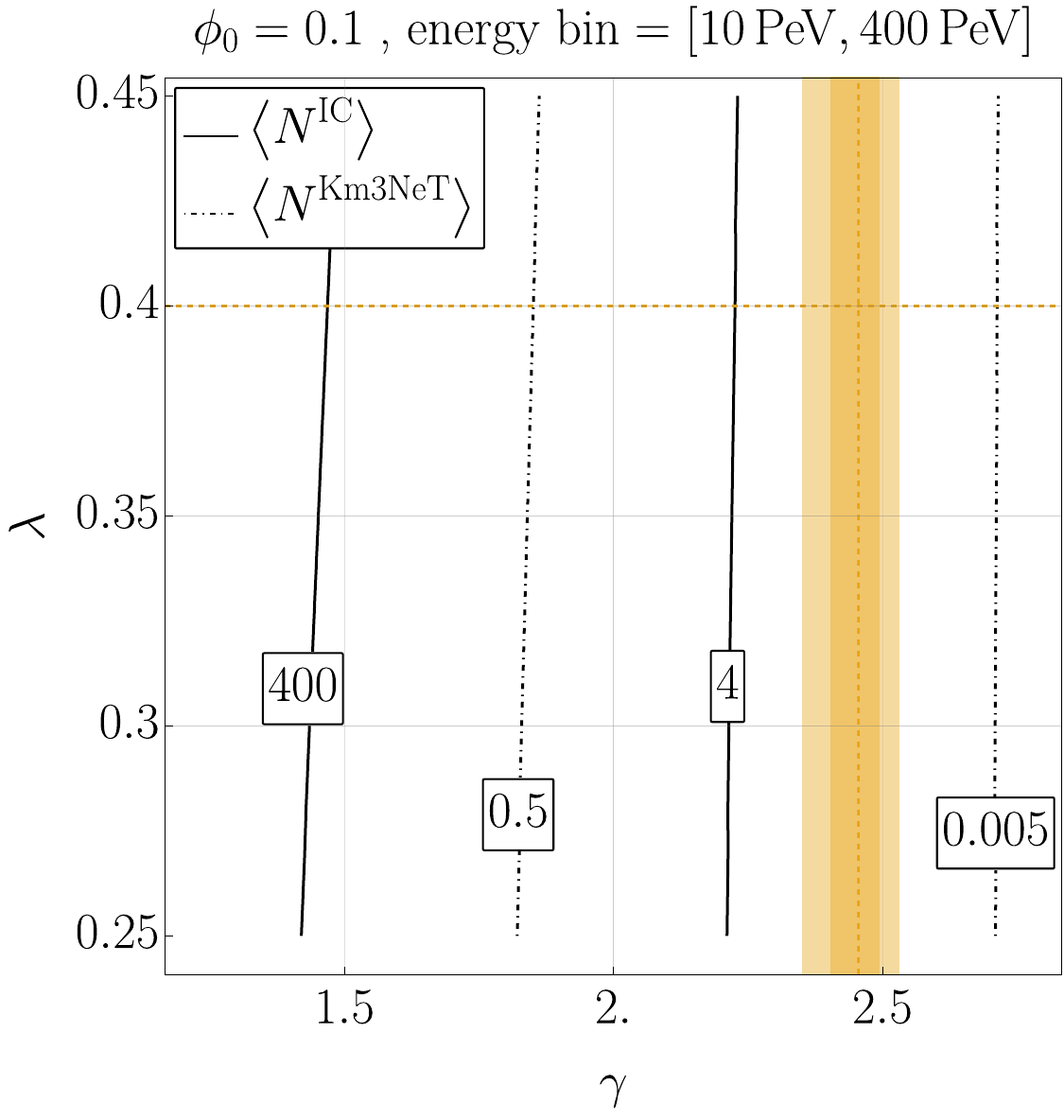}
    \hfill
    \includegraphics[width=.475\textwidth, height=.475\textwidth]{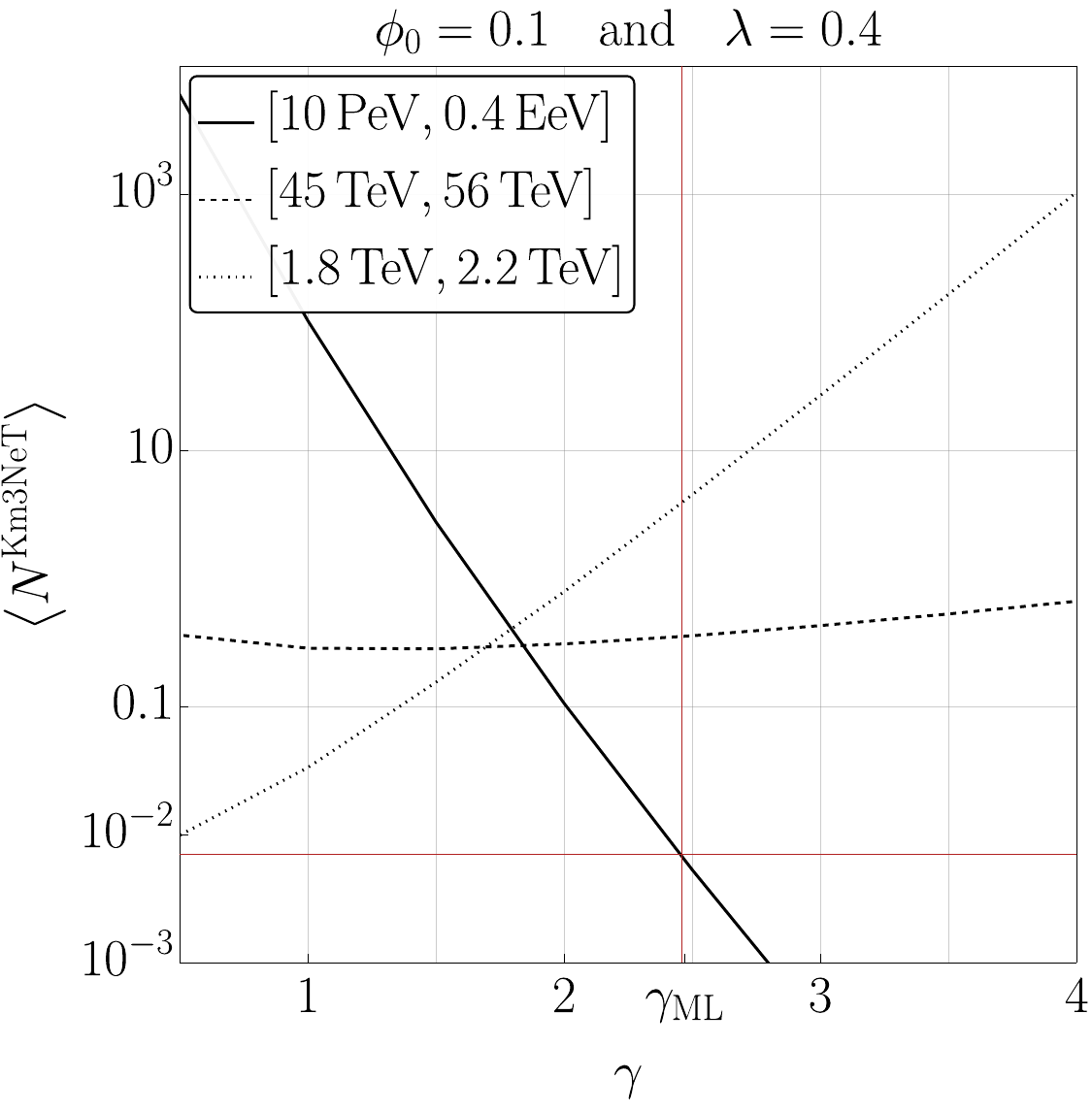}
    \caption{{\bf Left:} Expected number of events in the UHE bin $\left[10\,\PeV,400\,\PeV\right]$ (continuous line in the right panel) at \ic{} (continuous) and \kmn{} (dot-dashed) in the case of a diffuse, power-law flux, as a function of the spectral index $\gamma$ of \cref{eq:NeutrinoFlux} and the cross section high energy scaling $\lambda$ in \cref{eq:crossection-Above10PeV}. The flux normalization is fixed to the value maximizing the likelihood (see \cref{fig:diffuse-fit} and \cref{eq:best-fit}). The orange band shows the best fit and errors on the spectral index, while the horizontal dashed line is the value of $\lambda$ used by {\tt MadGraph}.
    {\bf Right}: Number of expected events at \kmn{} as a function of the spectral index $\gamma$, with fixed $\lambda=0.4$. The red vertical line shows the best fit for $\gamma$ \cref{eq:best-fit}. Recall that the flux is normalized at $E_{\text{min}} = 100\,\TeV$, which explains the decreasing, approximately constant, and increasing behavior with increasing $\gamma$ in the three bins, respectively.}
    \label{fig:Diffuse:Results}
\end{figure}

\begin{figure}[t]
    \includegraphics[width=\textwidth]{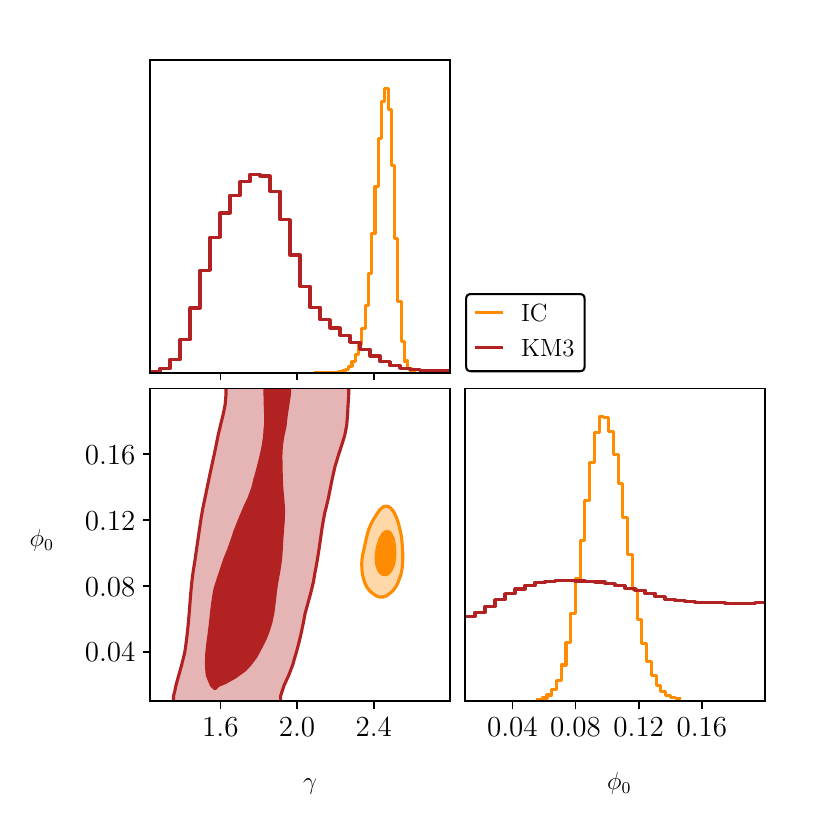}
    \caption{Results of the MCMC analysis \cite{emcee}. We show the one-dimensional and two-dimensional posterior distributions for the flux normalization $\phi_0$ and the spectral index $\gamma$, as per
    \cref{eq:NeutrinoFlux} with $E_\ast=100\,\TeV$, in the case of \ic{} alone (egg) and \kmn{} alone (vegan bacon), assuming flat priors on all parameters. }\label{fig:diffuse-fit}
\end{figure}

In \cref{fig:Diffuse:Results} we show how the expected number of events depends on the spectral index, $\gamma$, and on the cross section energy scaling, $\lambda$. In the left panel we show the expected number of events in the highest energy bin at both \ic{} and \kmn{}. The dependence on $\lambda$ is very mild and for this reason we fix $\lambda=0.4$ from now on. In the right panel we show the expected number of events at \kmn{} for the highest energy bin together with an intermediate bin and the lowest energy bin that we consider. Since the flux is normalized at $E_{\rm}=100\,\rm{TeV}$ we see how higher values of $\gamma$ suppress the number of events in the highest energy bin in favor of the lowest one.

The ratio of expected muon events between \ic{} and \kmn{} in a given energy bin is independent of the energy bin itself and depends mainly on the spectral index \(\gamma\), while the flux normalization cancels out. Explicitly, we find:
\be\label{eq:ratio_diff}
\begin{aligned}
\Delta_{\rm diffuse} = \frac{\left\langle N_\mu \right\rangle_{\rm diffuse}^{\rm IC}}{\left\langle N_\mu \right\rangle_{\rm diffuse}^{\rm KM3}}\approx & \frac{[T A]_{\rm IC}}{[T A]_{\rm KM}}\times \frac{\left(R_{\rm det}+\frac{1}{b_\mu(2+\gamma-\lambda)}\right)_{\rm IC}}{\left(R_{\rm det}+\frac{1}{b_\mu(2+\gamma-\lambda)}\right)_{\rm KM}} = \\
=&68 + 2.0 (\gamma - 2.46) - 0.16 (\gamma - 2.46)^2 +\cdots\,,
\end{aligned}
\ee
where in the second equality we used the numerical inputs from \cref{tab:ExpSetup}, the \(b_\mu\) factors for ice and water and $\lambda = 0.4$. The dependence on \(\gamma\) is not particularly strong, and the formula highlights an important point: the effective volume of the experiment is enhanced by \(\sim 1/b_\mu\), which reduces the ratio in favor of \kmn{}, whose geometrical volume is smaller: without this effect, the ratio would be closer to $100$. Properly accounting for this effective volume is hence key to understanding the tension between the experiments. We later discuss the impact of muon production from \(\tau\)-decays in this context, as well as the possibility of enhancing the effective volume via BSM physics.

Our strategy is to fit \kmn{} alone and \ic{} alone, and extract priors on $(\phi_0,\gamma)_{\rm IC}$ from the fit to \ic{}. This allows us to understand if the two datasets can be combined, since the possible tension between the two dataset is quantified according to the criterion in \cref{eq:bayes-D1D2} by referring to the baseline model determined by the \ic{} fit.

Before discussing the fits, let us derive an analytic understanding. The tension is quantified as in \cref{eq:bayes-D1D2}, where we use $\mathcal{L}_{\rm KM}$ as in \cref{eq:L-KM} and we compare two models - distinguished by a completely flat prior and a prior informed by the \ic{} fit - on the same \kmn{} data point. Given that the \ic{} prior suggests that the number of expected events at \kmn{} is $N\approx N^{\rm prior}_{\rm IC}/\Delta_{\rm diffuse}$, our Bayes factor can grow linearly with $\Delta_{\rm diffuse}$. 
This quantifies the tension and allows us to understand if the data can be combined for the diffuse model. Knowing the full posteriors of the two models, this can be done numerically.
With our approach, it is possible to perform the fit to \ic{} data, and we have no need to take the results of the collaborations as was done in Ref.~\cite{titans}. In particular, we do not expect our normalization of the flux to perfectly match that used by the collaboration. 
\rev{We 
take the reconstructed muon energies from the track events in the last $31$ bins in the right panel of Fig. 1 of Ref.~\cite{Abbasi:2021qfz}, using the quoted model for the atmospheric muon background, plus one empty high energy bin between $\SI{10}{\peta\electronvolt}$ and $\SI{400}{\peta\electronvolt}$ (where we assume negligible background).}
We take a flat prior on both the normalization $\phi_0$ and the spectral index $\gamma$ of the flux, and take the likelihood to be Poissonian in each of the bins as in \cref{eq:L-IC}. As far as \kmn{} is concerned, we take the same flat prior, but we only consider one bin between $\SI{10}{\peta\electronvolt}$ and $\SI{400}{\peta\electronvolt}$. We perform a Monte-Carlo Markov chain (MCMC) analysis \cite{emcee}, whose results we show in \cref{fig:diffuse-fit}. We refer to the neutrino flux of \cref{eq:NeutrinoFlux}, where $E_\ast$ is taken to be $100\,\TeV$. As expected, there is a large degeneracy between the normalization and the slope of the spectrum from the point of view of \kmn{}, as they have only observed one event. The values of the parameters that maximize the \ic{} likelihood are
\beq\label{eq:best-fit}
\gamma_{\rm ML} = 2.46\pm0.05 \quad \phi_{0,\,\rm ML} = 0.10\pm0.01\,,
\eeq

We are now in the position to quantify the tension. Using a flat prior encompassing a large portion of the \kmn{} likelihood and the normalized posterior of the \ic{} fit in the numerator and denominator of \cref{eq:bayes-D1D2}, respectively, we find a Bayes factor of about $20$. According to a common criterion \cite{Jeffrey}, this corresponds to a level of tension from substantial to strong, as summarized in \cref{tab:summary}. \rev{Alternatively, using \cref{eq:DeltaSigma}, we can quantify the tension to be $3.1\,\sigma$, in agreement with Ref.~\cite{titans}}.

\subsection{Point source}
\label{sec:PointSource}
Here we consider the possibility that the \eventname{} event originated from a PS, located at the coordinates  $(\text{RA}, \text{Dec})_{\rm{J }2000} =(94.3\degree, -7.8\degree)\pm3\degree$ (90\% CL), as inferred by the \kmn{} Collaboration, and shown as a red circle in \cref{fig:Sources}.

\begin{figure}[t]
    \centering
    \includegraphics[width=.85\textwidth]{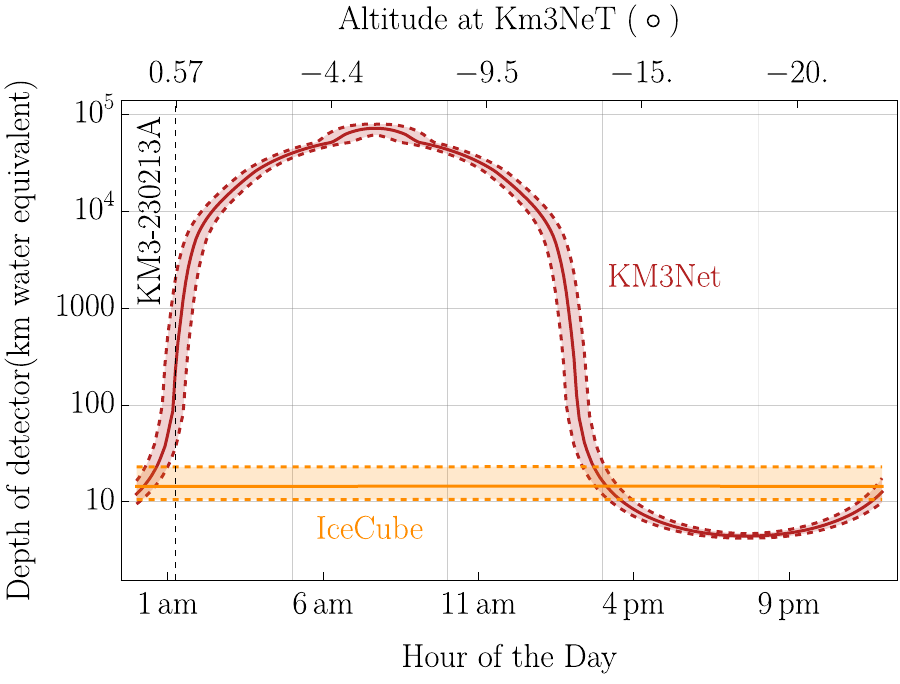}
\caption{\label{fig:PointSource:DistVSTime} Depth, in equivalent kilometers of water, of the \kmn{} (burgundy) and \ic{} (orange) detectors as seen from the point in the sky relative to the KM3-230213A observation, as a function of time (bottom axis) and corresponding altitude of the source as seen by \kmn{} (top axis). We compute the trajectory of the alleged PS using the python package \tt{pyephem} \cite{pyephem}.}

\end{figure}

In this case we can still use our master formulae of \cref{eq:dN:MasterFormula:Nmu} and \cref{eq:dN:MasterFormula:Ntau}, with an important caveat. The angular dependence $\Omega_p$ now changes in time during the day (and the year) and it is locked to the position in the sky of the PS as seen from the detector location. All the master formulae need to be used imposing at any time $\delta^{(2)}(\Omega_p-\Omega_{\rm PS, LAB}(t))$, and then average the rate over a day. We are then going to discuss day-averaged rates at each of the two experiments, neglecting the yearly revolution. 

\begin{figure}[t!]
    \centering
    \includegraphics[width=.9\textwidth]{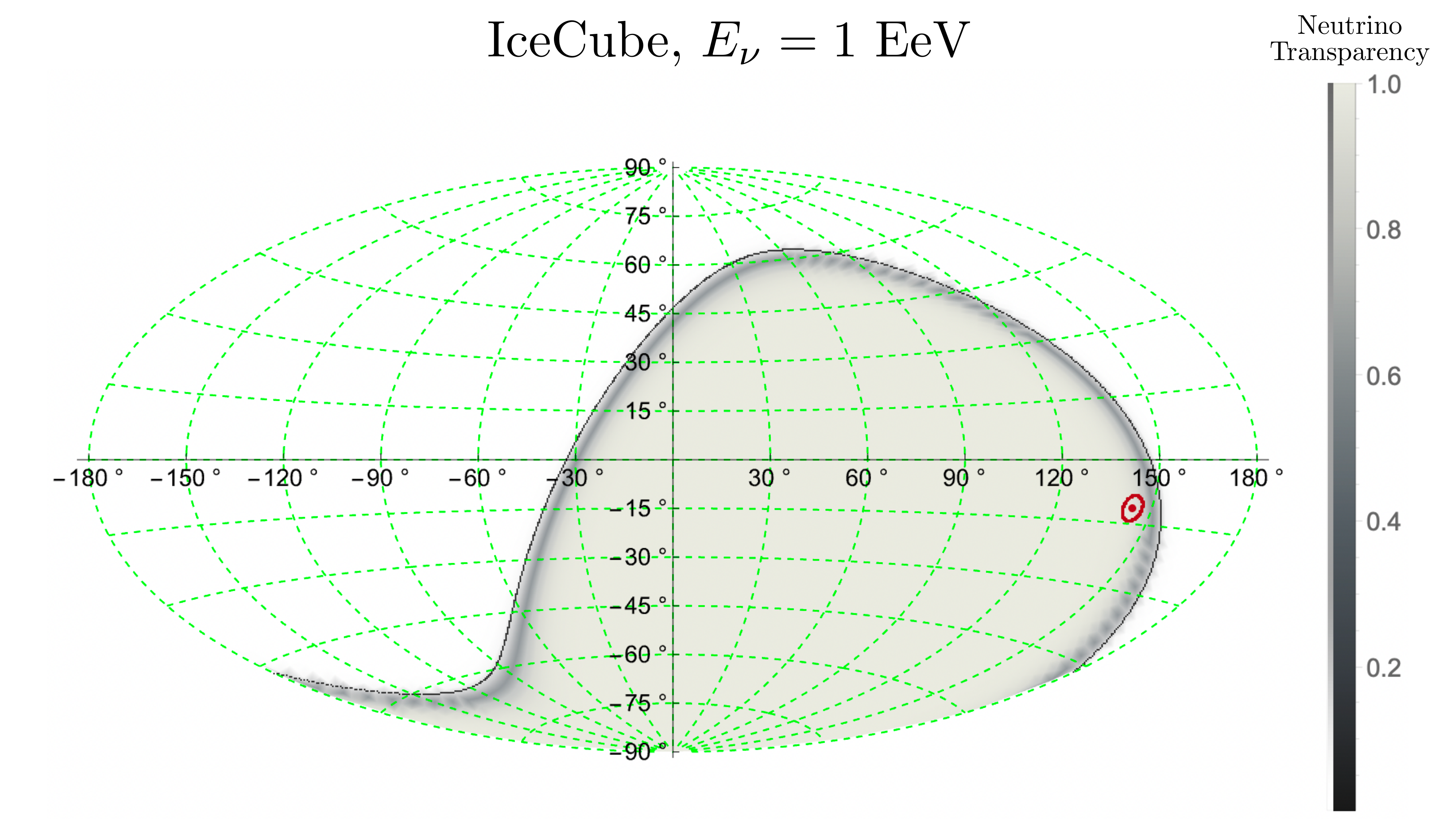}\vspace{1cm}
    \includegraphics[width=.9\textwidth]{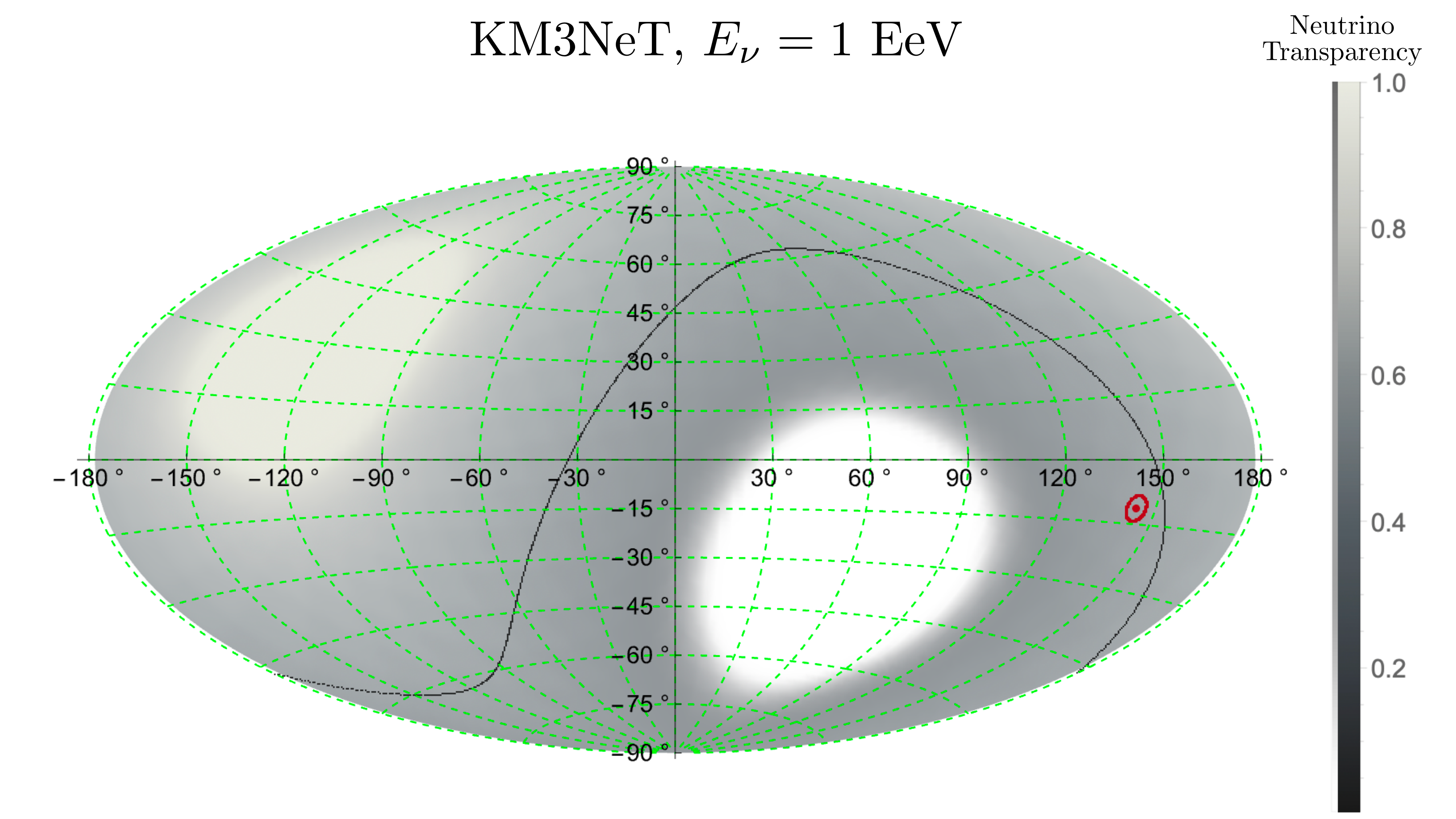}
\caption{Neutrino transparency at $E_\nu = 1$ EeV as a function of galactic latitude and longitude, for \ic{} (top) and for \kmn{} (bottom). The burgundy circle indicates the source inferred by Ref.~\cite{KM3NeT:2025npi}, while the thin black line shows the Earth equatorial plane projection. Translation to galactic coordinates has been obtained with AstroPy~\cite{astropy:2022}.}
\label{fig:Galactic:transparency}
\end{figure}

Crucially, such a source will be seen differently by \ic{} and \kmn{} during the day, due to the rotation of Earth. \cref{fig:PointSource:DistVSTime} shows the distance traveled inside the Earth by a neutrino/muon to reach each detector, as a function of the time of the day. \ic{} is located near the South Pole, thus the inclination with respect to the PS is constant with a very good approximation, as shown by the orange lines in \cref{fig:PointSource:DistVSTime}. Conversely, the \kmn{} field of view (burgundy line) changes during the day and for around 12 hours the source is located at zenith angles larger than $\pi/2$ (i.e. $d\gg\SI{1000}{\kilo\metre}$ of equivalent water). The average distance seen along the LOS by the two experiments is
\begin{equation}\label{eq:lunghezze}
    \left\langle L\right\rangle_{\rm KM}=\left(3769^{\scriptstyle + 134}_{\scriptstyle - 137}\right)\mathrm{km},\quad\quad \left\langle L\right\rangle_{\rm IC}=\left(14^{\scriptstyle + 5}_{\scriptstyle - 3}\right) \mathrm{km}\,.
\end{equation}
The effect of the rotation on the rate of muon events can also be understood by computing the day-averaged neutrino transparency as a function of the incoming direction. For a neutrino with energy $E_\nu = 1$ EeV, the mean free path in water is $\lambda_\nu (E_\nu) \sim 1.7\times10^3$ km (fixing $\lambda = 0.4$), thus we expect \kmn{} to be effectively blind to such neutrinos for almost half of the day, whereas \ic{} notices no significant attenuation. This is explicit in \cref{fig:Galactic:transparency}, where the top and bottom panels show the day-averaged neutrino transparency for \ic{} and \kmn{}, respectively, projected in galactic coordinates. At UHE, \ic{} is sensitive to sources located in the southern sky, where also the \eventname{} source lies (red circle). Conversely, \kmn{} scans most of the sky during the day, albeit with larger attenuation (smaller transparency). We estimate the day-averaged transparency factors as
\be\label{attenuazioni_24}
\quad\langle e^{-L/L_\nu}\rangle_{\rm IC}\approx 0.99,\quad \quad\langle e^{-L/L_\nu}\rangle_{\rm KM}\approx 0.48\,,
\ee
for $E_\nu=1$ EeV, see also \cref{fig:Galactic:transparency}. Given the above discussion, we expect the ratio of events predicted at the two experiments to be even larger than in the case of a diffuse source. For PSs the approximate expression of the ratio depends indeed on the transparency factor, roughly
\be\label{eq:ratio_point}
\Delta_{\text{PS}}\approx \frac{[T A]_{\rm IC}}{[T A]_{\rm KM}}\times \frac{\langle e^{-L/L_\nu}\rangle_{\rm PS, IC}}{\langle e^{-L/L_\nu}\rangle_{\rm PS, KM}}\times \frac{\left(R_{\rm det}+\frac{1}{b_\mu}\right)_{\rm IC}}{\left(R_{\rm det}+\frac{1}{b_\mu}\right)_{\rm KM}}\,,
\ee
which now differs from \cref{eq:ratio_diff}. Note that we omitted $\mathcal{O}(1)$ factors accompanying $b_\mu$, which depend on the specific choice of the energy spectrum of the flux. Numerically, we obtain a ratio $\Delta_{\rm PS} \simeq 131 \approx 2 \, \Delta_{\rm diffuse}$.

\begin{figure}[t]
    \centering
    \includegraphics[width=\textwidth]{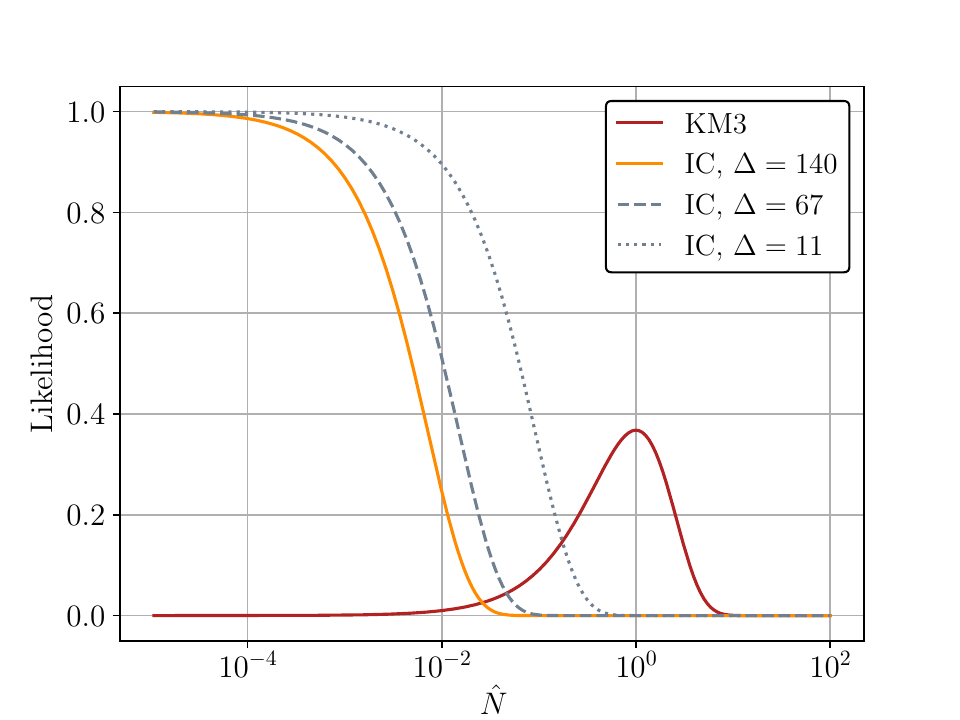}
    \caption{\label{fig:ps-fit}Likelihoods \cref{eq:likelihoods-ps} for \kmn{} and \ic{} for three values of the ratio $\Delta$ of expected number of events, as in \cref{tab:summary}. The fact that the overlap integral is increasing with decreasing $\Delta$ eventually translates into a decreasing tension in the same limit.}
\end{figure}

In order to assess the tension or compatibility between the two datasets, we consider only the UHE bin $\left[10\,\PeV, 400\,\PeV\right]$, in which \kmn{} observed one event, and \ic{} zero. Note that with only one observation in one experiment and zero in the other, the choice of the $0$-event bin is somewhat ambiguous. Where new data should be found at these energies (by either experiment or even a third one), one could refine the binning, gaining new information on the energy dependence of the putative source. In this case, we can take as stochastic variable just the expected number of events at \kmn{} in this bin originating from a PS, however it is computed as a function of the spectrum parameters. We dub it $\hat{N}$. We assume that, in the same bin, the expected number of events at \ic{} is $\Delta$ times larger, and write the following likelihoods
\beq\label{eq:likelihoods-ps}
\mathcal{L}_{\rm IC,0}(\hat{N}) = e^{- \Delta \times \hat{N}} \quad\text{and}\quad \mathcal{L}_{\rm KM3, 1} (\hat{N}) = \hat{N} e^{- \hat{N}}\,.
\eeq
We take a flat prior on $\hat{N}$ spanning over several orders of magnitude. The ratio of expected number of events $\Delta$ may be computed for any specific model. In particular note that while we are speaking now of a PS, the same language can apply for sources of any nature, if their spectra are localized in energy and only produce events in the UHE bin.

Unlike in the power-law diffuse analysis, when we look at sources whose spectrum is localized in energy, it is unclear a priori whether the best fit of \ic{} 0-event bin alone, or the best fit of \kmn{} alone is the best model to be prioritized.
We then compute two sets of Bayes factors: in one case we use the likelihood of \kmn{} and use the \ic{} 0-event bin fit as reference; in the other, we use the likelihood of \ic{} and use \kmn{} best fit as the reference model. In both cases, we take as priors the corresponding likelihoods normalized over the full parameter space, viz.
\beq\label{eq:normalized-likelihoods}
\pi_i = \frac{\mathcal{L}_i}{\int \mathcal{L}_i(\hat{N})\,d\hat{N}}\,.
\eeq
This is a quick way to have priors that are inspired by the maximum likelihood values. If $D_1$ refers to the \ic{} dataset and $D_2$ to the \kmn{} dataset, plugging \cref{eq:normalized-likelihoods} in \cref{eq:bayes-D1D2} we get 
\begin{equation}\label{eq:B-Btilda}
    B_{D_1/D_2} = B = \frac{\left(1+\Delta\right)^2}{4 \Delta} \sim \frac{\Delta}{4}\,,\quad\quad
B_{D_2/D_1} = \widetilde{B} = \frac{\left(1+\Delta\right)^2}{2}\sim{\frac{\Delta^2}{2}}\,,
\end{equation}
where in the last step we assumed $\Delta \gg 1$. We see that $B$ grows linearly with $\Delta$ as expected, while $\widetilde{B}$ grows quadratically with it. In both cases, then, the tension is more severe for larger values of $\Delta$, as to be expected. Furthermore, the tension is milder when the model is inspired by the fit to \ic{}.

\cref{fig:ps-fit} shows the likelihoods \cref{eq:likelihoods-ps} for \kmn{} and \ic{} for three values of $\Delta$. Smaller $\Delta$s result in larger overlaps between the likelihoods, leading to better agreement, as indicated by the Bayes factors.

A steadily emitting point source yields $\Delta \approx 140$, which is not able to reduce the tension between \ic{} and \kmn{} compared to the case of a diffuse source. This is in disagreement with \cite{titans}, in which it is suggested that a steady PS may partially alleviate the tension. We find increasing tension using the criterion in \cref{eq:bayes-D1D2}, while in \cite{titans} a different approach is used, namely, the tension is quantified by comparing with a hypothetical dataset of zero events at both IceCube and KM3NeT (see Sec. III.C of Ref.~\cite{titans}).

A diffuse UHE neutrino source, with a spectrum localized in energy, due for instance to dark matter decay~\cite{Borah:2025igh,Kohri:2025bsn,Jho:2025gaf}, gives $\Delta \approx 67$, showing similar tension to the case of the diffuse flux discussed in \cref{sec:Diffuse}. A transient PS emitting throughout the exposure of \kmn{}, like the one considered in \cite{Shimoda:2024qzw}, gives $\Delta \approx 12$, reducing the tension by a factor of about 12 compared to the steady point source. \cref{tab:summary} summarizes these cases and their Bayes factor values.

\subsection{Point source of \texorpdfstring{$\tau$}{tau}-neutrinos}\label{sec:PS-tau}
In this section, we consider a source emitting only $\tau$-neutrinos. It is intriguing that tau-induced events from a point source could be favored at \kmn{} compared to \ic{}, given the longer decay volume at \kmn{}. The number of tau-induced muons expected from directions with negligible neutrino attenuation scales as $1 - \exp(-L/d_\tau)$, where $d_\tau$ is the typical decay length of the $\tau$.
For energies around 1 EeV, where $d_\tau \approx 50$ km, the expected number of events from tau neutrinos at \kmn{} will be roughly unsuppressed since $L_{\rm KM} \gg d_\tau$. Meanwhile, the expected number of events at \ic{} will be suppressed as $N_\tau^{\rm IC} \propto L_{\rm IC}/d_\tau$, since $L_{\rm IC} \ll d_\tau$ (see \cref{fig:PointSource:DistVSTime}). As expected from \cref{fig:PointSource:DistVSTime}, \kmn{} has a larger $L$ during some parts of the day.

We entertain the idea that the neutrino flux at the Earth's crust from the \eventname{} point source consists purely of tau neutrinos. While producing a flavor-pure neutrino beam from an extragalactic source seems unlikely, it is tempting to consider a UV-dominated energy spectrum to reduce the factor $L_{\rm IC}/d_\tau$ while keeping $N_\tau^{\rm KM}$ large. However, accounting for tau energy loss complicates the picture. As shown in \cref{fig:RelevantScales}, at the energies of interest, $b_\tau^{-1}$ is nearly constant, while $d_\tau$ grows linearly. For $d_\tau < b_\tau^{-1}$, the tau does not lose energy in the medium but must be produced close enough to decay in the detector. Conversely, for $d_\tau > b_\tau^{-1}$, the tau is stable over longer distances, but if produced too far from the detector, it dissipates energy before decaying.
The smallest scale between $b_\tau^{-1}$ and $d_\tau$ sets the distance from the detector where muon events are generated. Since this scale can never be much larger than $L_{\rm IC}$, the na\"ively expected gain of \kmn{} over \ic{} from $\nu_\tau$-induced events is effectively lost.

For example, considering a monochromatic flux, the ratio of expected events at the two experiments is:
\begin{equation}
\Delta_{{\rm PS}\tau} \simeq 70,~80,~158,~250 ~~~~{\rm for} ~~~~ E_\nu = 6\,\PeV, ~20\,\PeV, ~200\,\PeV, ~2~{\rm EeV}\,.    
\end{equation}
In the context of a $\nu_\tau$-dominated flux, the only way to mildly reconcile the observations of \ic{} and \kmn{} would again be with a transient source.

\subsection{Point source of particles beyond the Standard Model}\label{sec:BSM}
Several attempts have been made to address the tension between \kmn{} and \ic{}, or to use \eventname{} to constrain BSM physics ~\cite{Boccia:2025hpm,Borah:2025igh,Brdar:2025azm,Narita:2025udw,Jho:2025gaf,Barman:2025,Murase:2025uwv,Khan:2025gxs,He:2025bex,Zhang:2025rqh,Baker:2025cff,Farzan:2025ydi,Dev:2025czz,Airoldi:2025bgr,Airoldi:2025opo, Zantedeschi2025, mondol2025}. No compelling BSM explanation has emerged so far. Our Standard Model-based analysis shows no clear geographic or local origin for the discrepancy. Motivated by the observation in \cref{sec:PointSource} and \cref{sec:PS-tau}, one possible direction is to find a way to make length scales of order $\mathcal{O}(10^2)$ km special. In the following, we briefly explore a BSM scenario to accomplish the latter, by building upon the intuition that tau leptons are able to propagate for long distances. Namely, we expect to have a preference for \kmn{} in all new physics scenarios where the lifespan $d_X$ of the decaying particle $X$ (barring energy losses, that is, $b_X\to 0$) is in the range
\be
\langle L\rangle_{\rm IC}\ll d_X \ll \langle L\rangle_{\rm KM}\,,
\ee
where the average depths of the experiments during the day are those of \cref{eq:lunghezze}.

We discuss two possibilities, both based on the idea of neutrino portal scenarios (see \cite{Abdullahi:2022jlv} for a status report). We extend the SM with one or more sterile neutrinos $N$, coupled to the $\tau$ lepton doublet, $ \sum_i \lambda_i L_3 H N_i - M_i N_i N_i+ \text{h.c}$. Upon EW symmetry breaking the left-handed $\nu_\tau$ mixes with the heavy neutral leptons $N_i$. This mixing has several effects, among which we focus on new $Z$-mediated interactions between the new states $X_i$, the mass eigenstates of the sterile neutrinos, and the $\nu_\tau$, as well as between the new states themselves. This will manifest itself in a modification of the invisible decay of the $Z$, which is bounded to the permille level.

First, we discuss the scenario in which only one new heavy neutral lepton is present. In this case a $\nu_\tau$ impinging on the surface of Earth has a non-zero probability of converting into $X$ through DIS with the nucleons. If we call $U_{\tau X}$ the respective mixing angle, the production rate scales as $|U_{\tau X}|^2 \sigma_{\rm NC}(E_\nu)$. For the same reason the new state decays to muons with a width similar to the one of the $\tau$, but rescaled by mixing and mass, $d_X(E)=|U_{\tau X}|^{-2}d_\tau(E)(m_\tau/m_X)^6$. The situation is completely similar to muon production from $\tau$-decay as in \cref{sec:tau-master} and \cref{sec:PS-tau}, with a very important difference. Indeed, given the extremely suppressed interaction with matter, $X$ will not lose energy when travelling inside Earth, realizing a scenario with $b_X=0$. This opens up the possibility -- when $d_X \gg \langle L_{\rm IC} \rangle$ as per \cref{eq:lunghezze} -- to explain an expected number of events at \kmn{} that is enhanced with respect to \ic{}, initiated by decays of $X$. We need at least $d_X\gtrsim 10^3$ km, which corresponds to $M_X \lesssim 100\, \mathrm{MeV} |U_{\tau X}/0.001|^{-1/3}$, to have a ratio $\Delta_X\approx O(1)$ between the two experiments. However, since the incoming flux of $\nu_\tau$ also contributes to $\tau$-decays, the BSM flux has to be compared with the one generated by the $\tau$ decays. We immediately see that the rate of produced taus is extremely large compared to the one of BSM particles because of the small mixing  $10^6 \sigma_{\rm CC}/\sigma_{\rm NC}\times|U_{\tau_X}/0.001|^{-2}$ in favor of $\tau$-decays. Concretely, for $m_X\approx 300\,\rm{MeV}$ CHARM bounds~\cite{CHARM:1985anb,Boiarska:2021yho} constrain $U_{\tau_X}<10^{-4}$, suppressing the BSM yield even further. Therefore, we expect few events in absolute magnitude from $X$-decay channel. 

Second, we consider a scenario with two new neutral states coupled to the $\tau$-doublet. We assume a hierarchical mass structure, such that one behaves as a very light sterile neutrino $\nu_X$, oscillating into active ones, and the other as an unstable heavy neutral lepton, $X$, which decays to muons via mixing, as per the first scenario considered above. To enhance the rate at \kmn{} and suppress it at \ic{}, we require again that $\langle L\rangle_{\rm IC}/d_X \ll 1$. However, in this scenario the overall number of events is instead dependent on the impinging flux of sterile neutrinos, $\phi_{\nu_X}^\oplus$, as $X$ is produced by their up-scattering. That is, we are effectively separating production and decay mechanisms. This channel will then be dominating the rate of muon events only if $\phi_{\nu X}^\oplus \gg \phi_{\nu_{\mu,\tau} }^\oplus\, \sigma_{\rm CC}/\sigma_{X}$. In a realistic astrophysical flux, where we expect the flavor composition of the flux at Earth to be democratic~\cite{Athar:2000yw}, $\nu_e:\nu_\mu:\nu_\tau:\nu_X=\frac{1}{4}:\frac{1}{4}:\frac{1}{4}:\frac{1}{4}$, the BSM rate will be suppressed compared to the SM one, making the former contribution negligible in the expected ratio of events at \kmn{} versus \ic. We believe that this challenge we find for realistic fluxes might be generic and not related to the specific model described here.

\section{Summary and outlook}
\label{sec:Conclusions}

In this work, we propose a strategy for making predictions for neutrino telescopes starting from first principles, by solving the Boltzmann equations for the distribution of muons at the location of the detector. We account for muons produced both from the up-scattering of particles such as muon neutrinos and from the decay of heavier particles, such as the tau lepton. Within this approach, it is straightforward to link the expected number of events to the parameters of the astrophysical flux (e.g. the normalization and the spectral index), as well as parameters of the particle physics process responsible for the production of muons and the attenuation of neutrinos. Notably, the latter suffers from large theoretical uncertainties in the UHE regime. With the approach outlined in this work, one can treat nuisance parameters in a straightforward fashion, or even infer their values. We provide an overview of the theoretical inputs, and their respective uncertainties, necessary to perform the computation of the expected number of events in a neutrino telescope. 

As a case study, we consider the recent observation of a UHE neutrino event at \kmn{}. This  has sparked interest in the community, as it has been suggested that it might be in tension with the non-observation of neutrinos of similarly high energy at \ic{}. We focused on the case of a diffuse source, as well as on localized point sources. The results of the comparison of the two experiments are summarized in \cref{tab:summary}. Indeed we find, in the case of a diffuse source with power-law flux, a tension between the two datasets that is compatible with Ref.~\cite{titans}. In the case of an energy-localized diffuse source or a point-source, the tension is somewhat stronger, while a point source of transient nature may alleviate the tension. \revnew{A word of caution is in order: recall that we are only considering one type of event, namely muon tracks. Other types include NC-initiated events from all neutrino flavors, as well as CC-initiated events from electron and tau neutrinos. Since both take place inside the detector, the effective volume coincides with the instrumented volume, and \ic{} is more sensitive to them than \kmn{}, at least until \kmn{} is fully instrumented. Thus, if one were to include them, one would find a larger tension (see \cite{titans}): in this sense the tensions we quote are to be taken as underestimations. In place of \cref{eq:effectiveradius}, the effective radius for events of all kinds should read, roughly
\beq
\left(R_{\rm det} + 1/b_\mu\right) + 3 \frac{\sigma^{\rm NC}}{\sigma^{\rm CC}} R_{\rm det} + 2 R_{\rm det}\,,
\eeq 
meaning that the expected total number of events in each experiment is larger than the expected number of muon tracks by a factor of order one. Note that this very simple estimate is in qualitative agreement with the \ic{} all-flavor effective area being about twice as large as the muon effective area (see, for instance, \cite{Abbasi_2025}). Correspondingly, the ratio $\Delta$ of expected number events at \ic{} versus \kmn{} increases by about $10\%$ in favor of \ic{}. In the case of a PS, by employing \cref{eq:B-Btilda,eq:DeltaSigma}, we estimate an increase of a few percent in the significance of the tension. In the case of a diffuse source, a precise assessment would require the inclusion of these event topologies in our formalism, and performing the fit to an extended dataset, which is beyond the scope of our work. We expect, however, a similar result.
}

\rev{Finally, let us mention that there are limits on the flux of UHE neutrinos
from the Pierre Auger Observatory (PAO) null results~\cite{PierreAuger:2023pjg}. We would arrive at similar conclusions regarding the tension between KM3NeT and PAO in the case of a diffuse flux, since the latter pose a
bound similar to that of IC (as can be seen also in \cite{KM3NeT:2025npi}), while in the case of a PS, on account of the limited exposure of PAO due to the Earth rotation and its reduced angular
sensitivity, no significant tension would arise.}

\begin{table}[h!]
\centering
\scalebox{1.1}{
\begin{tabular}{l c c c c c}
\toprule
\rowcolor{white}
\textbf{Model} & $\Delta_{\rm IC/KM}$ & \textbf{$B$ factor} & $ \Delta\sigma$  & \textbf{$\tilde{B}$ factor} & ${ \Delta\sigma}$ \\
\midrule
\rowcolor{white}
Diffuse (power-law) & 70 & \cellcolor{yellow!20}$21$ & \cellcolor{yellow!20}$3.1$ & -- & --\\
\midrule

\rowcolor{rowgray}
PS & 140 & \cellcolor{red!20}$\approx 35$ & \cellcolor{red!20}$3.8$ & \cellcolor{red!20}$\approx 10^4$ & \cellcolor{red!20}$\gtrsim 5$  \\

\rowcolor{rowgray} 
PS (transient) & 12 &\cellcolor{green!40} $\approx 2.2$ & \cellcolor{green!40}$1.6$ & \cellcolor{red!20}$\approx 2.5\times 10^3$ & \cellcolor{red!20}$\gtrsim 6$ \\

\rowcolor{rowgray}
Diffuse (energy localized) & 70 & \cellcolor{yellow!20}$\approx 3.4$ & \cellcolor{yellow!20}$2.4$ & \cellcolor{red!20}$\approx 85$ & \cellcolor{red!20}$4.2$\\
\bottomrule
\end{tabular}
}
\caption{\label{tab:summary} Comparison between models. The first (white) cell shows the fit to the SM diffuse power-law. The (gray) second cell shows the fit to energy-localized sources (point-like and diffuse). For each model we report the average value of the ratio of events $\Delta_{\rm IC/KM}$ in the UHE bin. Here $B$ is the Bayes factor using \kmn{} data, prioritized on \ic{} 0-event bin best fit, while $\tilde B$ is the Bayes factor using \ic{} 0-event bin, prioritized on the \kmn{} best fit (see \cref{eq:B-Btilda}). For this reason $\tilde B$ only applies to energy-localized models of the second cell.
The color coding is inspired by \cite{Jeffrey}, with: green, ``not worth more than a bare mention'' ; yellow, ``substantial/strong'' ; red, ``very strong/decisive''.}
\end{table}

We believe that our approach is complementary to the dedicated Monte Carlo simulations that are typically employed for the predictions at neutrino telescopes, and is useful for efficiently including both theoretical and systematic uncertainties, by virtue of its simplicity: all predictions are made from just one integral along the line of sight. There are several directions along which our analysis may be improved, in preparation for the next decade of physics at ultra high energy neutrino telescopes. It is straightforward, although rather lengthy, to include the full geometric and environmental effects in a numerical solution to \cref{eq:full-solution}, including the shape of the detector, its efficiency, or more sophisticated models of the Earth's composition.
Moreover, the formalism allows for the inclusion of new physics effects, as we briefly outline in \cref{sec:BSM}.

As far as the tension between \kmn{} and \ic{}, our results imply that only the future will tell: the best way to reconcile the two datasets within the SM is to recognize that the tension between \ic{} and \kmn{} is reducing as time passes, as we write and as you read. Any other detection of ultra high energy events at \kmn{} with no counterpart at \ic{} will point towards the existence of new physics - of a rather unexpected form - as the explanation of an even growing tension. We find both possibilities to be extremely exciting.

\subsubsection*{Acknowledgments}
{\small We thank Michele Redi, Marko Simonovi\'c and Manuel Szewc for discussions. We would also like to thank the anonymous Referee for carefully reading our manuscript. Their insightful feedback gave us the opportunity to strengthen our comprehension of the subject. The work of AT and SP is
supported by the Italian Ministry of University and Research
(MUR) through the PRIN 2022 project n. 20228WHTYC (CUP:I53C24002320006). The work of DR is supported in part by the European Union-Next Generation EU through the PRIN2022 Grant n. 202289JEW4. MT acknowledges support by Next Generation EU, as part of Piano Nazionale di Ripresa e Resilienza (PNRR), Missione 4, Componente 2, Investimento 1.2 - CUP I13C25000150006.}

\begin{appendix}

\section{Kernel functions for the transport equation}\label{app:boltzmann}
The derivation of \cref{eq:transport} goes as follows. Since we are after the evolution in time of the distribution of muons, the collisional term of the transport equation captures the change of muon number per unit volume and phase space induced by the inclusive scattering process $a + N \to \mu + X$ (see main text for notation). Because of this we have
\be
\begin{split}
K^a_{\text{scattering}}(t,\vec x, \vec p_\mu) &= \int \frac{\dd^3p_a}{(2\pi)^3 2 E_a}\frac{\dd^3p_N}{(2\pi)^3 2 E_N} f_a(t, \vec x, \vec p_a) f_N(t, \vec x, \vec p_T)\\
&\times(2\pi)^4 \delta^{(4)}(p_a + p_N - p_{\mu} - p_X) \left|\mathcal{M}_{a N\to \mu X}\right|^2 \frac{\dd^3 p_X}{(2\pi)^3 2 E_X}\frac{1}{2 E_\mu}\,,
\end{split}
\ee
which is the transition probability weighed with the initial distributions at location $\vec x$ and time $t$. We neglect final state Pauli blocking factors.

The transition amplitude $|\mathcal{M}_{a N\to \mu X}|^2$ is integrated upon the full phase space of the final state $X$ (inclusively), while being completely differential in the muon momentum $\vec p$, with the constraints of energy and momentum conservation. It is straightforward to rewrite part of the integrand as the differential muon production cross section $\dd\sigma/\dd^3p$, reducing the above expression to
\be
K^a_{\text{scattering}}(t,\vec x, \vec p_\mu) = \int \frac{\dd^3p_a}{(2\pi)^3}\frac{\dd^3p_N}{(2\pi)^3} f_a(t, \vec x, \vec p_a) f_N(t, \vec x, \vec p_N) \times (2\pi)^3\, \frac{\dd\sigma}{\dd^3p_\mu}\, v\,.
\ee
where $v$ is the relative velocity of $a$ and $N$. At this point, noticing that in the laboratory frame the targets are at rest, we also impose that $f_N(t,\vec x, \vec p_N)=(2\pi)^3\delta^{(3)}(\vec p_N) n_N(\vec x)$, which leads to \cref{eq:transport}.

An analogous derivation holds for $K_{\text{decay}}$, which is given by
\be
K^a_{\text{decay}}(t,\vec x, \vec p_\mu)=\int \frac{\dd^3p_a}{(2\pi)^3 2 E_a} f_a(t,\vec x, \vec p_a) (2\pi)^4 \delta^{(4)}(p_a - p_{\mu} - p_X) \left|\mathcal{M}_{a \to \mu X}\right|^2 \frac{\dd^3p_X}{(2\pi)^3 2E_X}\frac{1}{2E_\mu}\,.
\ee
Also in this case it is possible to express the integrand as the fully differential boosted decay with $\dd\Gamma(\vec p_a)/\dd^3p$, leading to the expression
\be
K^a_{\text{decay}}(t,\vec x, \vec p_\mu)= \int \dd^3p_a f_a(t,\vec x, \vec p_a) \frac{\dd\Gamma_{a\to \mu + X}(\vec p_X)}{\dd^3p_\mu}\,.
\ee

\section{Deep inelastic scattering kinematic}\label{app:kin}
We briefly summarize the key aspects of deep inelastic scattering (DIS) kinematics relevant to various detectors and neutrino experiments. This provides context for \cref{fig:DIS}.    
\begin{itemize}
\item {\bf HERA:} 
The electron-proton scattering process allows a full kinematic reconstruction using the initial electron and proton energies, \( E_e \) and \( E_p \), along with the final energy of the scattered electron \( E_e' \), and its scattering angle relative to the beam direction \( \theta_e \). The standard DIS variables in the laboratory frame are given by  
\begin{equation}
s = 2 E_e E_p\,,\quad Q^2 = 4 E_e' E_e \sin^2\frac{\theta_e}{2}\,,\quad y = 1 - \frac{E_e'}{E_e} \cos^2\frac{\theta_e}{2}\,,\quad x = \frac{Q^2}{y s} \ .\label{eq:xB}
\end{equation}
At HERA, the beam energies are \( E_e = 27.5\,\mathrm{GeV} \) and \( E_p = 920\,\mathrm{GeV} \). The typical acceptance of ZEUS and H1 experiments covers scattering angles roughly between \( 5^\circ \lesssim \theta_e \lesssim 175^\circ \), with a minimal detectable electron energy \( E_e' \gtrsim 5\,\mathrm{GeV} \).  

In the \((Q^2, x)\) plane, the upper range is slightly constrained by the angular acceptance limit, while the lower range is restricted by the forward acceptance.

The minimal normalized \( Q^2 \) can be approximated as  
\begin{equation}
\left.\frac{Q^2}{s}\right\vert_{\mathrm{min}} \approx 2 \times 10^{-5} \left(\frac{920\,\mathrm{GeV}}{E_p}\right)\left(\frac{E_e}{27.5\,\mathrm{GeV}}\right) \left(\frac{E'_{e, \mathrm{min}}}{5\,\mathrm{GeV}}\right) \left(\frac{\theta_{e ,\mathrm{min}}}{5^\circ}\right)^2,    
\end{equation}
which, inserted into \cref{eq:xB}, gives the minimal achievable \( x \) assuming \( y \to 1 \).

\item {\bf LHC:} In hadronic collisions, it is useful to define the transverse momentum \(p_T\) and rapidity \(y\) with respect to the beam axis. These are related to the scattering angle \(\theta\) (measured from the beam direction) as
\be
p_T = p \sin\theta\,, \quad p_{\parallel} = p \cos\theta\,,\quad 
y = \frac{1}{2} \log\left(\frac{E + p_{\parallel}}{E - p_{\parallel}}\right) \simeq -\log\tan\left(\frac{\theta}{2}\right)\,,
\ee
where the last approximation holds for highly relativistic particles (\(m \ll p\)) and defines the pseudorapidity \(\eta\), commonly used at the LHC.

For a generic \(2 \to 2\) process, the partonic momentum fractions \(x_1\) and \(x_2\) determine the hard scale of the interaction through the relation $x_1 x_2=Q^2/s$, where \(s\) is the squared center-of-mass energy, and \(Q^2\) is the momentum transfer.  This scale is related to the transverse momentum and rapidity separation \(\Delta y = y_1 - y_2\) of the final-state particles:
\begin{equation}
Q^2 = 4p_T^2 \cosh^2\left(\frac{\Delta y}{2}\right)\ .
\end{equation}
The minimum accessible \(Q^2\) corresponds to collimated events (\(\Delta y \approx 0\)), giving
\begin{equation}
Q^2_{\text{min}} \approx 4p_T^2|_{\text{min}}\ .
\end{equation}

The parton momentum fractions can also be expressed in terms of the total rapidity \(y_{\text{tot}} = y_1 + y_2\):
\begin{equation}
x_{1,2} = \frac{2Q}{\sqrt{s}} \exp\left(\pm \frac{y_{\text{tot}}}{2}\right)\ .
\end{equation}
This shows that the smallest and largest values of \(x\) are accessed in events with the largest total rapidity \(y_{\text{tot}}\), i.e., those with forward or backward final-state particles.

The kinematically accessible regions differ across LHC experiments. ATLAS and CMS have tracking coverage up to \(|\eta| \lesssim 2.5\) and calorimetry extending to \(|\eta| \lesssim 5\). However, due to triggering limitations, they typically require jets with \(p_T \gtrsim \mathcal{O}(100)\,\mathrm{GeV}\).

In contrast, LHCb is optimized for low-\(p_T\) physics in the forward region, with tracking acceptance in \(2 \lesssim \eta \lesssim 5\) and sensitivity to particles with \(p_T \sim \mathcal{O}(1)\,\mathrm{GeV}\).

Summary of kinematic reach:
\begin{align}
&\text{ATLAS/CMS:} &\quad (100\,\mathrm{GeV})^2 < Q^2 < (13\,\mathrm{TeV})^2\ , &\quad 10^{-4} \lesssim x \lesssim 1\ , \\
&\text{LHCb:}      &\quad (1\,\mathrm{GeV})^2 < Q^2 < (13\,\mathrm{TeV})^2\ , &\quad 10^{-6} \lesssim x \lesssim 1\ .
\end{align}

\item {\bf KM3NET and IceCube}

Large-volume neutrino detectors are optimized to detect high-energy and ultra-high-energy neutrinos, which produce high-energy muons via deep inelastic scattering through charged current interactions. Unlike accelerator-based experiments such as HERA---where both the incoming and outgoing particles are measured---these detectors observe only the energy and direction of the outgoing muon. From this limited information, the original neutrino energy must be inferred by modeling the propagation of a hypothetical neutrino flux through the Earth.

This inference problem is addressed explicitly in \cref{sec:Diffuse} for a diffuse neutrino flux with a power-law energy spectrum, defined in \cref{eq:NeutrinoFlux}. For convenience, we repeat it here (recall that we take $\gamma > 0$):
\begin{equation}
\phi_\nu^\oplus(E_\nu) = \phi_0 \left(10^{18}\,\si{\giga\electronvolt\centi\metre\squared\second\steradian}\right)^{-1} \left( \frac{E_\nu}{E_\ast} \right)^{-\gamma}\,. 
\end{equation}

Using the posterior distribution of the spectral index $\gamma$ from \cref{fig:diffuse-fit}, we compute the corresponding posterior for the average neutrino energy:
\begin{equation}
\langle E_\nu \rangle = \frac{1}{\Phi_\nu^\oplus} \int \dd E_\nu\, E_\nu\, \phi_\nu^\oplus(E_\nu),
\end{equation}
where $\phi_\nu^\oplus$ is the total flux $\Phi_\nu^\oplus=\int \dd E_\nu \phi_\nu^\oplus(E_\nu)$. From this, we derive the center-of-mass energy $s = 2 m_p \langle E_\nu \rangle$ and the mean Bjorken scaling variable $\langle x \rangle = \frac{Q^2}{y s}$, using the average inelasticity $\langle y \rangle$ obtained in \cref{neutrinoxsec}.

The IceCube and KM3NeT bands in \cref{fig:DIS} represent the $95\%$ confidence level contours for $\gamma$ from \cref{fig:diffuse-fit}. Notably, KM3NeT data favor a smaller $\gamma$, indicating a harder spectrum and thus higher average neutrino energies. This shifts the relevant DIS kinematics into a regime not currently probed by existing experiments. In contrast, the IceCube best-fit flux corresponds to a kinematic region well-covered by current data, where parton distribution functions (PDFs) can be reliably extracted from experiments.

\end{itemize}

\end{appendix}

\bibliographystyle{JHEP}
\bibliography{biblio}

\end{document}